\documentclass[reprint]{revtex4-1}

\usepackage{graphicx}
\usepackage{epstopdf}
\usepackage{mathptmx}
\usepackage{verbatim}
\usepackage{graphicx}
\usepackage{amsmath}
\usepackage{amssymb}
\usepackage{amsfonts}
\usepackage{mathrsfs}
\usepackage{amsmath} 
\usepackage{color}
\usepackage{graphicx}
\usepackage{bm} 
\usepackage{amssymb}
\usepackage{xspace}
\usepackage{eucal}
\usepackage{float}
\usepackage{tabu}
\usepackage{textcomp}
\usepackage{anysize}
\usepackage{dsfont}
\usepackage{caption}
\marginsize{2cm}{2cm}{1cm}{1cm}
\usepackage{dcolumn}

\newcommand{\angstrom}{\mbox{\normalfont\AA}}
\newcommand{\mat}[1]{\widehat{\bf{#1}} }

\newcolumntype{L}[1]{>{\raggedright\arraybackslash}m{#1}}
\newcolumntype{C}[1]{>{\centering\arraybackslash}m{#1}}
\newcolumntype{R}[1]{>{\raggedleft\arraybackslash}m{#1}}
\newcolumntype{N}{@{}m{0pt}@{}}

\begin{document}

\title{Dynamic band structure tuning of graphene moir\'e superlattices with pressure}

\author{Matthew Yankowitz$^{1}$}
\author{Jeil Jung$^{2}$}
\author{Evan Laksono$^{3,4}$} 
\author{Nicolas Leconte$^{2}$}
\author{Bheema L. Chittari$^{2}$}
\author{K. Watanabe$^{5}$} 
\author{T. Taniguchi$^{5}$} 
\author{Shaffique Adam$^{3,4,6}$} 
\author{David Graf$^{7}$} 
\author{Cory R. Dean$^{1}$}
\email{dean@phys.columbia.edu}

\affiliation{$^{1}$Department of Physics, Columbia University, New York, NY, USA}
\affiliation{$^{2}$Department of Physics, University of Seoul, Seoul 02504, Korea}
\affiliation{$^{3}$Centre for Advanced 2D Materials, National University of Singapore, Singapore 117546, Singapore}
\affiliation{$^{4}$Deparment of Physics, Faculty of Science, National University of Singapore, Singapore 117542, Singapore}
\affiliation{$^{5}$National Institute for Materials Science, 1-1 Namiki, Tsukuba 305-0044, Japan}
\affiliation{$^{6}$Yale-NUS College, Singapore 138527, Singapore}
\affiliation{$^{7}$National High Magnetic Field Laboratory, Tallahassee, FL, USA}

\date{\today}

\begin{abstract}

Heterostructures of atomically-thin materials have attracted significant interest owing to their ability to host novel electronic properties fundamentally distinct from their constituent layers~\cite{Geim2013}. In the case of graphene on boron nitride, the closely-matched lattices yield a moir\'e superlattice that modifies the graphene electron dispersion~\cite{Yankowitz2012} and opens gaps both at the primary Dirac point (DP) and the moir\'e-induced secondary Dirac point (SDP) in the valence band~\cite{Hunt2013,Woods2014,Wang2015}. While significant effort has focused on controlling the superlattice period via the rotational stacking order~\cite{Ponomarenko2013,Woods2014,Wang2015,Kim2016,Woods2016,Wang2016}, the role played by the magnitude of the interlayer coupling has received comparatively little attention. Here, we modify the interaction between graphene and boron nitride by tuning their separation with hydrostatic pressure.  We observe a dramatic enhancement of the DP gap with increasing pressure, but little change in the SDP gap. Our surprising results identify the critical role played by atomic-scale structural deformations of the graphene lattice and reveal new opportunities for band structure engineering in van der Waals heterostructures.

\end{abstract}

\maketitle

Heterostructures fabricated from the mechanical assembly of atomically-thin van der Waals (vdW) crystals represent an exciting new paradigm in materials design. Owing to weak interlayer bonding, 2D crystals with wide ranging characteristics and composition -- such as graphene, boron nitride (BN), and the transition metal dichalcogenides -- can be readily mixed and matched, without the usual interfacial constraints of conventional crystal growth~\cite{Wang2013}. Moreover, atomic-scale crystalline alignment between the layers often plays a critical role in the resulting device characteristics leading to additional and controllable degrees of freedom. For example, electronic coupling processes that are sensitive to momentum mismatch, such as interlayer tunneling~\cite{Mishchenko2014} or exciton binding~\cite{Yu2015}, can be sensitively tuned by varying the rotational order.  For crystals with closely matched lattice constants, moir\'e interference at zero-angle alignment can additionally result in long-range superlattice potentials, which in turn can lead to entirely new electronic device characteristics~\cite{Yankowitz2012,Ponomarenko2013,Dean2013,Hunt2013,Woods2014,Gorbachev2014,Wang2015,Wang2016,Spanton2017}. 
  
BN-encapsulated graphene provides a model example of the variety of electronic properties that can be realized on-demand in a single type of vdW heterostructure. At large relative twist angles, BN acts as a featureless dielectric for graphene, minimizing coupling to extrinsic disorder but otherwise remaining effectively inert~\cite{Dean2010}. However, at small twist angle, coupling to the resulting moir\'e superlattice (MSL) profoundly alters the graphene bandstructure, giving rise to secondary Dirac cones at finite energy~\cite{Park2008,Yankowitz2012} while also modifying the Fermi velocity near the Dirac point~\cite{Park2008}. As a consequence several unusual electronic properties have been observed, such as density-dependent topological valley currents at zero magnetic field~\cite{Gorbachev2014} and the fractal Hofstadter butterfly spectrum at high field~\cite{Ponomarenko2013,Dean2013,Hunt2013} -- recently identified to host integer and fractional Chern insulating states~\cite{Spanton2017} and charge density waves~\cite{Wang2015,Chen2017}. Additionally, the MSL opens band gaps at both the primary and secondary graphene Dirac points, which are of particular interest for graphene-based digital logic applications.   

\begin{figure*}[ht]
\includegraphics[width=0.9\linewidth]{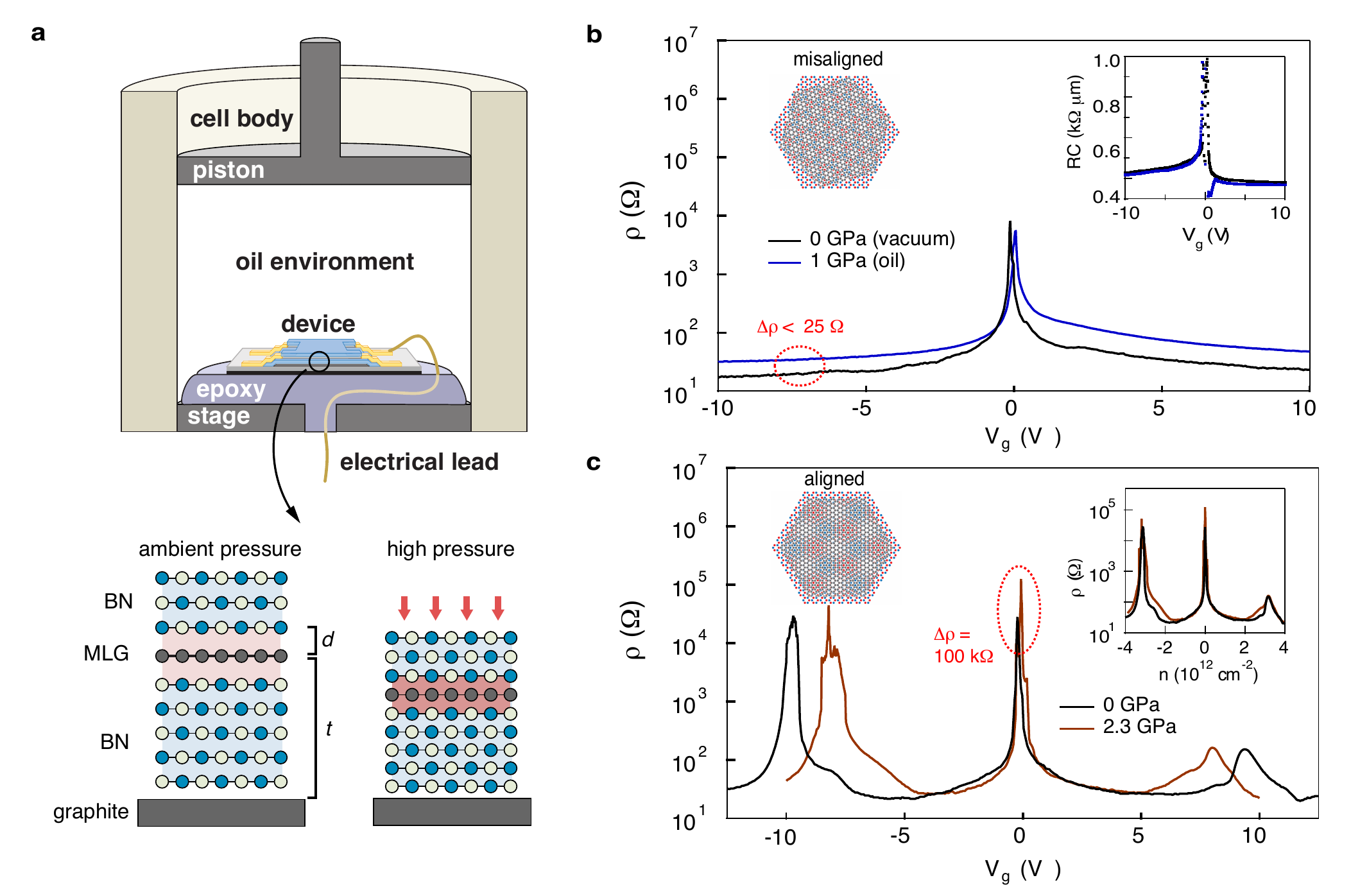} 
\caption{\textbf{| Experiment schematic and transport characterization under pressure.} \textbf{a}, (top) The sample sits inside a Teflon cup filled with oil, and wiring to the sample leaves the cell through an epoxy junction in the stage. The cup sits in the hollowed center of a metal pressure cell cylinder. A piston compresses the Teflon cup, loading the sample environment to the desired hydrostatic pressure. (bottom) Schematic of the sample of graphene encapsulated between two flakes of BN with a graphite gate (the number of atomic layers is reduced for clarity). Each layer sits a characteristic distance $z$ from its neighbor. The distance between the top graphite layer and the graphene $t$ sets the dielectric thickness. On the right, the same structure is shown under pressure, with each layer compressed towards its neighbor, decreasing both $z$ and $t$ (the magnitude of compression is exaggerated for clarity). The hue of the red coloring between the graphene and neighboring BN layers schematically represents the relative strength of the interlayer interaction. \textbf{b}, Resistivity of a misaligned device (Device P2) at $B$ = 0 T and $T$ = 2 K in vacuum and at high pressure (1 GPa) in oil, exhibiting little pressure dependence. (inset) The contact resistance (taken as the difference between the 2-terminal and 4-terminal resistances) also exhibits virtually no dependence on pressure. \textbf{c}, Resistivity of an aligned device (Device P1) at $B$ = 0 T and $T$ = 2 K. The SDPs appear at different gate voltages as a function of pressure, however they are perfectly aligned when plotted against $n$ (inset), indicating the moir\'e period does not change. The DP becomes significantly more insulating with pressure. Cartoon insets in (\textbf{b}) and (\textbf{c}) show a schematic of the moir\'e arising from the relative alignment of the graphene and BN.}
\label{fig:schematic}
\end{figure*}

Numerous techniques have been developed to control the rotational alignment in graphene/BN and related vdW heterostructures~\cite{Ponomarenko2013,Kim2016,Wang2015,Woods2016,Wang2016,Wang2016,Chari2016,Koren2016}. Equally important is the spacing between the layers, which dictates the magnitude of interlayer interactions. However, little  experimental work has been done to characterize or control this parameter, which is often not well-known and considered invariable. Here we demonstrate that by applying hydrostatic pressure to BN-encapsulated graphene, we are able to decrease the interlayer spacing by more than 5\%. At small rotation angles the resulting increase in the effective MSL potential substantially modifies the electronic device characteristics. Most dramatically, we observe a divergence in the moir\'e-induced band gap at the DP with increasing pressure and near doubling over the pressure ranges studied, yielding the largest gap so far demonstrated in pristine monolayer graphene. By contrast, the SDP gap shows little change with pressure. This unexpected result provides new insight into the precise influence of the MSL on the graphene layer and suggests that in addition to electrostatic coupling, lattice-scale deformations play an important role. Our findings reveal that interlayer spacing in vdW heterostructures is both an important and tunable degree of freedom that provides a new route to bandstructure engineering.

\begin{figure*}[t]
\includegraphics[width=0.95\linewidth]{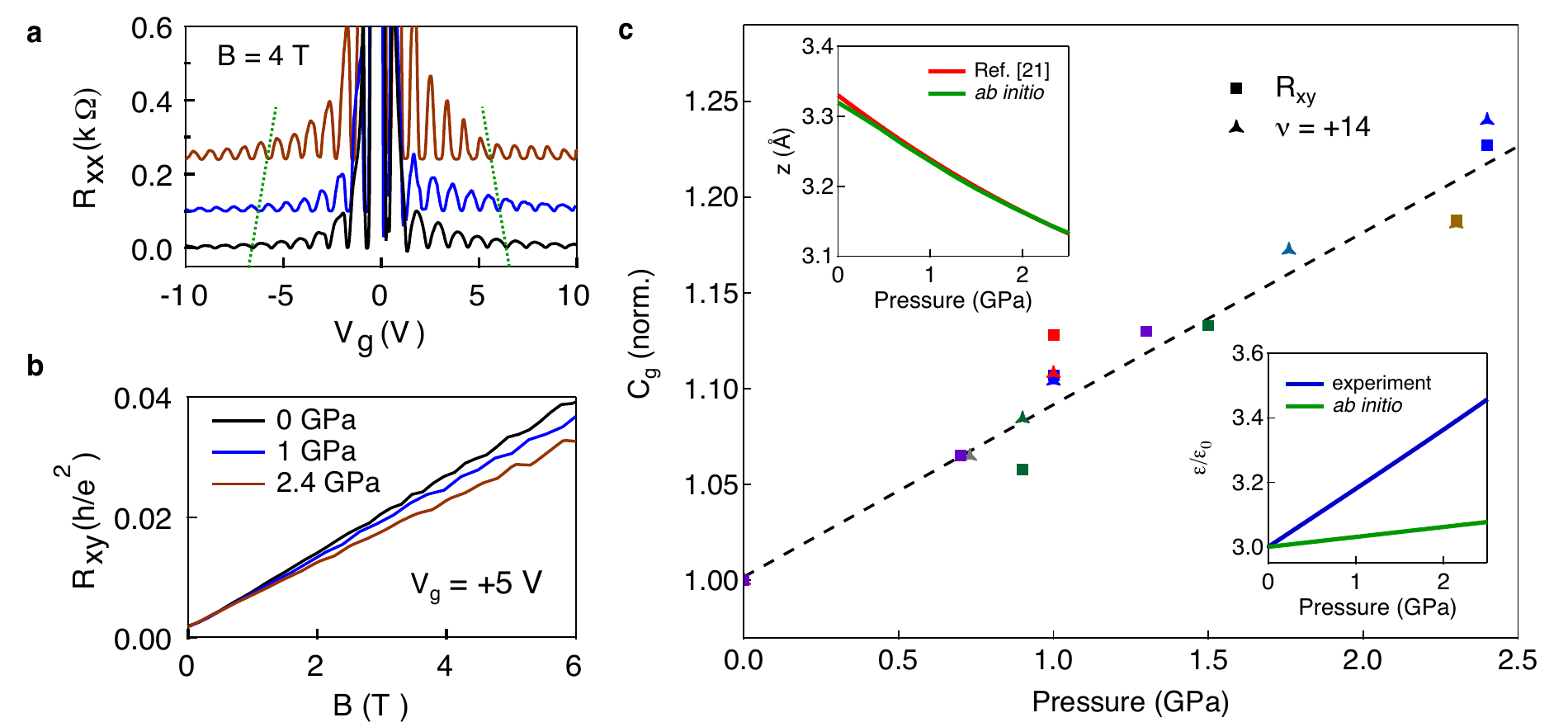}  
\caption{\textbf{| Modification of gate capacitance under pressure.} \textbf{a}, Longitudinal resistance of Device P2 at $B$ = 4 T and $T$ = 2 K. The dotted green dotted lines track the same quantum Hall feature with pressure. Curves are vertically offset proportional to the pressure applied. \textbf{b}, $R_{xy}$ as a function of pressure in the same device at $V_g$ = +5 V. \textbf{c}, Gate capacitance $C_g$ as a function of pressure, normalized to its value at 0 GPa for each device. Colors represent measurements on different devices, both aligned and misaligned. The square markers calculate $C_g$ using $n$ extracted from the low-field $R_{xy}$. The triangle markers use $n$ extracted from the high-field dispersion of $\nu$ = 14. (top inset) Change in BN interlayer spacing per layer as a function of pressure predicted by LDA \textit{ab initio} modeling (green curve). This nearly exactly matches previous measurements by X-ray diffraction in Ref.~\cite{Solozhenko1995} (reproduced in the red curve), assuming an equilibrium $z$ = 3.3 \AA. (bottom inset) The remaining increase in $C_g$ with pressure is attributed to an enhancement of the BN dielectric constant $\epsilon$, assuming $\epsilon$ = 3 at 0 GPa as taken from experiment (blue curve). The green curve shows the LDA \textit{ab initio} modeling of bulk BN.}
\label{fig:capacitance}
\end{figure*}

Fig.~1a shows a cartoon schematic of our experimental setup. We fabricate BN-encapsulated graphene devices using the vdW assembly technique~\cite{Wang2013} and mount them into a piston-cylinder pressure cell with electrical feedthroughs, capable of reaching temperatures below 1 K and magnetic fields above 18 T (see Methods and Supplementary Information 1 for details). The pressure cell sample space is filled with an oil solution which results in a uniform transfer of pressure to the sample. However, since the Young's modulus in the the out-of-plane stacking direction (c-axis) of the vdW crystal is typically a few orders of magnitude smaller than in-plane~\cite{Blakslee1970}, the pressure primarily results in a c-axis compression. We first examine the effect of applying pressure on a misaligned heterostructure where no MSL effects are present, and find that the oil environment and application of pressure have virtually no effect on the electronic properties of the graphene. Fig.~1b shows low temperature transport acquired from a misaligned device ($>$ 2$^\circ$ relative alignment) at zero magnetic field in both vacuum (with no oil) and under high pressure (1 GPa). The high-density resistance grows by small amount (less than 25 $\Omega$) with pressure in this device. Other similar devices showed no measurable change or even slightly decreasing resistance under pressure, indicating the field effect mobility of the encapsulated device is largely insensitive to the application of pressure (see Supplementary Information 2). The DP resistance is also not strongly modified, and there is no significant pressure dependence on the contact resistance (Fig.~1b inset). 

\begin{figure*}[t]
\includegraphics[width=0.95\linewidth]{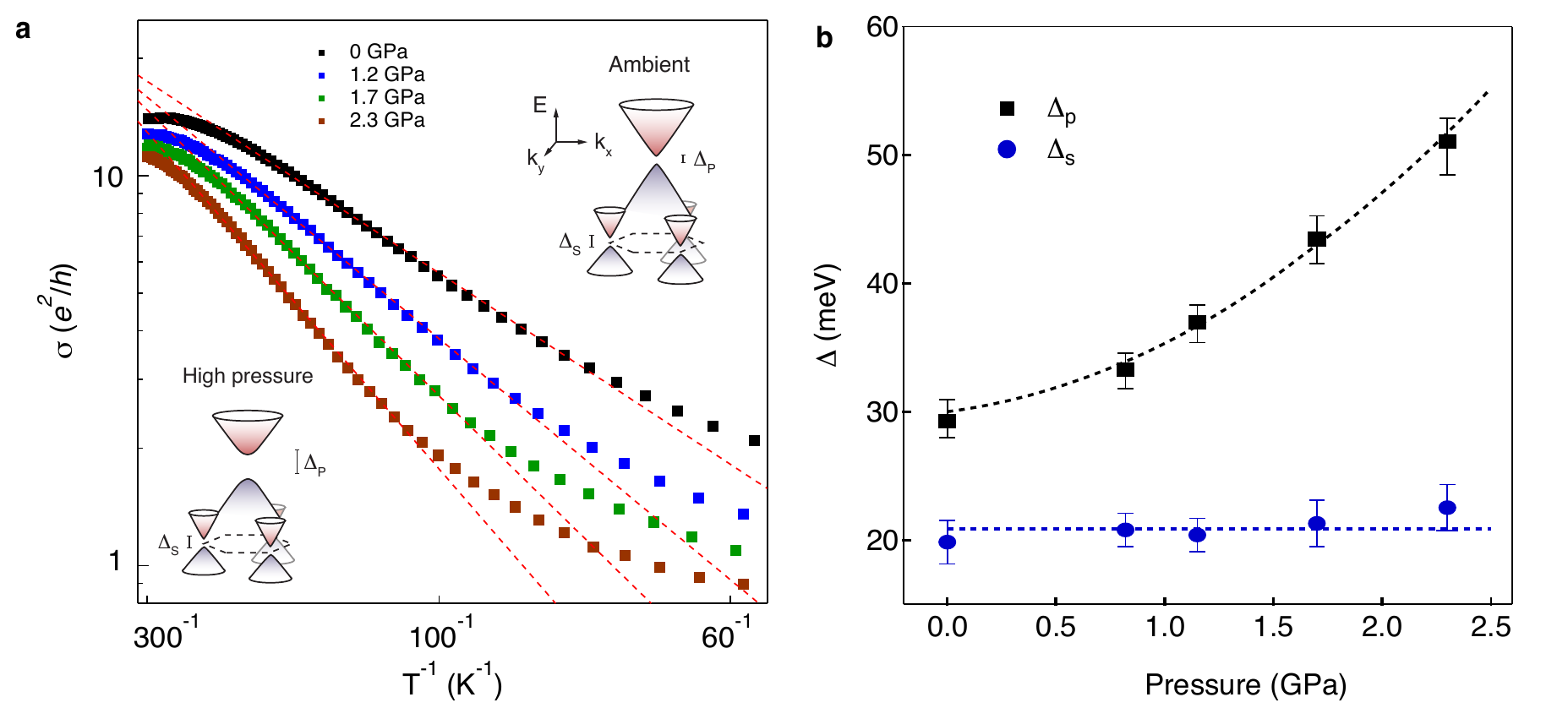} 
\caption{\textbf{| Band gaps as a function of pressure}. \textbf{a}, Arrhenius plot of the conductivity of the DP as a function of inverse temperature for Device P1 at various pressures. The slope of the linear fits (red dotted lines) give the band gap $\Delta$ at each pressure via $\sigma_{DP}(T) \propto e^{-\frac{\Delta}{2 k T}}$. (insets) Sketch of the aligned graphene on BN band structure, indicating $\Delta_p$ and $\Delta_s$ and illustrating their dependence on pressure. Our model predicts SDPs at only one of the two valleys of the superlattice Brillouin zone. \textbf{b}, $\Delta$ as a function of pressure. $\Delta_p$ grows with pressure, while $\Delta_s$ is relatively insensitive to pressure. Error bars in the gaps represent the uncertainty in determining the linear regime for the fit. The uncertainty determining the pressure is smaller than the marker size. The dashed curves show the gaps predicted by the theoretical model using appropriately tuned deformations.}
\label{fig:gaps}
\end{figure*}

Fig.~1c shows similar transport measurements but acquired from an aligned MSL device. Under ambient pressure the device shows excellent transport characteristics with moir\'e coupling evident by the appearance of two resistance peaks symmetrically located about the DP at roughly $\pm$3.3$\times 10^{12}$ cm$^{-2}$, corresponding to a relative rotation angle of $\sim$ 0.6$^\circ$. Moreover, the large resistivities exceeding 25 k$\Omega$ at the DP and SDP suggest sizable band gaps at those points in the band structure. Notably, as we apply high pressure up to 2.3 GPa, we find that the positions of the secondary peaks move symmetrically towards the DP. 

\begin{figure}[h!]
\includegraphics[width=1\linewidth]{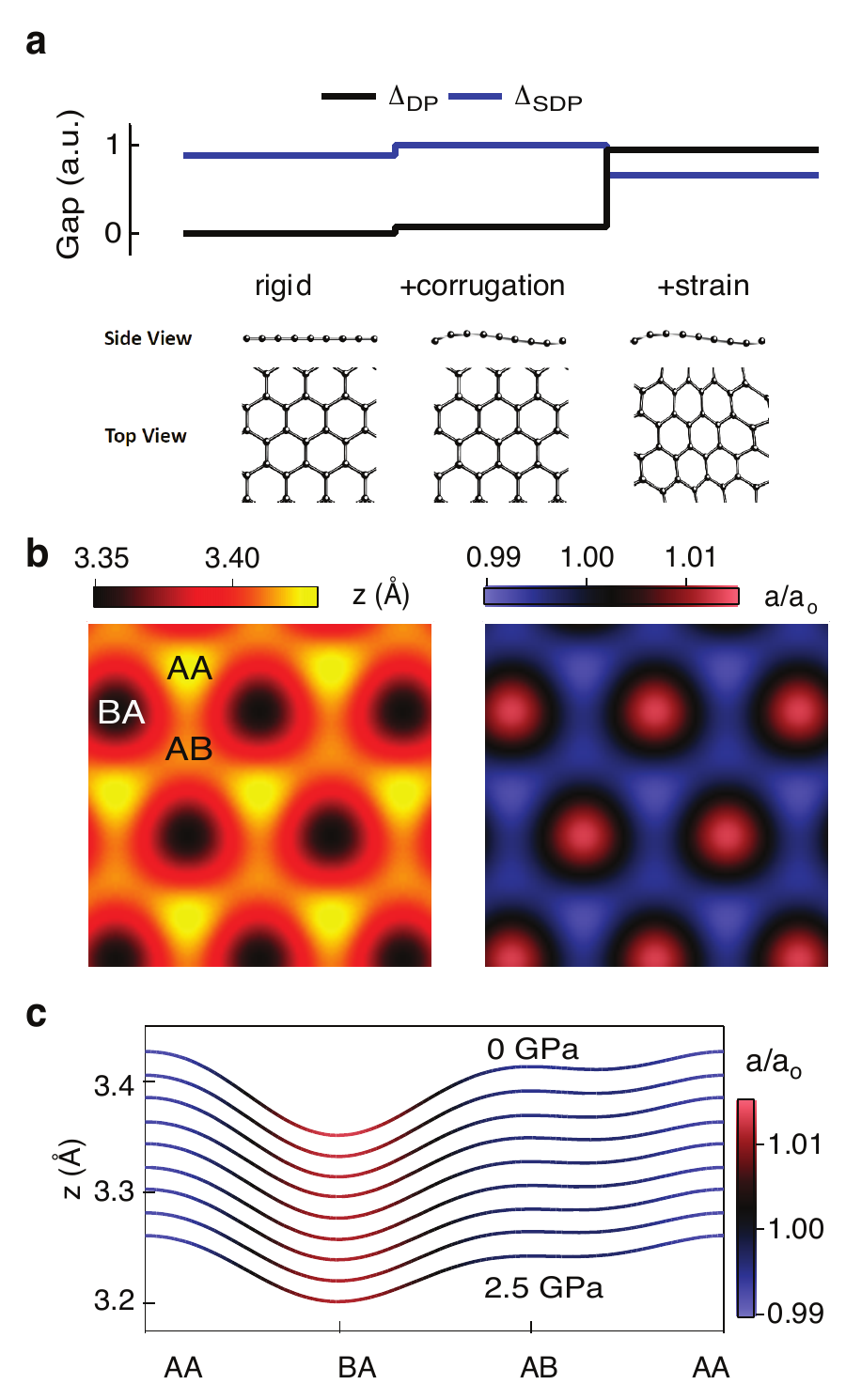} 
\caption{\textbf{| Modeling of the band gaps.} \textbf{a}, Relative contributions to $\Delta_p$ and $\Delta_s$ from lattice corrugations and strains. Strain is necessary to open a sizable $\Delta_p$, while deformations can substantially reduce $\Delta_s$. Schematics of the graphene lattice structure for each case are illustrated below. \textbf{b}, Maps of the out-of-plane corrugations (left) and in-plane lattice strains (top) at 0 GPa used to model the gaps in Fig.~3b. \textbf{c}, Models for lattice corrugations cut through the high symmetry points of the moir\'e as a function of pressure based on \textit{ab initio} calculation inputs in BN-encapsulated graphene with one aligned interface. The color scale represents the magnitude of in-plane lattice strains needed to model the gaps in Fig.~3b.}
\label{fig:gaps}
\end{figure}

To understand this effect, we track the back gate capacitance per unit area $C_g = \frac{en}{V_g} = \frac{\epsilon \epsilon_0}{t}$, where $e$ is the charge of the electron, $\epsilon$ is dielectric constant of BN, $\epsilon_0$ is the vacuum permittivity, $t$ is the thickness of the BN, $V_g$ is the applied gate bias, and $n$ is the carrier density. We measure the density in two ways; both from the Hall effect and magnetoresistance oscillations (Figs.~2a and b, see Methods). Notably, all devices exhibit a universal increase of $C_g$ with pressure of roughly 9\% per GPa independent of their relative alignment (Fig.~2c), which must arise due to a decrease in $t$ and/or an increase in $\epsilon$. To deconvolve the two, we have performed LDA \textit{ab initio} simulations of bulk BN multilayers under pressure and find approximately 2.5\% compression per GPa (green curve in the top inset of Fig.~2c), in remarkable quantitative agreement with previous X-ray diffraction measurements from Ref.~\cite{Solozhenko1995} (reproduced in the red curve). The remaining increase in gate capacitance is therefore attributed to an increase in the dielectric constant of the BN of roughly 6\% per GPa (blue curve in the bottom inset of Fig.~2c). Our \textit{ab initio} simulations also confirm that the BN dielectric constant should grow with pressure, though the effect is predicted to be a few times smaller (green curve). Taken together, this suggests that pressure is able to sensitively tune both the interlayer spacing between layered 2D materials and their dielectric properties~\cite{Chen2017b}. Returning to the transport measurements of the MSL device, we find that the three resistance peaks align exactly at all pressures when replotted against charge carrier density $n$ (Fig.~1c inset), suggesting that while the graphene and BN layers move closer together, the relative rotation angle and moir\'e period remains fixed under pressure.

A second notable feature of the MSL transport is that the resistance at the DP grows strongly with pressure, increasing by roughly 100 k$\Omega$ between 0 GPa and 2.3 GPa. We investigate the DP response in more detail by measuring its temperature dependence. Fig. 3a shows an Arrhenius plot of the DP conductivity, $\sigma_{DP}$, versus inverse temperature for various pressures, where for each pressure a linear fit to the simply activated regime (red dashed lines) gives a measure of the activation gap (see Methods). The resulting DP gap, $\Delta_{p}$, is shown versus pressure in Fig. 3b (square markers). The gap is found to diverge with increasing pressure, and is enhanced by nearly a factor of 2 at the highest pressure studied for the $\sim$ 0.6$^\circ$ device presented here. Similar gap enhancement was observed in other nearly-aligned samples as well, and we find that the order and direction of pressure cycling do not influence the measurement (see Supplementary Figure 3a). We similarly measure the pressure dependence of the valence band SDP gap, $\Delta_{s}$, plotted with circle markers in Fig.~3b. In significant contrast to the DP gap, the SDP gap is nearly unresponsive to pressure. The insets of Fig.~3a schematically illustrate the inferred band structure modifications with interlayer spacing, showing a growing $\Delta_p$ but a fixed $\Delta_s$. 

As a simple approximation, we may consider the increasing DP gap with pressure to result from increasing MSL coupling owing to the decreasing interlayer spacing. In this case, we expect that the SDP behavior might respond in a similar way, and from this perspective its insensitivity to pressure is surprising. We note, however, that despite considerable effort~\cite{Hunt2013,Woods2014,Chen2014,Wang2015,Wang2016, Song2013,Bokdam2014,Moon2014,Wallbank2015,Jung2015,SanJose2014,Slotman2015,Jung2017} consensus is still lacking as to the exact origin of these band gaps. Lattice scale deformations (in-plane strains and out-of-plane corrugations) of the graphene layer are expected to play an important role~\cite{Woods2014,Jung2015,SanJose2014,Slotman2015,Yankowitz2016,Jung2017}, but the exact equilibrium structure of graphene in contact with BN remains poorly understood, including whether these deformations even exist in fully BN-encapsulated devices~\cite{Woods2014}. In an effort to understand our observed behavior, we therefore first consider a rigid graphene lattice and examine the effects of pressure on the heterostructure using a combination of \textit{ab initio} and analytical models (see Supplementary Information 4 - 6). The interlayer electronic coupling between the graphene and BN, $V_0 = \widetilde{V}e^{-\beta (z_{0}-z_{r})}$, is highly sensitive to pressure, with \textit{ab initio} predictions indicating that the average interlayer spacing $z_0$ should decrease by $\sim$0.07 \AA/GPa. Here $\widetilde{V} \approx$ 18 meV is the interlayer electronic coupling, $z_r \approx$ 3.35 \AA~ is the equilibrium average interlayer spacing between the graphene and BN, and $\beta \approx$ 3.2 \AA$^{-1}$ quantifies the rate of increase of interlayer coupling when the spacing is reduced. Generically, both $\Delta_p$ and $\Delta_s$ should scale proportionally to $V_0$ under applied pressure, however this is in stark contrast to our experimental observations where $\Delta_s$ in particular exhibits little pressure dependence. Even at 0 GPa, a rigid model does not properly capture the experimentally observed hierarchy of the gaps, predicting a large $\Delta_s$ and a $\Delta_p \approx 0$ (Fig.~4a).

Next we consider the potential role of atomic-scale graphene deformations. By considering realistic values of the graphene elastic deformation potential, our atomic structure models suggest that it is favorable for the graphene lattice to both corrugate out-of-plane and strain in-plane on the moir\'e scale to minimize the overall stacking potential with the BN substrate (Fig.~4b). These deformations break the sublattice symmetry of the graphene, resulting in a finite average mass term in the Hamiltonian which opens a sizable $\Delta_p$. There are also moir\'e-induced electrostatic potentials and pseudomagnetic fields which contribute to $\Delta_s$ in addition to the mass term. Including corrugations alone opens a small $\Delta_p$ at 0 GPa, but still does not recover the observed gap hierarchy of $\Delta_p > \Delta_s$ (Fig.~4a). By additionally including strains, we are able to recover the observed 0 GPa gap hierarchy (Fig.~4a) as well as the diverging $\Delta_p$ and flat $\Delta_s$ response with pressure. One such example, whose corrugations and strains are obtained for BN-encapsulated graphene by a model partly informed by \textit{ab initio} calculation inputs (see Supplementary Information 4 - 6), is shown in Fig.~4c, and predicts gaps in remarkably good quantitative agreement with our experimental data (dashed curves in Fig.~3b).

While more experimental and theoretical work is necessary to understand the exact equilibrium structure of aligned graphene on BN, the varying evolution of the gaps with pressure directly rule out the possibility of a rigid graphene lattice in these structures, and moreover demonstrates that these gaps are of fundamentally different origin as the SDP cone is not a true replica of the DP cone. This suggests the possibility to independently control the magnitude of the two gaps, as well as other features of the moir\'e band structure and magnetoresponse, by selectively engineering specific lattice deformations. Additionally, sufficiently strong enhancement of the interlayer coupling could drive a phase-transition to the fully commensurate (lattice-matched) stacking configuration~\cite{Yankowitz2016}, marked by the absence of a MSL but the emergence of strong sublattice-symmetry breaking in the graphene and a gap many times room temperature at the Dirac point~\cite{Bokdam2014}. More generally, our results indicate that a wide variety of vdW heterostructure properties may be tunable by controlling the interlayer coupling strength with pressure. For example, the band structures of Bernal-stacked and twisted bilayer graphene depend critically on interlayer hopping terms~\cite{Laissardiere2010}, as do the strength of proximity-induced spin-orbit interactions in graphene resting on heavy transition metal dichalcogenides~\cite{Wang2015b}.

\section*{Methods}

\subsection*{Application of pressure}

In order to control the interlayer spacing in vdW heterostructures, we fabricate graphene devices encapsulated in BN with a graphite bottom gate on Si/SiO$_2$ wafers (Fig.~1a) and make one-dimensional electrical contact using standard reactive ion etching and electron beam patterning and deposition techniques~\cite{Wang2013}. Alignment to one of the encapsulating BN layers is achieved through either optical matching of crystalline edges~\cite{Ponomarenko2013} or through thermal self-rotation~\cite{Wang2015,Woods2016}. We then mount the device on a stage with a pre-wired feed-through and affix wires onto the electrodes by hand using silver paint. The stage is then enclosed by a Teflon cup filled with a hydrostatic pressure medium (Daphne 7373 or 7474 oil)~\cite{Murata1997,Murata2008}, and the cup is fit into the inner bore of a piston-cylinder pressure cell and loaded to the desired pressure using a hydraulic press (Fig.~1a). Finally, the pressure cell is affixed to a probe for electrical characterization at low temperature and high magnetic field (see Supplementary Information 1 for further details of the experimental setup). The pressure was determined by measuring the photoluminescence of a ruby crystal at both room- and low-temperature. 

\subsection*{Extraction of capacitance and band gaps}

We extract $n(V_g)$ for each pressure through either the dispersion of the quantum Hall states in high magnetic field as $n = \frac{\nu e B}{h}$, where $\nu$ is the filling factor (Fig.~2a), or from the low field Hall resistance $R_{xy}$ before the onset of strong Shubnikov-de Haas oscillations as $n = \frac{1}{eR_{xy}}$ (Fig.~2b). Quantum Hall states move symmetrically closer to the DP with increasing pressure, and the slope of the Hall resistance decreases with pressure, both implying a growing $n(V_g)$ with increasing pressure. In Fig.~2c, $C_g$ is normalized to its measured value at 0 GPa to account for the different thicknesses of the bottom BNs across the different devices. Additionally, the \textit{ab initio} value of $\epsilon$ at 0 GPa is normalized to match the average experimental value of $\approx$ 3. 

We extract the band gaps $\Delta$ of the DP and SDP at each pressure according to the thermally activated response where $\sigma_{DP}(T) \propto e^{-\frac{\Delta}{2 k T}}$, where $k$ is the Boltzmann constant.

\section*{Acknowledgements}

We thank Pablo San-Jose, Justin Song, Andrey Shytov, Leonid Levitov, John Wallbank, and Pilkyung Moon for valuable theoretical discussions. This work was supported by the National Science Foundation (DMR-1462383). CRD acknowledges partial support from the David and Lucille Packard foundation. We acknowledge Stan Tozer for use of his 16 T PPMS which is partially supported as part of the Center for Actinide Science and Technology (CAST), an Energy Frontier Research Center (EFRC) funded by the Department of Energy, Office of Science, Basic Energy Sciences under Award Number DE-SC0016568. A portion of this work was performed at the National High Magnetic Field Laboratory, which is supported by National Science Foundation Cooperative Agreement No. DMR-0654118, the State of Florida and the U.S. Department of Energy and additionally provided support for pressure cell development through User Collaboration Grant Program (UCGP) funding. JJ and NL have been supported by the Korean NRF through the grant NRF-2016R1A2B4010105 and the Korean Research Fellowship grant NRF-2016H1D3A1023826, and BLC has been supported by the grant NRF-2017R1D1A1B03035932. EL and SA are supported by the National Research Foundation of Singapore under its Fellowship program (NRF-NRFF2012-01). K.W. and T.T. acknowledge support from the Elemental Strategy Initiative
conducted by the MEXT, Japan and JSPS KAKENHI Grant Numbers JP15K21722.

\section*{Author contributions}

M.Y. and C.R.D. conceived the experiment. M.Y. fabricated the samples, analyzed the data and wrote the paper. M.Y. and D.G. performed the experiments. J.J., E.L. and S.A. developed the theory. N.L. and B.L.C. calculated the \textit{ab initio} potentials. K.W. and T.T. grew the hBN crystals. C.R.D. advised on the experiments.

\bibliography{references}

\begin{thebibliography}{47}%
\makeatletter
\providecommand \@ifxundefined [1]{%
 \@ifx{#1\undefined}
}%
\providecommand \@ifnum [1]{%
 \ifnum #1\expandafter \@firstoftwo
 \else \expandafter \@secondoftwo
 \fi
}%
\providecommand \@ifx [1]{%
 \ifx #1\expandafter \@firstoftwo
 \else \expandafter \@secondoftwo
 \fi
}%
\providecommand \natexlab [1]{#1}%
\providecommand \enquote  [1]{``#1''}%
\providecommand \bibnamefont  [1]{#1}%
\providecommand \bibfnamefont [1]{#1}%
\providecommand \citenamefont [1]{#1}%
\providecommand \href@noop [0]{\@secondoftwo}%
\providecommand \href [0]{\begingroup \@sanitize@url \@href}%
\providecommand \@href[1]{\@@startlink{#1}\@@href}%
\providecommand \@@href[1]{\endgroup#1\@@endlink}%
\providecommand \@sanitize@url [0]{\catcode `\\12\catcode `\$12\catcode
  `\&12\catcode `\#12\catcode `\^12\catcode `\_12\catcode `\%12\relax}%
\providecommand \@@startlink[1]{}%
\providecommand \@@endlink[0]{}%
\providecommand \url  [0]{\begingroup\@sanitize@url \@url }%
\providecommand \@url [1]{\endgroup\@href {#1}{\urlprefix }}%
\providecommand \urlprefix  [0]{URL }%
\providecommand \Eprint [0]{\href }%
\providecommand \doibase [0]{http://dx.doi.org/}%
\providecommand \selectlanguage [0]{\@gobble}%
\providecommand \bibinfo  [0]{\@secondoftwo}%
\providecommand \bibfield  [0]{\@secondoftwo}%
\providecommand \translation [1]{[#1]}%
\providecommand \BibitemOpen [0]{}%
\providecommand \bibitemStop [0]{}%
\providecommand \bibitemNoStop [0]{.\EOS\space}%
\providecommand \EOS [0]{\spacefactor3000\relax}%
\providecommand \BibitemShut  [1]{\csname bibitem#1\endcsname}%
\let\auto@bib@innerbib\@empty
\bibitem [{\citenamefont {Geim}\ and\ \citenamefont
  {Grigorieva}(2013)}]{Geim2013}%
  \BibitemOpen
  \bibfield  {author} {\bibinfo {author} {\bibfnamefont {A.~K.}\ \bibnamefont
  {Geim}}\ and\ \bibinfo {author} {\bibfnamefont {I.~V.}\ \bibnamefont
  {Grigorieva}},\ }\href@noop {} {\bibfield  {journal} {\bibinfo  {journal}
  {Nature}\ }\textbf {\bibinfo {volume} {499}},\ \bibinfo {pages} {419}
  (\bibinfo {year} {2013})}\BibitemShut {NoStop}%
\bibitem [{\citenamefont {Yankowitz}\ \emph {et~al.}(2012)\citenamefont
  {Yankowitz}, \citenamefont {Xue}, \citenamefont {Cormode}, \citenamefont
  {Sanchez-Yamagishi}, \citenamefont {Watanabe}, \citenamefont {Taniguchi},
  \citenamefont {Jarillo-Herrero}, \citenamefont {Jacquod},\ and\ \citenamefont
  {LeRoy}}]{Yankowitz2012}%
  \BibitemOpen
  \bibfield  {author} {\bibinfo {author} {\bibfnamefont {M.}~\bibnamefont
  {Yankowitz}}, \bibinfo {author} {\bibfnamefont {J.}~\bibnamefont {Xue}},
  \bibinfo {author} {\bibfnamefont {D.}~\bibnamefont {Cormode}}, \bibinfo
  {author} {\bibfnamefont {J.~D.}\ \bibnamefont {Sanchez-Yamagishi}}, \bibinfo
  {author} {\bibfnamefont {K.}~\bibnamefont {Watanabe}}, \bibinfo {author}
  {\bibfnamefont {T.}~\bibnamefont {Taniguchi}}, \bibinfo {author}
  {\bibfnamefont {P.}~\bibnamefont {Jarillo-Herrero}}, \bibinfo {author}
  {\bibfnamefont {P.}~\bibnamefont {Jacquod}}, \ and\ \bibinfo {author}
  {\bibfnamefont {B.~J.}\ \bibnamefont {LeRoy}},\ }\href@noop {} {\bibfield
  {journal} {\bibinfo  {journal} {Nature Physics}\ }\textbf {\bibinfo {volume}
  {8}},\ \bibinfo {pages} {382} (\bibinfo {year} {2012})}\BibitemShut {NoStop}%
\bibitem [{\citenamefont {Hunt}\ \emph {et~al.}(2013)\citenamefont {Hunt},
  \citenamefont {Sanchez-Yamagishi}, \citenamefont {Young}, \citenamefont
  {Yankowitz}, \citenamefont {LeRoy}, \citenamefont {Watanabe}, \citenamefont
  {Taniguchi}, \citenamefont {Moon}, \citenamefont {Koshino}, \citenamefont
  {Jarillo-Herrero},\ and\ \citenamefont {Ashoori}}]{Hunt2013}%
  \BibitemOpen
  \bibfield  {author} {\bibinfo {author} {\bibfnamefont {B.}~\bibnamefont
  {Hunt}}, \bibinfo {author} {\bibfnamefont {J.~D.}\ \bibnamefont
  {Sanchez-Yamagishi}}, \bibinfo {author} {\bibfnamefont {A.~F.}\ \bibnamefont
  {Young}}, \bibinfo {author} {\bibfnamefont {M.}~\bibnamefont {Yankowitz}},
  \bibinfo {author} {\bibfnamefont {B.~J.}\ \bibnamefont {LeRoy}}, \bibinfo
  {author} {\bibfnamefont {K.}~\bibnamefont {Watanabe}}, \bibinfo {author}
  {\bibfnamefont {T.}~\bibnamefont {Taniguchi}}, \bibinfo {author}
  {\bibfnamefont {P.}~\bibnamefont {Moon}}, \bibinfo {author} {\bibfnamefont
  {M.}~\bibnamefont {Koshino}}, \bibinfo {author} {\bibfnamefont
  {P.}~\bibnamefont {Jarillo-Herrero}}, \ and\ \bibinfo {author} {\bibfnamefont
  {R.~C.}\ \bibnamefont {Ashoori}},\ }\href@noop {} {\bibfield  {journal}
  {\bibinfo  {journal} {Science}\ }\textbf {\bibinfo {volume} {340}},\ \bibinfo
  {pages} {1427} (\bibinfo {year} {2013})}\BibitemShut {NoStop}%
\bibitem [{\citenamefont {Woods}\ \emph {et~al.}(2014)\citenamefont {Woods},
  \citenamefont {Britnell}, \citenamefont {Eckmann}, \citenamefont {Ma},
  \citenamefont {Lu}, \citenamefont {Guo}, \citenamefont {Lin}, \citenamefont
  {Yu}, \citenamefont {Cao}, \citenamefont {Gorbachev}, \citenamefont
  {Kretinin}, \citenamefont {Park}, \citenamefont {Ponomarenko}, \citenamefont
  {Katsnelson}, \citenamefont {Gornostyrev}, \citenamefont {Watanabe},
  \citenamefont {Taniguchi}, \citenamefont {Casiraghi}, \citenamefont {Gao},
  \citenamefont {Geim},\ and\ \citenamefont {Novoselov}}]{Woods2014}%
  \BibitemOpen
  \bibfield  {author} {\bibinfo {author} {\bibfnamefont {C.~R.}\ \bibnamefont
  {Woods}}, \bibinfo {author} {\bibfnamefont {L.}~\bibnamefont {Britnell}},
  \bibinfo {author} {\bibfnamefont {A.}~\bibnamefont {Eckmann}}, \bibinfo
  {author} {\bibfnamefont {R.~S.}\ \bibnamefont {Ma}}, \bibinfo {author}
  {\bibfnamefont {J.~C.}\ \bibnamefont {Lu}}, \bibinfo {author} {\bibfnamefont
  {H.~M.}\ \bibnamefont {Guo}}, \bibinfo {author} {\bibfnamefont
  {X.}~\bibnamefont {Lin}}, \bibinfo {author} {\bibfnamefont {G.~L.}\
  \bibnamefont {Yu}}, \bibinfo {author} {\bibfnamefont {Y.}~\bibnamefont
  {Cao}}, \bibinfo {author} {\bibfnamefont {R.~V.}\ \bibnamefont {Gorbachev}},
  \bibinfo {author} {\bibfnamefont {A.~V.}\ \bibnamefont {Kretinin}}, \bibinfo
  {author} {\bibfnamefont {J.}~\bibnamefont {Park}}, \bibinfo {author}
  {\bibfnamefont {L.~A.}\ \bibnamefont {Ponomarenko}}, \bibinfo {author}
  {\bibfnamefont {M.~I.}\ \bibnamefont {Katsnelson}}, \bibinfo {author}
  {\bibfnamefont {Y.~N.}\ \bibnamefont {Gornostyrev}}, \bibinfo {author}
  {\bibfnamefont {K.}~\bibnamefont {Watanabe}}, \bibinfo {author}
  {\bibfnamefont {T.}~\bibnamefont {Taniguchi}}, \bibinfo {author}
  {\bibfnamefont {C.}~\bibnamefont {Casiraghi}}, \bibinfo {author}
  {\bibfnamefont {H.-J.}\ \bibnamefont {Gao}}, \bibinfo {author} {\bibfnamefont
  {A.~K.}\ \bibnamefont {Geim}}, \ and\ \bibinfo {author} {\bibfnamefont
  {K.~S.}\ \bibnamefont {Novoselov}},\ }\href@noop {} {\bibfield  {journal}
  {\bibinfo  {journal} {Nature Physics}\ }\textbf {\bibinfo {volume} {10}},\
  \bibinfo {pages} {451} (\bibinfo {year} {2014})}\BibitemShut {NoStop}%
\bibitem [{\citenamefont {Wang}\ \emph
  {et~al.}(2015{\natexlab{a}})\citenamefont {Wang}, \citenamefont {Gao},
  \citenamefont {Wen}, \citenamefont {Han}, \citenamefont {Taniguchi},
  \citenamefont {Watanabe}, \citenamefont {Koshino}, \citenamefont {Hone},\
  and\ \citenamefont {Dean}}]{Wang2015}%
  \BibitemOpen
  \bibfield  {author} {\bibinfo {author} {\bibfnamefont {L.}~\bibnamefont
  {Wang}}, \bibinfo {author} {\bibfnamefont {Y.}~\bibnamefont {Gao}}, \bibinfo
  {author} {\bibfnamefont {B.}~\bibnamefont {Wen}}, \bibinfo {author}
  {\bibfnamefont {Z.}~\bibnamefont {Han}}, \bibinfo {author} {\bibfnamefont
  {T.}~\bibnamefont {Taniguchi}}, \bibinfo {author} {\bibfnamefont
  {K.}~\bibnamefont {Watanabe}}, \bibinfo {author} {\bibfnamefont
  {M.}~\bibnamefont {Koshino}}, \bibinfo {author} {\bibfnamefont
  {J.}~\bibnamefont {Hone}}, \ and\ \bibinfo {author} {\bibfnamefont {C.~R.}\
  \bibnamefont {Dean}},\ }\href@noop {} {\bibfield  {journal} {\bibinfo
  {journal} {Science}\ }\textbf {\bibinfo {volume} {350}},\ \bibinfo {pages}
  {1231} (\bibinfo {year} {2015}{\natexlab{a}})}\BibitemShut {NoStop}%
\bibitem [{\citenamefont {Ponomarenko}\ \emph {et~al.}(2013)\citenamefont
  {Ponomarenko}, \citenamefont {Gorbachev}, \citenamefont {Yu}, \citenamefont
  {Elias}, \citenamefont {Jalil}, \citenamefont {Patel}, \citenamefont
  {Mishchenko}, \citenamefont {Mayorov}, \citenamefont {Woods}, \citenamefont
  {Wallbank}, \citenamefont {Mucha-Kruczynski}, \citenamefont {Pio},
  \citenamefont {Potemski}, \citenamefont {Grigorieva}, \citenamefont
  {Novoselov}, \citenamefont {Guinea}, \citenamefont {Fal'ko},\ and\
  \citenamefont {Geim}}]{Ponomarenko2013}%
  \BibitemOpen
  \bibfield  {author} {\bibinfo {author} {\bibfnamefont {L.~A.}\ \bibnamefont
  {Ponomarenko}}, \bibinfo {author} {\bibfnamefont {R.~V.}\ \bibnamefont
  {Gorbachev}}, \bibinfo {author} {\bibfnamefont {G.~L.}\ \bibnamefont {Yu}},
  \bibinfo {author} {\bibfnamefont {D.~C.}\ \bibnamefont {Elias}}, \bibinfo
  {author} {\bibfnamefont {R.}~\bibnamefont {Jalil}}, \bibinfo {author}
  {\bibfnamefont {A.~A.}\ \bibnamefont {Patel}}, \bibinfo {author}
  {\bibfnamefont {A.}~\bibnamefont {Mishchenko}}, \bibinfo {author}
  {\bibfnamefont {A.~S.}\ \bibnamefont {Mayorov}}, \bibinfo {author}
  {\bibfnamefont {C.~R.}\ \bibnamefont {Woods}}, \bibinfo {author}
  {\bibfnamefont {J.~R.}\ \bibnamefont {Wallbank}}, \bibinfo {author}
  {\bibfnamefont {M.}~\bibnamefont {Mucha-Kruczynski}}, \bibinfo {author}
  {\bibfnamefont {B.~A.}\ \bibnamefont {Pio}}, \bibinfo {author} {\bibfnamefont
  {M.}~\bibnamefont {Potemski}}, \bibinfo {author} {\bibfnamefont {I.~V.}\
  \bibnamefont {Grigorieva}}, \bibinfo {author} {\bibfnamefont {K.~S.}\
  \bibnamefont {Novoselov}}, \bibinfo {author} {\bibfnamefont {F.}~\bibnamefont
  {Guinea}}, \bibinfo {author} {\bibfnamefont {V.~I.}\ \bibnamefont {Fal'ko}},
  \ and\ \bibinfo {author} {\bibfnamefont {A.~K.}\ \bibnamefont {Geim}},\
  }\href@noop {} {\bibfield  {journal} {\bibinfo  {journal} {Nature}\ }\textbf
  {\bibinfo {volume} {497}},\ \bibinfo {pages} {594} (\bibinfo {year}
  {2013})}\BibitemShut {NoStop}%
\bibitem [{\citenamefont {Kim}\ \emph {et~al.}(2016)\citenamefont {Kim},
  \citenamefont {Yankowitz}, \citenamefont {Fallahazad}, \citenamefont {Kang},
  \citenamefont {Movva}, \citenamefont {Huang}, \citenamefont {Larentis},
  \citenamefont {Corbet}, \citenamefont {Taniguchi}, \citenamefont {Watanabe},
  \citenamefont {Banerjee}, \citenamefont {LeRoy},\ and\ \citenamefont
  {Tutuc}}]{Kim2016}%
  \BibitemOpen
  \bibfield  {author} {\bibinfo {author} {\bibfnamefont {K.}~\bibnamefont
  {Kim}}, \bibinfo {author} {\bibfnamefont {M.}~\bibnamefont {Yankowitz}},
  \bibinfo {author} {\bibfnamefont {B.}~\bibnamefont {Fallahazad}}, \bibinfo
  {author} {\bibfnamefont {S.}~\bibnamefont {Kang}}, \bibinfo {author}
  {\bibfnamefont {H.~C.~P.}\ \bibnamefont {Movva}}, \bibinfo {author}
  {\bibfnamefont {S.}~\bibnamefont {Huang}}, \bibinfo {author} {\bibfnamefont
  {S.}~\bibnamefont {Larentis}}, \bibinfo {author} {\bibfnamefont {C.~M.}\
  \bibnamefont {Corbet}}, \bibinfo {author} {\bibfnamefont {T.}~\bibnamefont
  {Taniguchi}}, \bibinfo {author} {\bibfnamefont {K.}~\bibnamefont {Watanabe}},
  \bibinfo {author} {\bibfnamefont {S.~K.}\ \bibnamefont {Banerjee}}, \bibinfo
  {author} {\bibfnamefont {B.~J.}\ \bibnamefont {LeRoy}}, \ and\ \bibinfo
  {author} {\bibfnamefont {E.}~\bibnamefont {Tutuc}},\ }\href@noop {}
  {\bibfield  {journal} {\bibinfo  {journal} {Nano Letters}\ }\textbf {\bibinfo
  {volume} {16}},\ \bibinfo {pages} {1989} (\bibinfo {year}
  {2016})}\BibitemShut {NoStop}%
\bibitem [{\citenamefont {Woods}\ \emph {et~al.}(2016)\citenamefont {Woods},
  \citenamefont {Withers}, \citenamefont {Zhu}, \citenamefont {Cao},
  \citenamefont {Yu}, \citenamefont {Kozikov}, \citenamefont {Shalom},
  \citenamefont {Morozov}, \citenamefont {van Wijk}, \citenamefont {Fasolino},
  \citenamefont {Katsnelson}, \citenamefont {Watanabe}, \citenamefont
  {Taniguchi}, \citenamefont {Geim}, \citenamefont {Mishchenko},\ and\
  \citenamefont {Novoselov}}]{Woods2016}%
  \BibitemOpen
  \bibfield  {author} {\bibinfo {author} {\bibfnamefont {C.~R.}\ \bibnamefont
  {Woods}}, \bibinfo {author} {\bibfnamefont {F.}~\bibnamefont {Withers}},
  \bibinfo {author} {\bibfnamefont {M.~J.}\ \bibnamefont {Zhu}}, \bibinfo
  {author} {\bibfnamefont {Y.}~\bibnamefont {Cao}}, \bibinfo {author}
  {\bibfnamefont {G.}~\bibnamefont {Yu}}, \bibinfo {author} {\bibfnamefont
  {A.}~\bibnamefont {Kozikov}}, \bibinfo {author} {\bibfnamefont {M.~B.}\
  \bibnamefont {Shalom}}, \bibinfo {author} {\bibfnamefont {S.~V.}\
  \bibnamefont {Morozov}}, \bibinfo {author} {\bibfnamefont {M.~M.}\
  \bibnamefont {van Wijk}}, \bibinfo {author} {\bibfnamefont {A.}~\bibnamefont
  {Fasolino}}, \bibinfo {author} {\bibfnamefont {M.~I.}\ \bibnamefont
  {Katsnelson}}, \bibinfo {author} {\bibfnamefont {K.}~\bibnamefont
  {Watanabe}}, \bibinfo {author} {\bibfnamefont {T.}~\bibnamefont {Taniguchi}},
  \bibinfo {author} {\bibfnamefont {A.~K.}\ \bibnamefont {Geim}}, \bibinfo
  {author} {\bibfnamefont {A.}~\bibnamefont {Mishchenko}}, \ and\ \bibinfo
  {author} {\bibfnamefont {K.~S.}\ \bibnamefont {Novoselov}},\ }\href@noop {}
  {\bibfield  {journal} {\bibinfo  {journal} {Nature Communications}\ }\textbf
  {\bibinfo {volume} {7}},\ \bibinfo {pages} {10800} (\bibinfo {year}
  {2016})}\BibitemShut {NoStop}%
\bibitem [{\citenamefont {Wang}\ \emph {et~al.}(2016)\citenamefont {Wang},
  \citenamefont {Lu}, \citenamefont {Ding}, \citenamefont {Yao}, \citenamefont
  {Yan}, \citenamefont {Wan}, \citenamefont {Deng}, \citenamefont {Wang},
  \citenamefont {Chen}, \citenamefont {Ma}, \citenamefont {Jung}, \citenamefont
  {Fedorov}, \citenamefont {Zhang}, \citenamefont {Zhang},\ and\ \citenamefont
  {Zhou}}]{Wang2016}%
  \BibitemOpen
  \bibfield  {author} {\bibinfo {author} {\bibfnamefont {E.}~\bibnamefont
  {Wang}}, \bibinfo {author} {\bibfnamefont {X.}~\bibnamefont {Lu}}, \bibinfo
  {author} {\bibfnamefont {S.}~\bibnamefont {Ding}}, \bibinfo {author}
  {\bibfnamefont {W.}~\bibnamefont {Yao}}, \bibinfo {author} {\bibfnamefont
  {M.}~\bibnamefont {Yan}}, \bibinfo {author} {\bibfnamefont {G.}~\bibnamefont
  {Wan}}, \bibinfo {author} {\bibfnamefont {K.}~\bibnamefont {Deng}}, \bibinfo
  {author} {\bibfnamefont {S.}~\bibnamefont {Wang}}, \bibinfo {author}
  {\bibfnamefont {G.}~\bibnamefont {Chen}}, \bibinfo {author} {\bibfnamefont
  {L.}~\bibnamefont {Ma}}, \bibinfo {author} {\bibfnamefont {J.}~\bibnamefont
  {Jung}}, \bibinfo {author} {\bibfnamefont {A.~V.}\ \bibnamefont {Fedorov}},
  \bibinfo {author} {\bibfnamefont {Y.}~\bibnamefont {Zhang}}, \bibinfo
  {author} {\bibfnamefont {G.}~\bibnamefont {Zhang}}, \ and\ \bibinfo {author}
  {\bibfnamefont {S.}~\bibnamefont {Zhou}},\ }\href@noop {} {\bibfield
  {journal} {\bibinfo  {journal} {Nature Physics}\ }\textbf {\bibinfo {volume}
  {12}},\ \bibinfo {pages} {1111} (\bibinfo {year} {2016})}\BibitemShut
  {NoStop}%
\bibitem [{\citenamefont {Wang}\ \emph {et~al.}(2013)\citenamefont {Wang},
  \citenamefont {Meric}, \citenamefont {Huang}, \citenamefont {Gao},
  \citenamefont {Gao}, \citenamefont {Tran}, \citenamefont {Taniguchi},
  \citenamefont {Watanabe}, \citenamefont {Campos}, \citenamefont {Muller},
  \citenamefont {Guo}, \citenamefont {Kim}, \citenamefont {Hone}, \citenamefont
  {Shepard},\ and\ \citenamefont {Dean}}]{Wang2013}%
  \BibitemOpen
  \bibfield  {author} {\bibinfo {author} {\bibfnamefont {L.}~\bibnamefont
  {Wang}}, \bibinfo {author} {\bibfnamefont {I.}~\bibnamefont {Meric}},
  \bibinfo {author} {\bibfnamefont {P.~Y.}\ \bibnamefont {Huang}}, \bibinfo
  {author} {\bibfnamefont {Q.}~\bibnamefont {Gao}}, \bibinfo {author}
  {\bibfnamefont {Y.}~\bibnamefont {Gao}}, \bibinfo {author} {\bibfnamefont
  {H.}~\bibnamefont {Tran}}, \bibinfo {author} {\bibfnamefont {T.}~\bibnamefont
  {Taniguchi}}, \bibinfo {author} {\bibfnamefont {K.}~\bibnamefont {Watanabe}},
  \bibinfo {author} {\bibfnamefont {L.~M.}\ \bibnamefont {Campos}}, \bibinfo
  {author} {\bibfnamefont {D.~A.}\ \bibnamefont {Muller}}, \bibinfo {author}
  {\bibfnamefont {J.}~\bibnamefont {Guo}}, \bibinfo {author} {\bibfnamefont
  {P.}~\bibnamefont {Kim}}, \bibinfo {author} {\bibfnamefont {J.}~\bibnamefont
  {Hone}}, \bibinfo {author} {\bibfnamefont {K.~L.}\ \bibnamefont {Shepard}}, \
  and\ \bibinfo {author} {\bibfnamefont {C.~R.}\ \bibnamefont {Dean}},\
  }\href@noop {} {\bibfield  {journal} {\bibinfo  {journal} {Science}\ }\textbf
  {\bibinfo {volume} {342}},\ \bibinfo {pages} {614} (\bibinfo {year}
  {2013})}\BibitemShut {NoStop}%
\bibitem [{\citenamefont {Mishchenko}\ \emph {et~al.}(2014)\citenamefont
  {Mishchenko}, \citenamefont {Tu}, \citenamefont {Cao}, \citenamefont
  {Borbachev}, \citenamefont {Wallbank}, \citenamefont {Greenaway},
  \citenamefont {Morozon}, \citenamefont {Morozov}, \citenamefont {Zhu},
  \citenamefont {Wong}, \citenamefont {Withers}, \citenamefont {Woods},
  \citenamefont {Kim}, \citenamefont {Watanabe}, \citenamefont {Taniguchi},
  \citenamefont {Vdovin}, \citenamefont {Makarovsky}, \citenamefont {Fromhold},
  \citenamefont {Fal'ko}, \citenamefont {Geim}, \citenamefont {Eaves},\ and\
  \citenamefont {Novoselov}}]{Mishchenko2014}%
  \BibitemOpen
  \bibfield  {author} {\bibinfo {author} {\bibfnamefont {A.}~\bibnamefont
  {Mishchenko}}, \bibinfo {author} {\bibfnamefont {J.~S.}\ \bibnamefont {Tu}},
  \bibinfo {author} {\bibfnamefont {Y.}~\bibnamefont {Cao}}, \bibinfo {author}
  {\bibfnamefont {R.~V.}\ \bibnamefont {Borbachev}}, \bibinfo {author}
  {\bibfnamefont {J.~R.}\ \bibnamefont {Wallbank}}, \bibinfo {author}
  {\bibfnamefont {M.~T.}\ \bibnamefont {Greenaway}}, \bibinfo {author}
  {\bibfnamefont {V.~E.}\ \bibnamefont {Morozon}}, \bibinfo {author}
  {\bibfnamefont {S.~V.}\ \bibnamefont {Morozov}}, \bibinfo {author}
  {\bibfnamefont {M.~J.}\ \bibnamefont {Zhu}}, \bibinfo {author} {\bibfnamefont
  {S.~L.}\ \bibnamefont {Wong}}, \bibinfo {author} {\bibfnamefont
  {F.}~\bibnamefont {Withers}}, \bibinfo {author} {\bibfnamefont {C.~R.}\
  \bibnamefont {Woods}}, \bibinfo {author} {\bibfnamefont {Y.-J.}\ \bibnamefont
  {Kim}}, \bibinfo {author} {\bibfnamefont {K.}~\bibnamefont {Watanabe}},
  \bibinfo {author} {\bibfnamefont {T.}~\bibnamefont {Taniguchi}}, \bibinfo
  {author} {\bibfnamefont {E.~E.}\ \bibnamefont {Vdovin}}, \bibinfo {author}
  {\bibfnamefont {O.}~\bibnamefont {Makarovsky}}, \bibinfo {author}
  {\bibfnamefont {T.~M.}\ \bibnamefont {Fromhold}}, \bibinfo {author}
  {\bibfnamefont {V.~I.}\ \bibnamefont {Fal'ko}}, \bibinfo {author}
  {\bibfnamefont {A.~K.}\ \bibnamefont {Geim}}, \bibinfo {author}
  {\bibfnamefont {L.}~\bibnamefont {Eaves}}, \ and\ \bibinfo {author}
  {\bibfnamefont {K.~S.}\ \bibnamefont {Novoselov}},\ }\href@noop {} {\bibfield
   {journal} {\bibinfo  {journal} {Nature Nanotechnology}\ }\textbf {\bibinfo
  {volume} {9}},\ \bibinfo {pages} {808} (\bibinfo {year} {2014})}\BibitemShut
  {NoStop}%
\bibitem [{\citenamefont {Yu}\ \emph {et~al.}(2015)\citenamefont {Yu},
  \citenamefont {Wang}, \citenamefont {Tong}, \citenamefont {Xu},\ and\
  \citenamefont {Yao}}]{Yu2015}%
  \BibitemOpen
  \bibfield  {author} {\bibinfo {author} {\bibfnamefont {H.}~\bibnamefont
  {Yu}}, \bibinfo {author} {\bibfnamefont {Y.}~\bibnamefont {Wang}}, \bibinfo
  {author} {\bibfnamefont {Q.}~\bibnamefont {Tong}}, \bibinfo {author}
  {\bibfnamefont {X.}~\bibnamefont {Xu}}, \ and\ \bibinfo {author}
  {\bibfnamefont {W.}~\bibnamefont {Yao}},\ }\href@noop {} {\bibfield
  {journal} {\bibinfo  {journal} {Physical Review Letters}\ }\textbf {\bibinfo
  {volume} {115}},\ \bibinfo {pages} {187002} (\bibinfo {year}
  {2015})}\BibitemShut {NoStop}%
\bibitem [{\citenamefont {Dean}\ \emph {et~al.}(2013)\citenamefont {Dean},
  \citenamefont {Wang}, \citenamefont {Maher}, \citenamefont {Forsythe},
  \citenamefont {Ghahari}, \citenamefont {Gao}, \citenamefont {Katoch},
  \citenamefont {Ishigami}, \citenamefont {Moon}, \citenamefont {Koshino},
  \citenamefont {Taniguchi}, \citenamefont {Watanabe}, \citenamefont {Shepard},
  \citenamefont {Hone},\ and\ \citenamefont {Kim}}]{Dean2013}%
  \BibitemOpen
  \bibfield  {author} {\bibinfo {author} {\bibfnamefont {C.~R.}\ \bibnamefont
  {Dean}}, \bibinfo {author} {\bibfnamefont {L.}~\bibnamefont {Wang}}, \bibinfo
  {author} {\bibfnamefont {P.}~\bibnamefont {Maher}}, \bibinfo {author}
  {\bibfnamefont {C.}~\bibnamefont {Forsythe}}, \bibinfo {author}
  {\bibfnamefont {F.}~\bibnamefont {Ghahari}}, \bibinfo {author} {\bibfnamefont
  {Y.}~\bibnamefont {Gao}}, \bibinfo {author} {\bibfnamefont {J.}~\bibnamefont
  {Katoch}}, \bibinfo {author} {\bibfnamefont {M.}~\bibnamefont {Ishigami}},
  \bibinfo {author} {\bibfnamefont {P.}~\bibnamefont {Moon}}, \bibinfo {author}
  {\bibfnamefont {M.}~\bibnamefont {Koshino}}, \bibinfo {author} {\bibfnamefont
  {T.}~\bibnamefont {Taniguchi}}, \bibinfo {author} {\bibfnamefont
  {K.}~\bibnamefont {Watanabe}}, \bibinfo {author} {\bibfnamefont {K.~L.}\
  \bibnamefont {Shepard}}, \bibinfo {author} {\bibfnamefont {J.}~\bibnamefont
  {Hone}}, \ and\ \bibinfo {author} {\bibfnamefont {P.}~\bibnamefont {Kim}},\
  }\href@noop {} {\bibfield  {journal} {\bibinfo  {journal} {Nature}\ }\textbf
  {\bibinfo {volume} {497}},\ \bibinfo {pages} {598} (\bibinfo {year}
  {2013})}\BibitemShut {NoStop}%
\bibitem [{\citenamefont {Gorbachev}\ \emph {et~al.}(2014)\citenamefont
  {Gorbachev}, \citenamefont {Song}, \citenamefont {Yu}, \citenamefont
  {Kretinin}, \citenamefont {Withers}, \citenamefont {Cao}, \citenamefont
  {Mishchenko}, \citenamefont {Grigorieva}, \citenamefont {Novoselov},
  \citenamefont {Levitov},\ and\ \citenamefont {Geim}}]{Gorbachev2014}%
  \BibitemOpen
  \bibfield  {author} {\bibinfo {author} {\bibfnamefont {R.~V.}\ \bibnamefont
  {Gorbachev}}, \bibinfo {author} {\bibfnamefont {J.~C.~W.}\ \bibnamefont
  {Song}}, \bibinfo {author} {\bibfnamefont {G.~L.}\ \bibnamefont {Yu}},
  \bibinfo {author} {\bibfnamefont {A.~V.}\ \bibnamefont {Kretinin}}, \bibinfo
  {author} {\bibfnamefont {F.}~\bibnamefont {Withers}}, \bibinfo {author}
  {\bibfnamefont {Y.}~\bibnamefont {Cao}}, \bibinfo {author} {\bibfnamefont
  {A.}~\bibnamefont {Mishchenko}}, \bibinfo {author} {\bibfnamefont {I.~V.}\
  \bibnamefont {Grigorieva}}, \bibinfo {author} {\bibfnamefont {K.~S.}\
  \bibnamefont {Novoselov}}, \bibinfo {author} {\bibfnamefont {L.~S.}\
  \bibnamefont {Levitov}}, \ and\ \bibinfo {author} {\bibfnamefont {A.~K.}\
  \bibnamefont {Geim}},\ }\href@noop {} {\bibfield  {journal} {\bibinfo
  {journal} {Science}\ }\textbf {\bibinfo {volume} {346}},\ \bibinfo {pages}
  {448} (\bibinfo {year} {2014})}\BibitemShut {NoStop}%
\bibitem [{\citenamefont {Spanton}\ \emph {et~al.}(2017)\citenamefont
  {Spanton}, \citenamefont {Zibron}, \citenamefont {Zhou}, \citenamefont
  {Taniguchi}, \citenamefont {Watanabe}, \citenamefont {Zaletel},\ and\
  \citenamefont {Young}}]{Spanton2017}%
  \BibitemOpen
  \bibfield  {author} {\bibinfo {author} {\bibfnamefont {E.~M.}\ \bibnamefont
  {Spanton}}, \bibinfo {author} {\bibfnamefont {A.~A.}\ \bibnamefont {Zibron}},
  \bibinfo {author} {\bibfnamefont {H.}~\bibnamefont {Zhou}}, \bibinfo {author}
  {\bibfnamefont {T.}~\bibnamefont {Taniguchi}}, \bibinfo {author}
  {\bibfnamefont {K.}~\bibnamefont {Watanabe}}, \bibinfo {author}
  {\bibfnamefont {M.~P.}\ \bibnamefont {Zaletel}}, \ and\ \bibinfo {author}
  {\bibfnamefont {A.~F.}\ \bibnamefont {Young}},\ }\href@noop {} {\bibfield
  {journal} {\bibinfo  {journal} {arXiv:1706.06166}\ } (\bibinfo {year}
  {2017})}\BibitemShut {NoStop}%
\bibitem [{\citenamefont {Dean}\ \emph {et~al.}(2010)\citenamefont {Dean},
  \citenamefont {Young}, \citenamefont {Meric}, \citenamefont {Lee},
  \citenamefont {Wang}, \citenamefont {Sorgenfrei}, \citenamefont {Watanabe},
  \citenamefont {Taniguchi}, \citenamefont {Kim}, \citenamefont {Shepard},\
  and\ \citenamefont {Hone}}]{Dean2010}%
  \BibitemOpen
  \bibfield  {author} {\bibinfo {author} {\bibfnamefont {C.~R.}\ \bibnamefont
  {Dean}}, \bibinfo {author} {\bibfnamefont {A.~F.}\ \bibnamefont {Young}},
  \bibinfo {author} {\bibfnamefont {I.}~\bibnamefont {Meric}}, \bibinfo
  {author} {\bibfnamefont {C.}~\bibnamefont {Lee}}, \bibinfo {author}
  {\bibfnamefont {L.}~\bibnamefont {Wang}}, \bibinfo {author} {\bibfnamefont
  {S.}~\bibnamefont {Sorgenfrei}}, \bibinfo {author} {\bibfnamefont
  {K.}~\bibnamefont {Watanabe}}, \bibinfo {author} {\bibfnamefont
  {T.}~\bibnamefont {Taniguchi}}, \bibinfo {author} {\bibfnamefont
  {P.}~\bibnamefont {Kim}}, \bibinfo {author} {\bibfnamefont {K.~L.}\
  \bibnamefont {Shepard}}, \ and\ \bibinfo {author} {\bibfnamefont
  {J.}~\bibnamefont {Hone}},\ }\href@noop {} {\bibfield  {journal} {\bibinfo
  {journal} {Nature Nanotechnology}\ }\textbf {\bibinfo {volume} {5}},\
  \bibinfo {pages} {722} (\bibinfo {year} {2010})}\BibitemShut {NoStop}%
\bibitem [{\citenamefont {Park}\ \emph {et~al.}(2008)\citenamefont {Park},
  \citenamefont {Yang}, \citenamefont {Son}, \citenamefont {Cohen},\ and\
  \citenamefont {Louie}}]{Park2008}%
  \BibitemOpen
  \bibfield  {author} {\bibinfo {author} {\bibfnamefont {C.-H.}\ \bibnamefont
  {Park}}, \bibinfo {author} {\bibfnamefont {L.}~\bibnamefont {Yang}}, \bibinfo
  {author} {\bibfnamefont {Y.-W.}\ \bibnamefont {Son}}, \bibinfo {author}
  {\bibfnamefont {M.~L.}\ \bibnamefont {Cohen}}, \ and\ \bibinfo {author}
  {\bibfnamefont {S.~G.}\ \bibnamefont {Louie}},\ }\href@noop {} {\bibfield
  {journal} {\bibinfo  {journal} {Physical Review Letters}\ }\textbf {\bibinfo
  {volume} {101}},\ \bibinfo {pages} {126804} (\bibinfo {year}
  {2008})}\BibitemShut {NoStop}%
\bibitem [{\citenamefont {Chen}\ \emph
  {et~al.}(2017{\natexlab{a}})\citenamefont {Chen}, \citenamefont {Sui},
  \citenamefont {Wang}, \citenamefont {Wang}, \citenamefont {Jung},
  \citenamefont {Moon}, \citenamefont {Adam}, \citenamefont {Watanabe},
  \citenamefont {Taniguchi}, \citenamefont {Zhou}, \citenamefont {Koshino},
  \citenamefont {Zhang},\ and\ \citenamefont {Zhang}}]{Chen2017}%
  \BibitemOpen
  \bibfield  {author} {\bibinfo {author} {\bibfnamefont {G.}~\bibnamefont
  {Chen}}, \bibinfo {author} {\bibfnamefont {M.}~\bibnamefont {Sui}}, \bibinfo
  {author} {\bibfnamefont {D.}~\bibnamefont {Wang}}, \bibinfo {author}
  {\bibfnamefont {S.}~\bibnamefont {Wang}}, \bibinfo {author} {\bibfnamefont
  {J.}~\bibnamefont {Jung}}, \bibinfo {author} {\bibfnamefont {P.}~\bibnamefont
  {Moon}}, \bibinfo {author} {\bibfnamefont {S.}~\bibnamefont {Adam}}, \bibinfo
  {author} {\bibfnamefont {K.}~\bibnamefont {Watanabe}}, \bibinfo {author}
  {\bibfnamefont {T.}~\bibnamefont {Taniguchi}}, \bibinfo {author}
  {\bibfnamefont {S.}~\bibnamefont {Zhou}}, \bibinfo {author} {\bibfnamefont
  {M.}~\bibnamefont {Koshino}}, \bibinfo {author} {\bibfnamefont
  {G.}~\bibnamefont {Zhang}}, \ and\ \bibinfo {author} {\bibfnamefont
  {Y.}~\bibnamefont {Zhang}},\ }\href@noop {} {\bibfield  {journal} {\bibinfo
  {journal} {arXiv:1702.03107}\ } (\bibinfo {year}
  {2017}{\natexlab{a}})}\BibitemShut {NoStop}%
\bibitem [{\citenamefont {Chari}\ \emph {et~al.}(2016)\citenamefont {Chari},
  \citenamefont {Ribeiro-Palau}, \citenamefont {Dean},\ and\ \citenamefont
  {Shepard}}]{Chari2016}%
  \BibitemOpen
  \bibfield  {author} {\bibinfo {author} {\bibfnamefont {T.}~\bibnamefont
  {Chari}}, \bibinfo {author} {\bibfnamefont {R.}~\bibnamefont
  {Ribeiro-Palau}}, \bibinfo {author} {\bibfnamefont {C.~R.}\ \bibnamefont
  {Dean}}, \ and\ \bibinfo {author} {\bibfnamefont {K.}~\bibnamefont
  {Shepard}},\ }\href@noop {} {\bibfield  {journal} {\bibinfo  {journal} {Nano
  Letters}\ }\textbf {\bibinfo {volume} {16}},\ \bibinfo {pages} {4477}
  (\bibinfo {year} {2016})}\BibitemShut {NoStop}%
\bibitem [{\citenamefont {Koren}\ \emph {et~al.}(2016)\citenamefont {Koren},
  \citenamefont {Leven}, \citenamefont {L{\"o}rtscher}, \citenamefont {Knoll},
  \citenamefont {Hod},\ and\ \citenamefont {Duerig}}]{Koren2016}%
  \BibitemOpen
  \bibfield  {author} {\bibinfo {author} {\bibfnamefont {E.}~\bibnamefont
  {Koren}}, \bibinfo {author} {\bibfnamefont {I.}~\bibnamefont {Leven}},
  \bibinfo {author} {\bibfnamefont {E.}~\bibnamefont {L{\"o}rtscher}}, \bibinfo
  {author} {\bibfnamefont {A.}~\bibnamefont {Knoll}}, \bibinfo {author}
  {\bibfnamefont {O.}~\bibnamefont {Hod}}, \ and\ \bibinfo {author}
  {\bibfnamefont {U.}~\bibnamefont {Duerig}},\ }\href@noop {} {\bibfield
  {journal} {\bibinfo  {journal} {Nature Nanotechnology}\ }\textbf {\bibinfo
  {volume} {11}},\ \bibinfo {pages} {752} (\bibinfo {year} {2016})}\BibitemShut
  {NoStop}%
\bibitem [{\citenamefont {Solozhenko}\ \emph {et~al.}(1995)\citenamefont
  {Solozhenko}, \citenamefont {Will},\ and\ \citenamefont
  {Elf}}]{Solozhenko1995}%
  \BibitemOpen
  \bibfield  {author} {\bibinfo {author} {\bibfnamefont {V.~L.}\ \bibnamefont
  {Solozhenko}}, \bibinfo {author} {\bibfnamefont {G.}~\bibnamefont {Will}}, \
  and\ \bibinfo {author} {\bibfnamefont {F.}~\bibnamefont {Elf}},\ }\href@noop
  {} {\bibfield  {journal} {\bibinfo  {journal} {Solid State Communications}\
  }\textbf {\bibinfo {volume} {96}},\ \bibinfo {pages} {1} (\bibinfo {year}
  {1995})}\BibitemShut {NoStop}%
\bibitem [{\citenamefont {Blakslee}\ \emph {et~al.}(1970)\citenamefont
  {Blakslee}, \citenamefont {Proctor}, \citenamefont {Seldin}, \citenamefont
  {Spence},\ and\ \citenamefont {Weng}}]{Blakslee1970}%
  \BibitemOpen
  \bibfield  {author} {\bibinfo {author} {\bibfnamefont {O.~L.}\ \bibnamefont
  {Blakslee}}, \bibinfo {author} {\bibfnamefont {D.~G.}\ \bibnamefont
  {Proctor}}, \bibinfo {author} {\bibfnamefont {E.~J.}\ \bibnamefont {Seldin}},
  \bibinfo {author} {\bibfnamefont {G.~B.}\ \bibnamefont {Spence}}, \ and\
  \bibinfo {author} {\bibfnamefont {T.}~\bibnamefont {Weng}},\ }\href@noop {}
  {\bibfield  {journal} {\bibinfo  {journal} {Journal of Applied Physics}\
  }\textbf {\bibinfo {volume} {41}},\ \bibinfo {pages} {3373} (\bibinfo {year}
  {1970})}\BibitemShut {NoStop}%
\bibitem [{\citenamefont {Chen}\ \emph
  {et~al.}(2017{\natexlab{b}})\citenamefont {Chen}, \citenamefont {Ke},
  \citenamefont {Ci}, \citenamefont {Ko}, \citenamefont {Park}, \citenamefont
  {Saremi}, \citenamefont {Liu}, \citenamefont {Lee}, \citenamefont {Suh},
  \citenamefont {Martin}, \citenamefont {Ager}, \citenamefont {Chen},\ and\
  \citenamefont {Wu}}]{Chen2017b}%
  \BibitemOpen
  \bibfield  {author} {\bibinfo {author} {\bibfnamefont {Y.}~\bibnamefont
  {Chen}}, \bibinfo {author} {\bibfnamefont {F.}~\bibnamefont {Ke}}, \bibinfo
  {author} {\bibfnamefont {P.}~\bibnamefont {Ci}}, \bibinfo {author}
  {\bibfnamefont {C.}~\bibnamefont {Ko}}, \bibinfo {author} {\bibfnamefont
  {T.}~\bibnamefont {Park}}, \bibinfo {author} {\bibfnamefont {S.}~\bibnamefont
  {Saremi}}, \bibinfo {author} {\bibfnamefont {H.}~\bibnamefont {Liu}},
  \bibinfo {author} {\bibfnamefont {Y.}~\bibnamefont {Lee}}, \bibinfo {author}
  {\bibfnamefont {J.}~\bibnamefont {Suh}}, \bibinfo {author} {\bibfnamefont
  {L.~W.}\ \bibnamefont {Martin}}, \bibinfo {author} {\bibfnamefont {J.~W.}\
  \bibnamefont {Ager}}, \bibinfo {author} {\bibfnamefont {B.}~\bibnamefont
  {Chen}}, \ and\ \bibinfo {author} {\bibfnamefont {J.}~\bibnamefont {Wu}},\
  }\href@noop {} {\bibfield  {journal} {\bibinfo  {journal} {Nano Letters}\
  }\textbf {\bibinfo {volume} {17}},\ \bibinfo {pages} {194} (\bibinfo {year}
  {2017}{\natexlab{b}})}\BibitemShut {NoStop}%
\bibitem [{\citenamefont {Chen}\ \emph {et~al.}(2014)\citenamefont {Chen},
  \citenamefont {Shi}, \citenamefont {Yang}, \citenamefont {Lu}, \citenamefont
  {Lai}, \citenamefont {Yan}, \citenamefont {Wang}, \citenamefont {Zhang},\
  and\ \citenamefont {Li}}]{Chen2014}%
  \BibitemOpen
  \bibfield  {author} {\bibinfo {author} {\bibfnamefont {Z.-G.}\ \bibnamefont
  {Chen}}, \bibinfo {author} {\bibfnamefont {Z.}~\bibnamefont {Shi}}, \bibinfo
  {author} {\bibfnamefont {W.}~\bibnamefont {Yang}}, \bibinfo {author}
  {\bibfnamefont {X.}~\bibnamefont {Lu}}, \bibinfo {author} {\bibfnamefont
  {Y.}~\bibnamefont {Lai}}, \bibinfo {author} {\bibfnamefont {H.}~\bibnamefont
  {Yan}}, \bibinfo {author} {\bibfnamefont {F.}~\bibnamefont {Wang}}, \bibinfo
  {author} {\bibfnamefont {G.}~\bibnamefont {Zhang}}, \ and\ \bibinfo {author}
  {\bibfnamefont {Z.}~\bibnamefont {Li}},\ }\href@noop {} {\bibfield  {journal}
  {\bibinfo  {journal} {Nature Communications}\ }\textbf {\bibinfo {volume}
  {5}},\ \bibinfo {pages} {4461} (\bibinfo {year} {2014})}\BibitemShut
  {NoStop}%
\bibitem [{\citenamefont {Song}\ \emph {et~al.}(2013)\citenamefont {Song},
  \citenamefont {Shytov},\ and\ \citenamefont {Levitov}}]{Song2013}%
  \BibitemOpen
  \bibfield  {author} {\bibinfo {author} {\bibfnamefont {J.~C.~W.}\
  \bibnamefont {Song}}, \bibinfo {author} {\bibfnamefont {A.~V.}\ \bibnamefont
  {Shytov}}, \ and\ \bibinfo {author} {\bibfnamefont {L.~S.}\ \bibnamefont
  {Levitov}},\ }\href@noop {} {\bibfield  {journal} {\bibinfo  {journal}
  {Physical Review Letters}\ }\textbf {\bibinfo {volume} {111}},\ \bibinfo
  {pages} {266801} (\bibinfo {year} {2013})}\BibitemShut {NoStop}%
\bibitem [{\citenamefont {Bokdam}\ \emph {et~al.}(2014)\citenamefont {Bokdam},
  \citenamefont {Amlaki}, \citenamefont {Brocks},\ and\ \citenamefont
  {Kelly}}]{Bokdam2014}%
  \BibitemOpen
  \bibfield  {author} {\bibinfo {author} {\bibfnamefont {M.}~\bibnamefont
  {Bokdam}}, \bibinfo {author} {\bibfnamefont {T.}~\bibnamefont {Amlaki}},
  \bibinfo {author} {\bibfnamefont {G.}~\bibnamefont {Brocks}}, \ and\ \bibinfo
  {author} {\bibfnamefont {P.~J.}\ \bibnamefont {Kelly}},\ }\href@noop {}
  {\bibfield  {journal} {\bibinfo  {journal} {Physical Review B}\ }\textbf
  {\bibinfo {volume} {89}},\ \bibinfo {pages} {201404(R)} (\bibinfo {year}
  {2014})}\BibitemShut {NoStop}%
\bibitem [{\citenamefont {Moon}\ and\ \citenamefont
  {Koshino}(2014)}]{Moon2014}%
  \BibitemOpen
  \bibfield  {author} {\bibinfo {author} {\bibfnamefont {P.}~\bibnamefont
  {Moon}}\ and\ \bibinfo {author} {\bibfnamefont {M.}~\bibnamefont {Koshino}},\
  }\href@noop {} {\bibfield  {journal} {\bibinfo  {journal} {Physical Reveiw
  B}\ }\textbf {\bibinfo {volume} {90}},\ \bibinfo {pages} {155406} (\bibinfo
  {year} {2014})}\BibitemShut {NoStop}%
\bibitem [{\citenamefont {Wallbank}\ \emph
  {et~al.}(2015{\natexlab{a}})\citenamefont {Wallbank}, \citenamefont
  {Mucha-Kruczynski}, \citenamefont {Chen},\ and\ \citenamefont
  {Fal'ko}}]{Wallbank2015}%
  \BibitemOpen
  \bibfield  {author} {\bibinfo {author} {\bibfnamefont {J.~R.}\ \bibnamefont
  {Wallbank}}, \bibinfo {author} {\bibfnamefont {M.}~\bibnamefont
  {Mucha-Kruczynski}}, \bibinfo {author} {\bibfnamefont {X.}~\bibnamefont
  {Chen}}, \ and\ \bibinfo {author} {\bibfnamefont {V.~I.}\ \bibnamefont
  {Fal'ko}},\ }\href@noop {} {\bibfield  {journal} {\bibinfo  {journal}
  {Annalen der Physik}\ }\textbf {\bibinfo {volume} {527}},\ \bibinfo {pages}
  {359} (\bibinfo {year} {2015}{\natexlab{a}})}\BibitemShut {NoStop}%
\bibitem [{\citenamefont {Jung}\ \emph
  {et~al.}(2015{\natexlab{a}})\citenamefont {Jung}, \citenamefont {DaSilva},
  \citenamefont {MacDonald},\ and\ \citenamefont {Adam}}]{Jung2015}%
  \BibitemOpen
  \bibfield  {author} {\bibinfo {author} {\bibfnamefont {J.}~\bibnamefont
  {Jung}}, \bibinfo {author} {\bibfnamefont {A.~M.}\ \bibnamefont {DaSilva}},
  \bibinfo {author} {\bibfnamefont {A.~H.}\ \bibnamefont {MacDonald}}, \ and\
  \bibinfo {author} {\bibfnamefont {S.}~\bibnamefont {Adam}},\ }\href@noop {}
  {\bibfield  {journal} {\bibinfo  {journal} {Nature Communications}\ }\textbf
  {\bibinfo {volume} {6}},\ \bibinfo {pages} {6308} (\bibinfo {year}
  {2015}{\natexlab{a}})}\BibitemShut {NoStop}%
\bibitem [{\citenamefont {San-Jose}\ \emph {et~al.}(2014)\citenamefont
  {San-Jose}, \citenamefont {Guti\'errez-Rubio}, \citenamefont {Sturla},\ and\
  \citenamefont {Guinea}}]{SanJose2014}%
  \BibitemOpen
  \bibfield  {author} {\bibinfo {author} {\bibfnamefont {P.}~\bibnamefont
  {San-Jose}}, \bibinfo {author} {\bibfnamefont {A.}~\bibnamefont
  {Guti\'errez-Rubio}}, \bibinfo {author} {\bibfnamefont {M.}~\bibnamefont
  {Sturla}}, \ and\ \bibinfo {author} {\bibfnamefont {F.}~\bibnamefont
  {Guinea}},\ }\href@noop {} {\bibfield  {journal} {\bibinfo  {journal}
  {Physical Review B}\ }\textbf {\bibinfo {volume} {90}},\ \bibinfo {pages}
  {075428} (\bibinfo {year} {2014})}\BibitemShut {NoStop}%
\bibitem [{\citenamefont {Slotman}\ \emph {et~al.}(2015)\citenamefont
  {Slotman}, \citenamefont {van Wijk}, \citenamefont {Zhao}, \citenamefont
  {Fasolino}, \citenamefont {Katsnelson},\ and\ \citenamefont
  {Yuan}}]{Slotman2015}%
  \BibitemOpen
  \bibfield  {author} {\bibinfo {author} {\bibfnamefont {G.}~\bibnamefont
  {Slotman}}, \bibinfo {author} {\bibfnamefont {M.}~\bibnamefont {van Wijk}},
  \bibinfo {author} {\bibfnamefont {P.-L.}\ \bibnamefont {Zhao}}, \bibinfo
  {author} {\bibfnamefont {A.}~\bibnamefont {Fasolino}}, \bibinfo {author}
  {\bibfnamefont {M.}~\bibnamefont {Katsnelson}}, \ and\ \bibinfo {author}
  {\bibfnamefont {S.}~\bibnamefont {Yuan}},\ }\href@noop {} {\bibfield
  {journal} {\bibinfo  {journal} {Physical Review Letters}\ }\textbf {\bibinfo
  {volume} {115}},\ \bibinfo {pages} {186801} (\bibinfo {year}
  {2015})}\BibitemShut {NoStop}%
\bibitem [{\citenamefont {Jung}\ \emph {et~al.}(2017)\citenamefont {Jung},
  \citenamefont {Laksono}, \citenamefont {DaSilva}, \citenamefont {MacDonald},
  \citenamefont {Mucha-Kruchy{\'n}ski},\ and\ \citenamefont {Adam}}]{Jung2017}%
  \BibitemOpen
  \bibfield  {author} {\bibinfo {author} {\bibfnamefont {J.}~\bibnamefont
  {Jung}}, \bibinfo {author} {\bibfnamefont {E.}~\bibnamefont {Laksono}},
  \bibinfo {author} {\bibfnamefont {A.~M.}\ \bibnamefont {DaSilva}}, \bibinfo
  {author} {\bibfnamefont {A.~H.}\ \bibnamefont {MacDonald}}, \bibinfo {author}
  {\bibfnamefont {M.}~\bibnamefont {Mucha-Kruchy{\'n}ski}}, \ and\ \bibinfo
  {author} {\bibfnamefont {S.}~\bibnamefont {Adam}},\ }\href@noop {} {\bibfield
   {journal} {\bibinfo  {journal} {arXiv:1706.06016}\ } (\bibinfo {year}
  {2017})}\BibitemShut {NoStop}%
\bibitem [{\citenamefont {Yankowitz}\ \emph {et~al.}(2016)\citenamefont
  {Yankowitz}, \citenamefont {Watanabe}, \citenamefont {Taniguchi},
  \citenamefont {San-Jose},\ and\ \citenamefont {LeRoy}}]{Yankowitz2016}%
  \BibitemOpen
  \bibfield  {author} {\bibinfo {author} {\bibfnamefont {M.}~\bibnamefont
  {Yankowitz}}, \bibinfo {author} {\bibfnamefont {K.}~\bibnamefont {Watanabe}},
  \bibinfo {author} {\bibfnamefont {T.}~\bibnamefont {Taniguchi}}, \bibinfo
  {author} {\bibfnamefont {P.}~\bibnamefont {San-Jose}}, \ and\ \bibinfo
  {author} {\bibfnamefont {B.~J.}\ \bibnamefont {LeRoy}},\ }\href@noop {}
  {\bibfield  {journal} {\bibinfo  {journal} {Nature Communications}\ }\textbf
  {\bibinfo {volume} {7}},\ \bibinfo {pages} {193168} (\bibinfo {year}
  {2016})}\BibitemShut {NoStop}%
\bibitem [{\citenamefont {Trambly De~Laissardi\`ere}\ \emph
  {et~al.}(2010)\citenamefont {Trambly De~Laissardi\`ere}, \citenamefont
  {Mayou},\ and\ \citenamefont {Magaud}}]{Laissardiere2010}%
  \BibitemOpen
  \bibfield  {author} {\bibinfo {author} {\bibfnamefont {G.}~\bibnamefont
  {Trambly De~Laissardi\`ere}}, \bibinfo {author} {\bibfnamefont
  {D.}~\bibnamefont {Mayou}}, \ and\ \bibinfo {author} {\bibfnamefont
  {L.}~\bibnamefont {Magaud}},\ }\href@noop {} {\bibfield  {journal} {\bibinfo
  {journal} {Nano Letters}\ }\textbf {\bibinfo {volume} {10}},\ \bibinfo
  {pages} {804} (\bibinfo {year} {2010})}\BibitemShut {NoStop}%
\bibitem [{\citenamefont {Wang}\ \emph
  {et~al.}(2015{\natexlab{b}})\citenamefont {Wang}, \citenamefont {Ki},
  \citenamefont {Chen}, \citenamefont {Berger}, \citenamefont {MacDonald},\
  and\ \citenamefont {Morpurgo}}]{Wang2015b}%
  \BibitemOpen
  \bibfield  {author} {\bibinfo {author} {\bibfnamefont {Z.}~\bibnamefont
  {Wang}}, \bibinfo {author} {\bibfnamefont {D.}~\bibnamefont {Ki}}, \bibinfo
  {author} {\bibfnamefont {H.}~\bibnamefont {Chen}}, \bibinfo {author}
  {\bibfnamefont {H.}~\bibnamefont {Berger}}, \bibinfo {author} {\bibfnamefont
  {A.~H.}\ \bibnamefont {MacDonald}}, \ and\ \bibinfo {author} {\bibfnamefont
  {A.~F.}\ \bibnamefont {Morpurgo}},\ }\href@noop {} {\bibfield  {journal}
  {\bibinfo  {journal} {Nature Communications}\ }\textbf {\bibinfo {volume}
  {6}},\ \bibinfo {pages} {8339} (\bibinfo {year}
  {2015}{\natexlab{b}})}\BibitemShut {NoStop}%
\bibitem [{\citenamefont {Murata}\ \emph {et~al.}(1997)\citenamefont {Murata},
  \citenamefont {Yoshina}, \citenamefont {Yadav}, \citenamefont {Honda},\ and\
  \citenamefont {Shirakawa}}]{Murata1997}%
  \BibitemOpen
  \bibfield  {author} {\bibinfo {author} {\bibfnamefont {K.}~\bibnamefont
  {Murata}}, \bibinfo {author} {\bibfnamefont {H.}~\bibnamefont {Yoshina}},
  \bibinfo {author} {\bibfnamefont {H.~O.}\ \bibnamefont {Yadav}}, \bibinfo
  {author} {\bibfnamefont {Y.}~\bibnamefont {Honda}}, \ and\ \bibinfo {author}
  {\bibfnamefont {N.}~\bibnamefont {Shirakawa}},\ }\href@noop {} {\bibfield
  {journal} {\bibinfo  {journal} {Review of Scientific Instruments}\ }\textbf
  {\bibinfo {volume} {68}},\ \bibinfo {pages} {2490} (\bibinfo {year}
  {1997})}\BibitemShut {NoStop}%
\bibitem [{\citenamefont {Murata}\ \emph {et~al.}(2008)\citenamefont {Murata},
  \citenamefont {Yokogawa}, \citenamefont {Yoshino}, \citenamefont {Klotz},
  \citenamefont {Munsch}, \citenamefont {Irizawa}, \citenamefont {Nishiyama},
  \citenamefont {Iizuka}, \citenamefont {Nanba}, \citenamefont {Okada},
  \citenamefont {Shiraga},\ and\ \citenamefont {Aoyama}}]{Murata2008}%
  \BibitemOpen
  \bibfield  {author} {\bibinfo {author} {\bibfnamefont {K.}~\bibnamefont
  {Murata}}, \bibinfo {author} {\bibfnamefont {K.}~\bibnamefont {Yokogawa}},
  \bibinfo {author} {\bibfnamefont {H.}~\bibnamefont {Yoshino}}, \bibinfo
  {author} {\bibfnamefont {S.}~\bibnamefont {Klotz}}, \bibinfo {author}
  {\bibfnamefont {P.}~\bibnamefont {Munsch}}, \bibinfo {author} {\bibfnamefont
  {A.}~\bibnamefont {Irizawa}}, \bibinfo {author} {\bibfnamefont
  {M.}~\bibnamefont {Nishiyama}}, \bibinfo {author} {\bibfnamefont
  {K.}~\bibnamefont {Iizuka}}, \bibinfo {author} {\bibfnamefont
  {T.}~\bibnamefont {Nanba}}, \bibinfo {author} {\bibfnamefont
  {T.}~\bibnamefont {Okada}}, \bibinfo {author} {\bibfnamefont
  {Y.}~\bibnamefont {Shiraga}}, \ and\ \bibinfo {author} {\bibfnamefont
  {S.}~\bibnamefont {Aoyama}},\ }\href@noop {} {\bibfield  {journal} {\bibinfo
  {journal} {Review of Scientific Instruments}\ }\textbf {\bibinfo {volume}
  {79}},\ \bibinfo {pages} {085101} (\bibinfo {year} {2008})}\BibitemShut
  {NoStop}%
\bibitem [{\citenamefont {Weis}\ and\ \citenamefont {von
  Klitzing}(2011)}]{Weis2011}%
  \BibitemOpen
  \bibfield  {author} {\bibinfo {author} {\bibfnamefont {J.}~\bibnamefont
  {Weis}}\ and\ \bibinfo {author} {\bibfnamefont {K.}~\bibnamefont {von
  Klitzing}},\ }\href@noop {} {\bibfield  {journal} {\bibinfo  {journal}
  {Philosophical Transactions of the Royal Society A}\ }\textbf {\bibinfo
  {volume} {369}},\ \bibinfo {pages} {3954} (\bibinfo {year}
  {2011})}\BibitemShut {NoStop}%
\bibitem [{\citenamefont {Maher}\ \emph {et~al.}(2014)\citenamefont {Maher},
  \citenamefont {Wang}, \citenamefont {Gao}, \citenamefont {Forsythe},
  \citenamefont {Taniguchi}, \citenamefont {Watanabe}, \citenamefont {Abanin},
  \citenamefont {Papi\'{c}}, \citenamefont {Cadden-Zimansky}, \citenamefont
  {Hone}, \citenamefont {Kim},\ and\ \citenamefont {Dean}}]{Maher2014}%
  \BibitemOpen
  \bibfield  {author} {\bibinfo {author} {\bibfnamefont {P.}~\bibnamefont
  {Maher}}, \bibinfo {author} {\bibfnamefont {P.}~\bibnamefont {Wang}},
  \bibinfo {author} {\bibfnamefont {Y.}~\bibnamefont {Gao}}, \bibinfo {author}
  {\bibfnamefont {C.}~\bibnamefont {Forsythe}}, \bibinfo {author}
  {\bibfnamefont {T.}~\bibnamefont {Taniguchi}}, \bibinfo {author}
  {\bibfnamefont {K.}~\bibnamefont {Watanabe}}, \bibinfo {author}
  {\bibfnamefont {D.}~\bibnamefont {Abanin}}, \bibinfo {author} {\bibfnamefont
  {Z.}~\bibnamefont {Papi\'{c}}}, \bibinfo {author} {\bibfnamefont
  {P.}~\bibnamefont {Cadden-Zimansky}}, \bibinfo {author} {\bibfnamefont
  {J.}~\bibnamefont {Hone}}, \bibinfo {author} {\bibfnamefont {P.}~\bibnamefont
  {Kim}}, \ and\ \bibinfo {author} {\bibfnamefont {C.~R.}\ \bibnamefont
  {Dean}},\ }\href@noop {} {\bibfield  {journal} {\bibinfo  {journal}
  {Science}\ }\textbf {\bibinfo {volume} {345}},\ \bibinfo {pages} {61}
  (\bibinfo {year} {2014})}\BibitemShut {NoStop}%
\bibitem [{\citenamefont {Giannozzi}\ \emph {et~al.}(2009)\citenamefont
  {Giannozzi}, \citenamefont {Baroni}, \citenamefont {Bonini}, \citenamefont
  {Calandra}, \citenamefont {Car}, \citenamefont {Cavazzoni}, \citenamefont
  {Ceresoli}, \citenamefont {Chiarotti}, \citenamefont {Cococcioni},
  \citenamefont {Dabo}, \citenamefont {Corso}, \citenamefont {de~Gironcoli},
  \citenamefont {Fabris}, \citenamefont {Fratesi}, \citenamefont {Gebauer},
  \citenamefont {Gerstmann}, \citenamefont {Gougoussis}, \citenamefont
  {Kokalj}, \citenamefont {Lazzeri}, \citenamefont {Martin-Samos},
  \citenamefont {Marzari}, \citenamefont {Mauri}, \citenamefont {Mazzarello},
  \citenamefont {Paolini}, \citenamefont {Pasquarello}, \citenamefont
  {Paulatto}, \citenamefont {Sbraccia}, \citenamefont {Scandolo}, \citenamefont
  {Sclauzero}, \citenamefont {Seitsonen}, \citenamefont {Smogunov},
  \citenamefont {Umari},\ and\ \citenamefont
  {Wentzcovitch}}]{giannozzi2009quantum}%
  \BibitemOpen
  \bibfield  {author} {\bibinfo {author} {\bibfnamefont {P.}~\bibnamefont
  {Giannozzi}}, \bibinfo {author} {\bibfnamefont {S.}~\bibnamefont {Baroni}},
  \bibinfo {author} {\bibfnamefont {N.}~\bibnamefont {Bonini}}, \bibinfo
  {author} {\bibfnamefont {M.}~\bibnamefont {Calandra}}, \bibinfo {author}
  {\bibfnamefont {R.}~\bibnamefont {Car}}, \bibinfo {author} {\bibfnamefont
  {C.}~\bibnamefont {Cavazzoni}}, \bibinfo {author} {\bibfnamefont
  {D.}~\bibnamefont {Ceresoli}}, \bibinfo {author} {\bibfnamefont {G.~L.}\
  \bibnamefont {Chiarotti}}, \bibinfo {author} {\bibfnamefont {M.}~\bibnamefont
  {Cococcioni}}, \bibinfo {author} {\bibfnamefont {I.}~\bibnamefont {Dabo}},
  \bibinfo {author} {\bibfnamefont {A.~D.}\ \bibnamefont {Corso}}, \bibinfo
  {author} {\bibfnamefont {S.}~\bibnamefont {de~Gironcoli}}, \bibinfo {author}
  {\bibfnamefont {S.}~\bibnamefont {Fabris}}, \bibinfo {author} {\bibfnamefont
  {G.}~\bibnamefont {Fratesi}}, \bibinfo {author} {\bibfnamefont
  {R.}~\bibnamefont {Gebauer}}, \bibinfo {author} {\bibfnamefont
  {U.}~\bibnamefont {Gerstmann}}, \bibinfo {author} {\bibfnamefont
  {C.}~\bibnamefont {Gougoussis}}, \bibinfo {author} {\bibfnamefont
  {A.}~\bibnamefont {Kokalj}}, \bibinfo {author} {\bibfnamefont
  {M.}~\bibnamefont {Lazzeri}}, \bibinfo {author} {\bibfnamefont
  {L.}~\bibnamefont {Martin-Samos}}, \bibinfo {author} {\bibfnamefont
  {N.}~\bibnamefont {Marzari}}, \bibinfo {author} {\bibfnamefont
  {F.}~\bibnamefont {Mauri}}, \bibinfo {author} {\bibfnamefont
  {R.}~\bibnamefont {Mazzarello}}, \bibinfo {author} {\bibfnamefont
  {S.}~\bibnamefont {Paolini}}, \bibinfo {author} {\bibfnamefont
  {A.}~\bibnamefont {Pasquarello}}, \bibinfo {author} {\bibfnamefont
  {L.}~\bibnamefont {Paulatto}}, \bibinfo {author} {\bibfnamefont
  {C.}~\bibnamefont {Sbraccia}}, \bibinfo {author} {\bibfnamefont
  {S.}~\bibnamefont {Scandolo}}, \bibinfo {author} {\bibfnamefont
  {G.}~\bibnamefont {Sclauzero}}, \bibinfo {author} {\bibfnamefont {A.~P.}\
  \bibnamefont {Seitsonen}}, \bibinfo {author} {\bibfnamefont {A.}~\bibnamefont
  {Smogunov}}, \bibinfo {author} {\bibfnamefont {P.}~\bibnamefont {Umari}}, \
  and\ \bibinfo {author} {\bibfnamefont {R.~M.}\ \bibnamefont {Wentzcovitch}},\
  }\href@noop {} {\bibfield  {journal} {\bibinfo  {journal} {J. Phys. Condens.
  Matter}\ }\textbf {\bibinfo {volume} {21}},\ \bibinfo {pages} {395502}
  (\bibinfo {year} {2009})}\BibitemShut {NoStop}%
\bibitem [{\citenamefont {Fox}(2001)}]{fox2001optical}%
  \BibitemOpen
  \bibfield  {author} {\bibinfo {author} {\bibfnamefont {M.}~\bibnamefont
  {Fox}},\ }\href@noop {} {\emph {\bibinfo {title} {Optical Properties of
  Solides}}}\ (\bibinfo  {publisher} {Oxford University Press},\ \bibinfo
  {year} {2001})\BibitemShut {NoStop}%
\bibitem [{\citenamefont {Kresse}\ and\ \citenamefont
  {Joubert}(1999)}]{kresse1999vasp}%
  \BibitemOpen
  \bibfield  {author} {\bibinfo {author} {\bibfnamefont {G.}~\bibnamefont
  {Kresse}}\ and\ \bibinfo {author} {\bibfnamefont {D.}~\bibnamefont
  {Joubert}},\ }\href {\doibase 10.1103/PhysRevB.59.1758} {\bibfield  {journal}
  {\bibinfo  {journal} {Phys. Rev. B}\ }\textbf {\bibinfo {volume} {59}},\
  \bibinfo {pages} {1758} (\bibinfo {year} {1999})}\BibitemShut {NoStop}%
\bibitem [{\citenamefont {Wallbank}\ \emph
  {et~al.}(2015{\natexlab{b}})\citenamefont {Wallbank}, \citenamefont
  {Mucha-Kruczyński}, \citenamefont {Chen},\ and\ \citenamefont
  {Fal'ko}}]{wallbank2015moire}%
  \BibitemOpen
  \bibfield  {author} {\bibinfo {author} {\bibfnamefont {J.~R.}\ \bibnamefont
  {Wallbank}}, \bibinfo {author} {\bibfnamefont {M.}~\bibnamefont
  {Mucha-Kruczyński}}, \bibinfo {author} {\bibfnamefont {X.}~\bibnamefont
  {Chen}}, \ and\ \bibinfo {author} {\bibfnamefont {V.~I.}\ \bibnamefont
  {Fal'ko}},\ }\href {\doibase 10.1002/andp.201400204} {\bibfield  {journal}
  {\bibinfo  {journal} {Ann. Phys. (Berlin)}\ }\textbf {\bibinfo {volume}
  {527}},\ \bibinfo {pages} {359} (\bibinfo {year}
  {2015}{\natexlab{b}})}\BibitemShut {NoStop}%
\bibitem [{\citenamefont {Jung}\ \emph {et~al.}(2014)\citenamefont {Jung},
  \citenamefont {Raoux}, \citenamefont {Qiao},\ and\ \citenamefont
  {MacDonald}}]{jung2014ab}%
  \BibitemOpen
  \bibfield  {author} {\bibinfo {author} {\bibfnamefont {J.}~\bibnamefont
  {Jung}}, \bibinfo {author} {\bibfnamefont {A.}~\bibnamefont {Raoux}},
  \bibinfo {author} {\bibfnamefont {Z.}~\bibnamefont {Qiao}}, \ and\ \bibinfo
  {author} {\bibfnamefont {A.~H.}\ \bibnamefont {MacDonald}},\ }\href@noop {}
  {\bibfield  {journal} {\bibinfo  {journal} {Phys. Rev. B}\ }\textbf {\bibinfo
  {volume} {89}},\ \bibinfo {pages} {205414} (\bibinfo {year}
  {2014})}\BibitemShut {NoStop}%
\bibitem [{\citenamefont {Jung}\ \emph
  {et~al.}(2015{\natexlab{b}})\citenamefont {Jung}, \citenamefont {DaSilva},
  \citenamefont {MacDonald},\ and\ \citenamefont {Adam}}]{jung2015origin}%
  \BibitemOpen
  \bibfield  {author} {\bibinfo {author} {\bibfnamefont {J.}~\bibnamefont
  {Jung}}, \bibinfo {author} {\bibfnamefont {A.~M.}\ \bibnamefont {DaSilva}},
  \bibinfo {author} {\bibfnamefont {A.~H.}\ \bibnamefont {MacDonald}}, \ and\
  \bibinfo {author} {\bibfnamefont {S.}~\bibnamefont {Adam}},\ }\href@noop {}
  {\bibfield  {journal} {\bibinfo  {journal} {Nat. Commun.}\ }\textbf {\bibinfo
  {volume} {6}} (\bibinfo {year} {2015}{\natexlab{b}})}\BibitemShut {NoStop}%
\bibitem [{\citenamefont {Ferone}\ \emph {et~al.}(2011)\citenamefont {Ferone},
  \citenamefont {Wallbank}, \citenamefont {Zolyomi}, \citenamefont {McCann},\
  and\ \citenamefont {Fal’ko}}]{ferone2011manifestation}%
  \BibitemOpen
  \bibfield  {author} {\bibinfo {author} {\bibfnamefont {R.}~\bibnamefont
  {Ferone}}, \bibinfo {author} {\bibfnamefont {J.~R.}\ \bibnamefont
  {Wallbank}}, \bibinfo {author} {\bibfnamefont {V.}~\bibnamefont {Zolyomi}},
  \bibinfo {author} {\bibfnamefont {E.}~\bibnamefont {McCann}}, \ and\ \bibinfo
  {author} {\bibfnamefont {V.~I.}\ \bibnamefont {Fal’ko}},\ }\href@noop {}
  {\bibfield  {journal} {\bibinfo  {journal} {Solid State Commun.}\ }\textbf
  {\bibinfo {volume} {151}},\ \bibinfo {pages} {1071} (\bibinfo {year}
  {2011})}\BibitemShut {NoStop}%
\bibitem [{\citenamefont {Wallbank}\ \emph {et~al.}(2013)\citenamefont
  {Wallbank}, \citenamefont {Patel}, \citenamefont {Mucha-Kruczyński},
  \citenamefont {Geim},\ and\ \citenamefont {Fal'ko}}]{wallbank2013generic}%
  \BibitemOpen
  \bibfield  {author} {\bibinfo {author} {\bibfnamefont {J.~R.}\ \bibnamefont
  {Wallbank}}, \bibinfo {author} {\bibfnamefont {A.~A.}\ \bibnamefont {Patel}},
  \bibinfo {author} {\bibfnamefont {M.}~\bibnamefont {Mucha-Kruczyński}},
  \bibinfo {author} {\bibfnamefont {A.~K.}\ \bibnamefont {Geim}}, \ and\
  \bibinfo {author} {\bibfnamefont {V.~I.}\ \bibnamefont {Fal'ko}},\
  }\href@noop {} {\bibfield  {journal} {\bibinfo  {journal} {Phys. Rev. B}\
  }\textbf {\bibinfo {volume} {87}},\ \bibinfo {pages} {245408} (\bibinfo
  {year} {2013})}\BibitemShut {NoStop}%
\end{thebibliography}%

\clearpage

\renewcommand{\thefigure}{S\arabic{figure}}
\renewcommand{\thesubsection}{S\arabic{subsection}}
\renewcommand{\theequation}{S\arabic{equation}}
\setcounter{figure}{0} 
\setcounter{equation}{0}

\section*{Supplementary Information}

\subsection{Details of pressure experiments}

\begin{figure*}
\centering
\includegraphics[width=17cm]{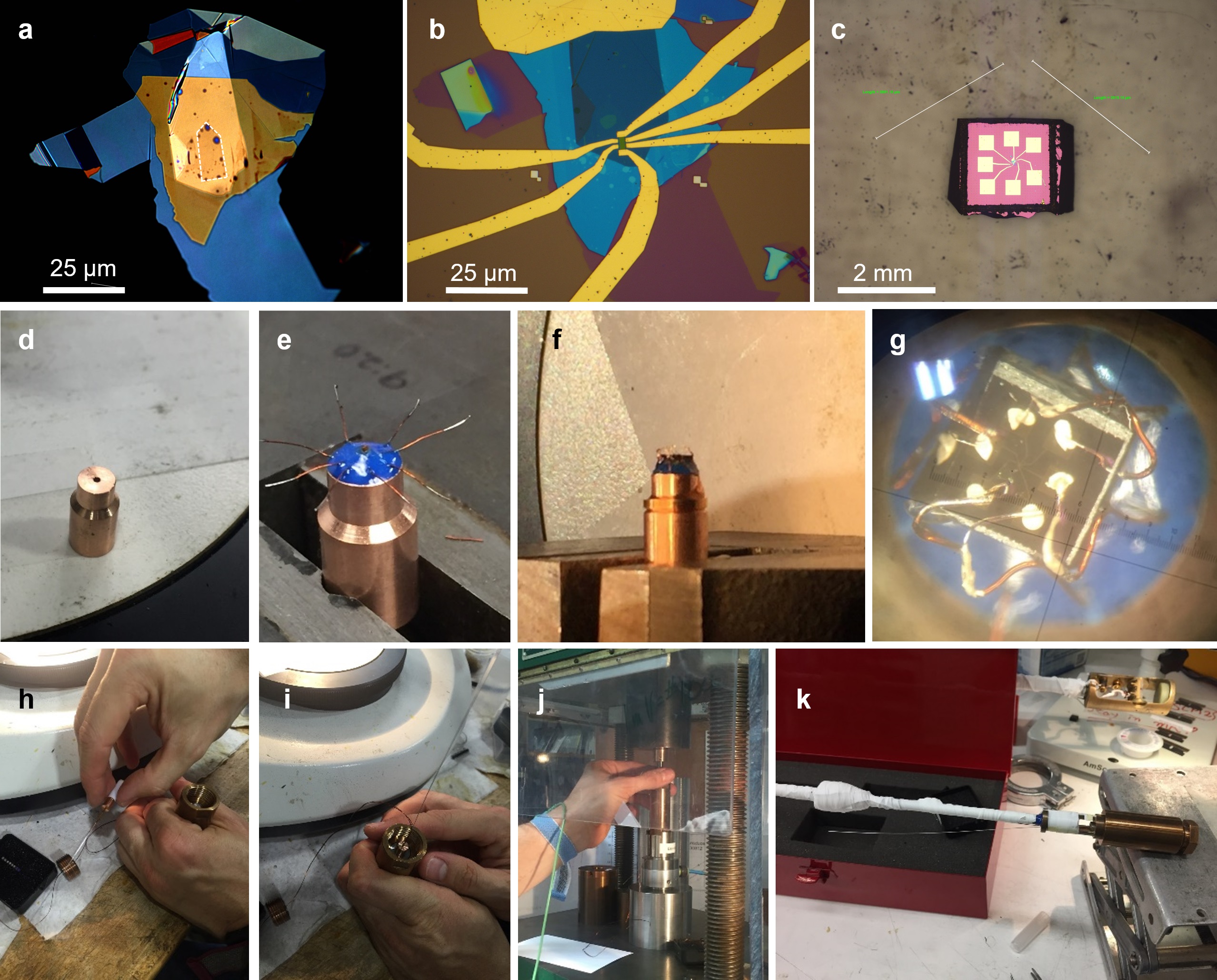} 
\caption{Pictures of the setup for pressure experiments (see text for details).}
\label{fig:experiment}
\end{figure*}

All devices in this study consist of monolayer graphene encapsulated between two layers of boron nitride. The BNs were generally between 20 - 60 nm thick, though our results did not depend on this in any noticeable way. The encapsulated stack sits on a flake of graphite which acts as a local back gate. Fig.~\ref{fig:experiment}a shows an image of a completed stack on a transfer slide. For aligned samples, the graphene (outlined with a dotted white line) was either intentionally aligned to one of the BNs using straight edges, or rotated to an aligned position with thermal heating during the transfer process. The final stack was partially etched into a Hall bar geometry, leaving some of the bottom BN unetched to prevent the metal contacts from shorting to the graphite gate (Fig.~\ref{fig:experiment}b). The Hall bar was intentionally kept small ($\approx$ 6 $\mu$m by 2 $\mu$m) to keep the pressure as uniform as possible across the entire device. The entire device sits on a Si/SiO$_2$ wafer, which must be diced to approximately 2 mm by 2 mm to fit into the inner bore of the pressure cell (Fig.~\ref{fig:experiment}c).

To prepare the pressure cell, a clean metal stage (Fig.~\ref{fig:experiment}d) is threaded with insulated copper 75 $\mu$m wires with tinned ends, which are epoxied in place using Stycast 2850 FT using 24LV catalyst (Fig.~\ref{fig:experiment}e). A ruby crystal is glued to the tip of a thin optical fiber that is fixed in place by the wires for \textit{in situ} pressure calibration. The sample is then glued above the fiber (Fig.~\ref{fig:experiment}f). Flexible 15 $\mu$m Pt wires are soldered to the copper wires and affixed by hand to the gold sample contacts using Dupong 4929N silver paste (Fig.~\ref{fig:experiment}g). Next, a Teflon cup is filled with the pressure medium (Daphne 7373 or 7474 oil) and carefully fitted over the sample and onto the stage (Fig.~\ref{fig:experiment}h). The sample is now completely encapsulated in oil. For the special case of vacuum measurements (0 GPa), no oil is loaded into the Teflon cup. The stage/teflon cup is then fitted into the inner bore of a piston cylinder cell (Fig.~\ref{fig:experiment}i), and a hydraulic press is used to compress the top of the Teflon cup (Fig.~\ref{fig:experiment}j). The oil medium hydrostatically increases in pressure as the cup is compressed. A top locking nut is tightened as more force is applied on the hydraulic press until the desired load is reached, at which point the hydraulic press is backed off, and the threads of the locking nut hold the pressure in the cell. The cell is then affixed to the end of a probe (Fig.~\ref{fig:experiment}k) for low-temperature, high-field measurements. Changing the pressure requires warming the cell to room temperature, adding or removing load with the hydraulic press, and re-cooling the cell.

Although the oil does not influence the electronic properties of the device, special care must be taken to account for its presence during measurements. The oil freezes at around 200 K as it is cooled at ambient pressure, and the freezing point moves to higher temperatures at higher starting loads. The pressure in the cell also drops as the oil cools inside the cell, with a larger relative drop in pressure at smaller initial loads. For example, a starting load below $\sim$ 0.3 GPa at room temperature will result in nearly ambient conditions at low temperature, while at pressures above 2 GPa there is virtually no change between the room temperature and low temperature pressures. We have found that the primary consequence of this effect is that starting at room temperature loads of roughly 0.5 GPa or less is dangerous for the device, as the stack is typically torn when the oil freezes below these pressures. However, above these pressures the devices always survive the cooling, and we have not noticed any effect of the oil freezing in transport measurements. Finally, special care is taken to account for the large thermal load of the pressure cell when performing temperature sweeps to measure band gaps. The temperature is swept slowly to keep the sample as close to equilibrium as possible, ideally at 0.5 K/min, and no faster than 1.5 K/min.

\subsection{Pressure dependent transport in other devices}

\begin{figure*}
\centering
\includegraphics[width=12cm]{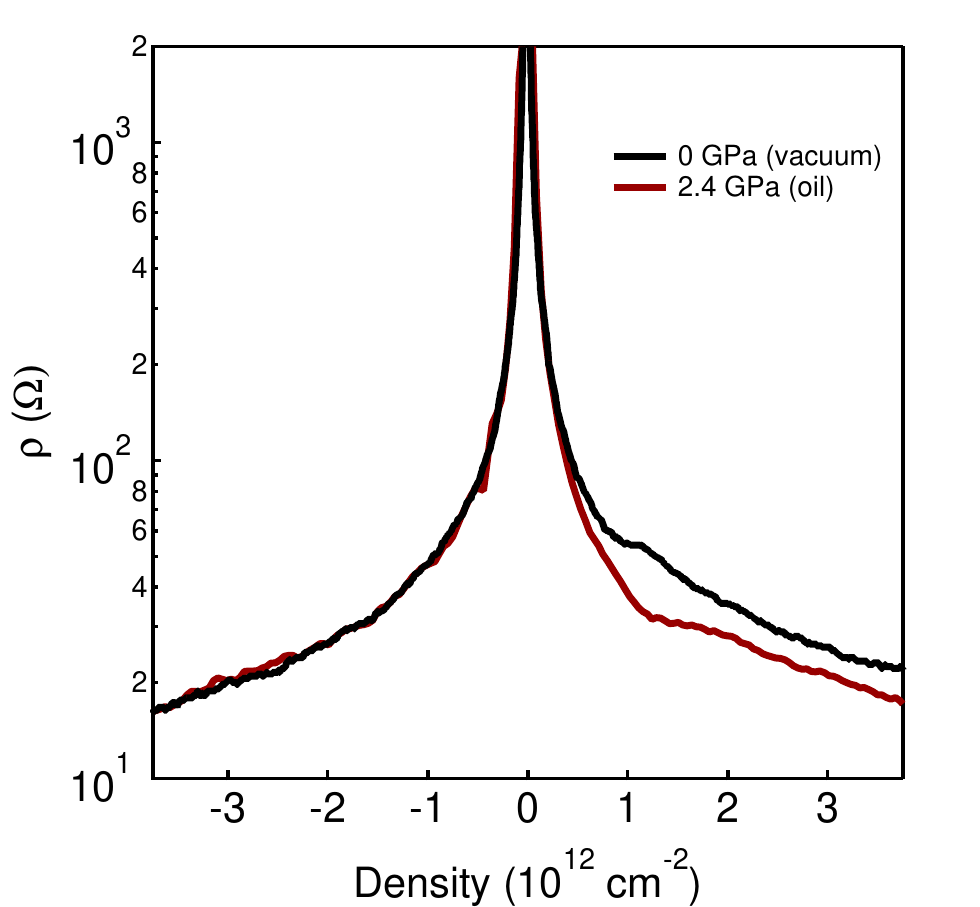} 
\caption{Resistance of Device P3 at $B$ = 0 T and $T$ = 2 K. Virtually no change in the device resistance is observed on the hole side of the device, and pressure serves to improve the resistance on the electron side.}
\label{fig:transport}
\end{figure*}

The behavior of high density resistance of devices under pressure varies slightly across devices. For Device P2 shown in Fig.~1b of the main text, an increase of 15-25 Ohms is observed at high density under pressure. However, this is not always the case, as is shown for Device P3 in Fig.~\ref{fig:transport}. This device exhibits virtually no pressure dependent hole-side resistance, and shows a decrease in the electron-side resistance at high pressure. 

We have measured a total of four gapped devices as a function of pressure. Fig.~\ref{fig:gaps}a plots the gap of the primary DP, where the data from Device P1 is copied from the main text. Error bars are left off for clarity, but are similar in magnitude to those in Fig. 3b of the main text. Devices P3 and P4 are slightly misaligned, such that the SDPs are outside the accessible density range in the device - therefore the rotation angle is unknown, but must be larger than $\sim$ 2$^\circ$. Previous work has demonstrated that devices with misalignment angles as large as $\sim$ 5$^\circ$ can exhibit band gaps in transport~\cite{Hunt2013}. In that study, the misalignment angle was determined by scanning tunneling topography measurements, however this is not possible in our devices due to the encapsulating top BN layer. Nevertheless, these devices show clear activation gaps, with insulating behavior at low temperatures and a decrease of the device conductivity by nearly an order of magnitude in the thermally activated regime at high temperatures (Fig.~\ref{fig:conductivity}a - c). The magnitude of the gap grows with pressure in all three samples.

\begin{figure*}
\centering
\includegraphics[width=17cm]{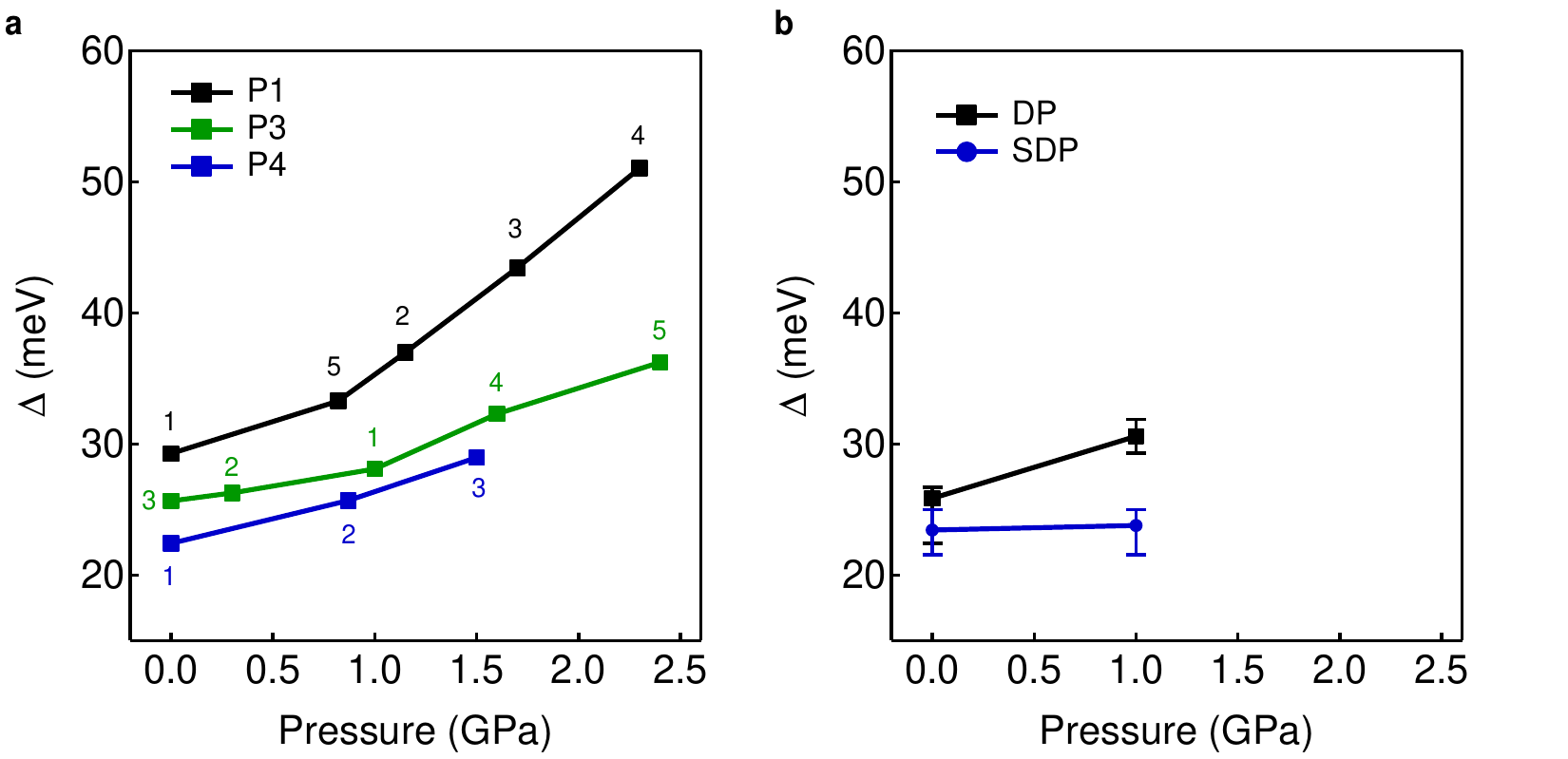} 
\caption{Band gaps as a function of pressure in other devices. (a) The numbered labels represent the order in which the gaps were acquired. The gap magnitude depends only only the pressure the device is under at the time of measurement, and not the history of the pressure that has been previously applied. (b) Gap dependence of a fourth gapped device (P5), which was nearly perfectly aligned. This device also exhibits a growing primary DP gap and constant SDP gap.}
\label{fig:gaps}
\end{figure*}

The fourth gapped device (P5) was very well aligned, with an estimated misalignment angle of $\sim$0.1$^{\circ}$. However, the device was significantly more disordered than all the other devices in this study, and as a result the onset of the variable range hopping regime occurred at significantly higher temperature. Consequentially, the simply-activated regime spanned only a small range of temperature, and therefore the gap extraction is significantly more uncertain than the other gapped devices examined. Nevertheless, the primary DP showed an unambiguous enhancement with pressure, while the SDP appeared relatively insensitive to pressure (Fig.~\ref{fig:gaps}b). This behavior is completely consistent with the clean aligned device (P1) presented in the main text, providing further evidence of the fundamental difference between the primary and secondary DPs. Fig.~\ref{fig:sdpconductivity} additionally shows the gap extraction for the SDP of the aligned device P1 in the main text Fig. 3b, demonstrating that this gap does not change with pressure.

\begin{figure*}
\centering
\includegraphics[width=17cm]{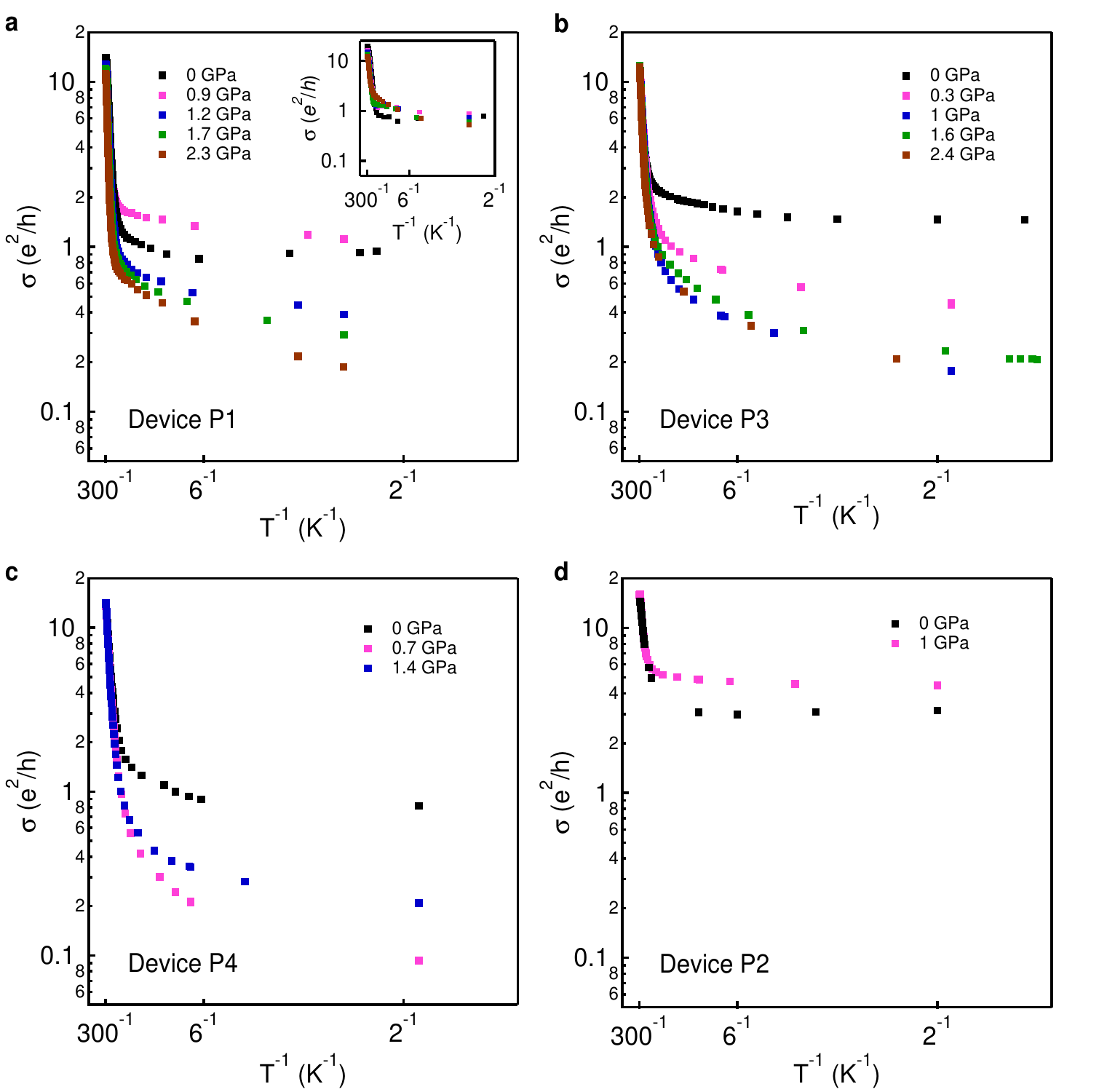} 
\caption{Conductivity of the DP over the full temperature range. Devices are (a) P1 and (inset) the SDP (b) P3, and (c) P4. (d) Similar plot for a misaligned device (P2) which does not exhibit strong insulating behavior, where pressure has minimal effect. While the DP generally grows more insulating with higher pressure at low temperatures, this is not universally true, suggesting that the details of the low temperature resistance depends more critically on the exact nature of disorder in the device.}
\label{fig:conductivity}
\end{figure*}

\begin{figure*}
\centering
\includegraphics[width=12cm]{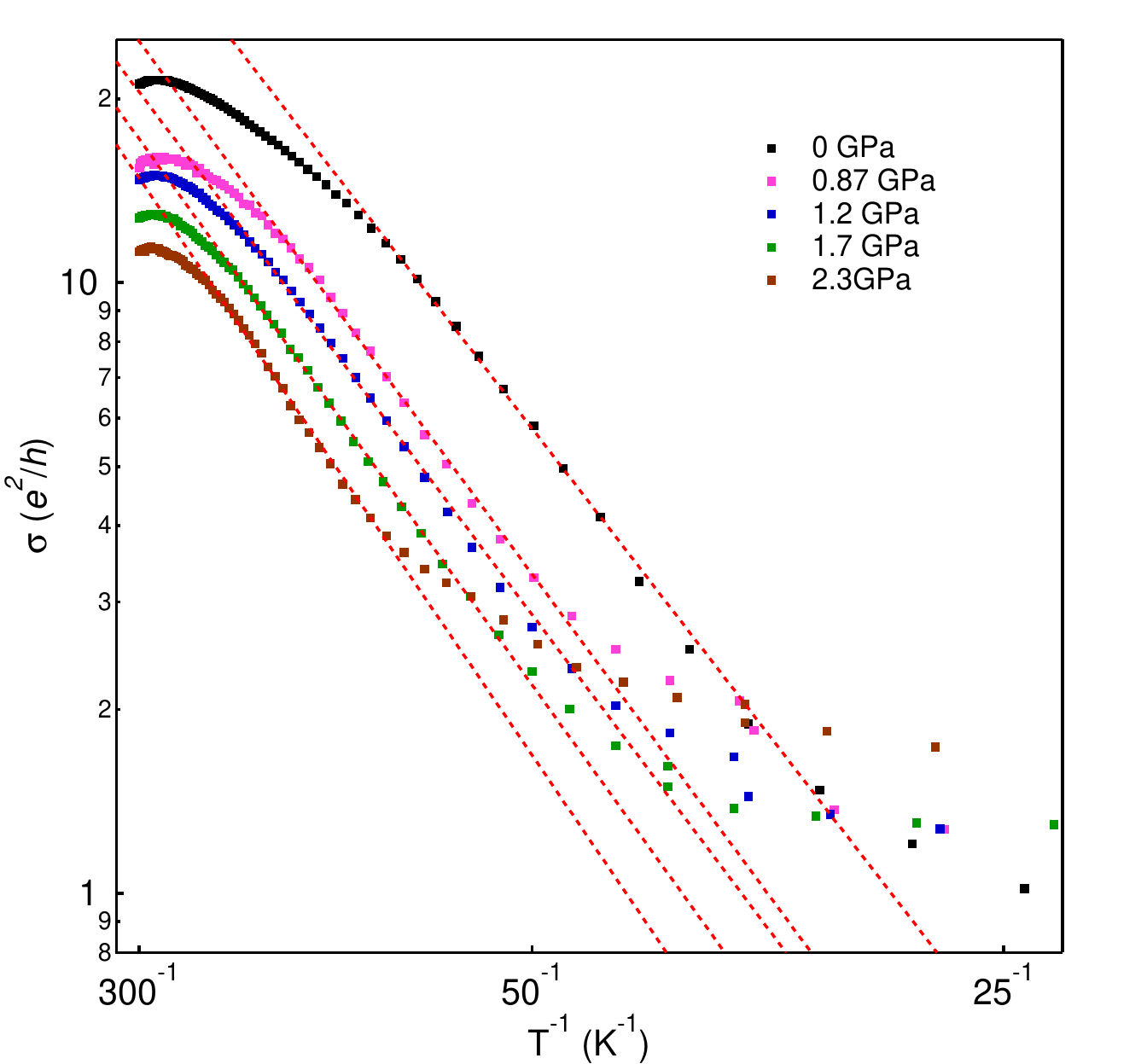} 
\caption{Arrhenius plots for the SDP in Device P1. The gap has virtually no dependence on pressure.}
\label{fig:sdpconductivity}
\end{figure*}

The enhancement of the band gap does not depend on the history of the pressure the device has been exposed to, rather it responds directly to the pressure the device is under at the time of measurement. To illustrate this, Fig.~\ref{fig:gaps}a annotates the order in which the gaps were acquired. In no case does the measured gap fall out of the anticipated sequence. This is an especially robust effect, as it does not depend on whether the device becomes more or less insulating with pressure. Fig.~\ref{fig:conductivity} tracks the CNP conductivity across a number of devices down to the lowest temperatures measured ($\sim$ 2 K). Typically the devices become more insulating at low temperatures for higher pressures, however this is not always the case. For Devices P1 and P3, the low temperature conductivity becomes out of sequence upon unloading pressure, or in reloading pressure after previously pressure cycling and returning to 0 GPa (see the annotations in Fig.~\ref{fig:gaps}a for the order of curve acquisition). However, even in Device P4, where the pressure was loaded up uniformly, the low temperature conductivity still depends non-monotonically on pressure. This suggests that small changes in the amount of disorder in the device and potentially even the details of its microscopic organization can ultimately influence how insulating the device becomes. However, no matter how the DP conductivity behaves at low temperature, the band gap is always enhanced by pressure, suggesting that the gap enhancement is a robust property of the heterostructure and depends most critically on the interlayer interaction strength between the graphene and the BN rather than small changes in device disorder.

\subsection{Effects of pressure in the quantum Hall regime}

\begin{figure*}
\centering
\includegraphics[width=12cm]{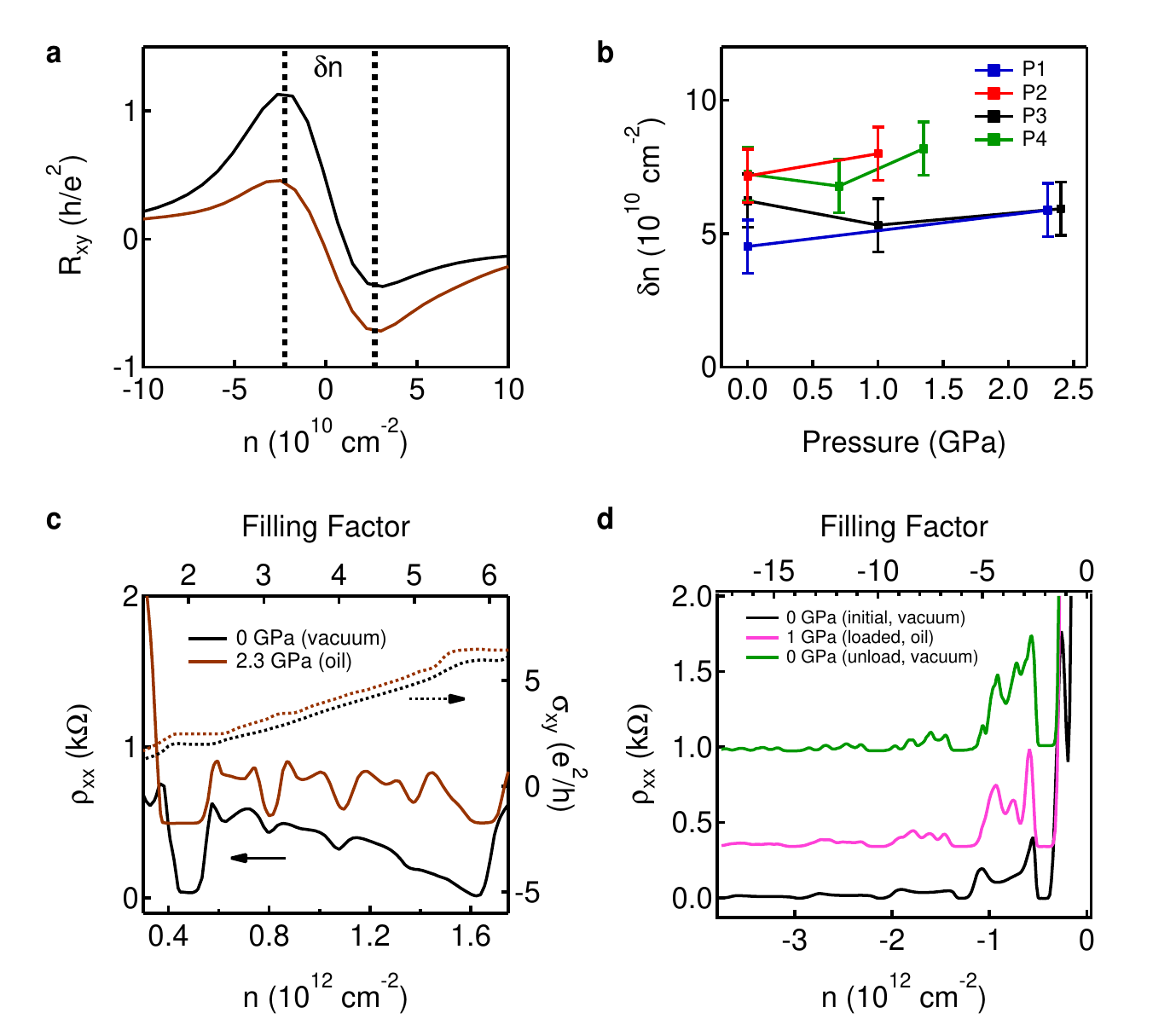} 
\caption{Disorder and the quantum Hall effect. (a) Device disorder $\delta n$ is measured as the $n$ necessary to dope between the maximum and minimum $R_{xy}$, averaged over $B$ = 0.25 T to 0.75 T. (b) Density fluctuations as a function of pressure across different devices. The error bars denote the uncertainty in picking the peak positions of $R_{xy}$. (c) $\rho_{xx}$ (solid lines, left axis) and $\sigma_{xy}$ (dotted lines, right axis) of Device P1 at $B$ = 12.5 T. Symmetry broken integer QH states become much more clearly developed with pressure, and fractional QH states begin to emerge as well. Both are offset for clarity. (d) $\rho_{xx}$ of Device P2 at $B$ = 9 T, similarly showing an improvement in the QH response. Surprisingly, the improvement persists even after the pressure is released back to vacuum and the device is cleaned in solvents (green curve). Curves are vertically offset for clarity.}
\label{fig:qhtransport}
\end{figure*}

The disorder in these devices can also be characterized by considering the amount of density $n$ necessary to switch the Hall resistance $R_{xy}$ from positive to negative (Fig.~\ref{fig:qhtransport}a), as this gives a measure of the effective magnitude of the electron-hole puddles in the bulk. Fig.~\ref{fig:qhtransport}b shows that the device disorder is not strongly dependent on pressure, growing by less than the extraction uncertainty over the applied pressure range. Curiously, pressure serves to qualitatively improve the quantum Hall response at high magnetic field. Fig.~\ref{fig:qhtransport}c shows one such example, where both integer and fractional quantum Hall states develop more clearly under pressure. This effect is observed in every device examined. It is even more surprising that this improvement persists even after the pressure is released (Fig.~\ref{fig:qhtransport}d), as this suggests the need for an explanation describing an irreversible effect. At present, we may only speculate as to the nature of this effect. 

First, it is important to note that despite the high electronic quality of the devices, only the main sequence of IQH gaps at $\nu$ = -2, -6, -10, etc., are typically clearly developed in initial measurements in vacuum, with the symmetry broken gaps still developing (Fig.~\ref{fig:qhtransport}d). This is not an intrinsic property of the graphene -- rather it is symptomatic of the metal contacts sitting above the graphite gate in the device geometry employed here (Fig.~\ref{fig:experiment}b). While the reason for this is currently not well understood, the situation may be analogous to previous observations in GaAs, where a partial reduction of the carrier concentration in the graphene just in front of the contacts can lead to a depletion region along the boundary, which may impede the ability to observe well-developed quantum Hall plateaus~\cite{Weis2011}. This problem has been addressed previously by leaving Si-gated regions of the graphene Hall bar leads to act as contacts, where the density at the boundary can vary more smoothly~\cite{Maher2014}. However, the devices in this study were intentionally kept small to ensure the pressure is as uniform as possible across the device, and so this geometry was not utilized.

Applying pressure may, for instance, provide a way to reduce this depletion region at the contacts, permitting better coupling to the QH edge modes. This could arise due to self-cleaning of contaminants along the contact boundaries, or if the metal making edge contact to the graphene forms a better bond to the graphene edge and lowers the work function mismatch at the boundary. Both of these effects could in principle persist even after the pressure is released. We find that the contact resistance is not significantly modified by pressure at $B$ = 0 T (Fig.~1b of the main text), pointing to an improvement arising from the detailed electrostatics at the contact barrier in high magnetic field, or to a reduction of non-local contamination which may be most relevant in the QH regime. Further study is necessary to understand the exact nature of this effect.

\begin{figure*}
\centering
\includegraphics[width=16cm]{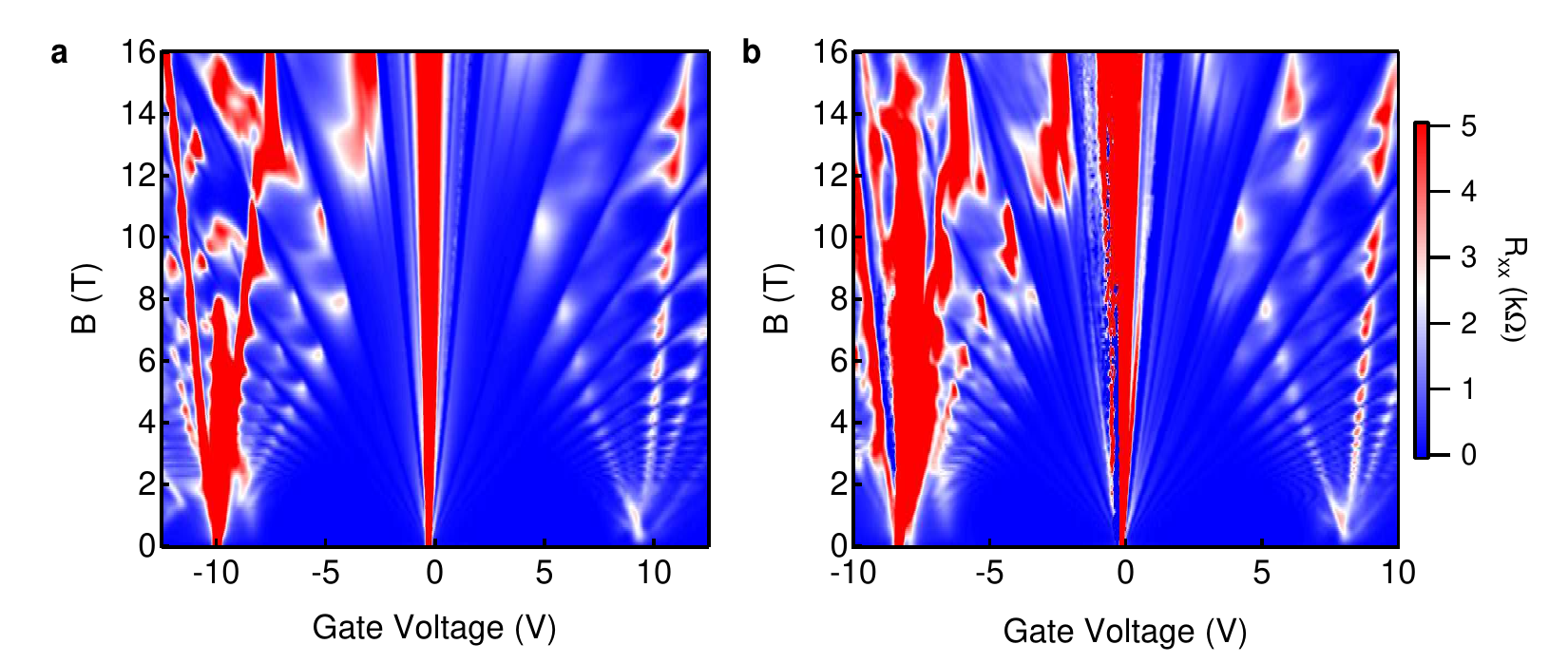} 
\caption{Hofstadter butterfly as a function of pressure at (a) 0 GPa and (b) 2.3 GPa.}
\label{fig:butterfly}
\end{figure*}

Finally, we examine the effects of pressure on the Hofstadter butterfly spectrum of the aligned Device P1. Fig.~\ref{fig:butterfly} shows the high field response at ambient conditions (panel a) and at 2.3 GPa (panel b). As pressure enhances the effective strength of the moir\'e potential, it should also influence the relative size of the LL gaps originating from the secondary DPs. The qualitative behavior of the two butterfly maps seems to be in qualitative agreement with this expectation, with stronger features originating from the secondary DPs under pressure. However, this must be deconvolved with the overall change in the QHE behavior with pressure, and a full investigation of this effect is outside the scope of this work. 

\subsection{First principle calculations and parametrizations}

The atomic and electronic structures of graphene and hexagonal boron nitride (G/BN) interfaces are captured using input from {\em ab-initio} density functional theory (DFT) calculations.
We rely on exact exchange and random phase approximation (EXX+RPA) input for atomic structure, while our treatment of the electronic structure is based on local-density approximation (LDA). 

\subsubsection*{Calculation of the dielectric constant from DFT}
\begin{figure*}
\begin{center}
\includegraphics[width=0.45\textwidth]{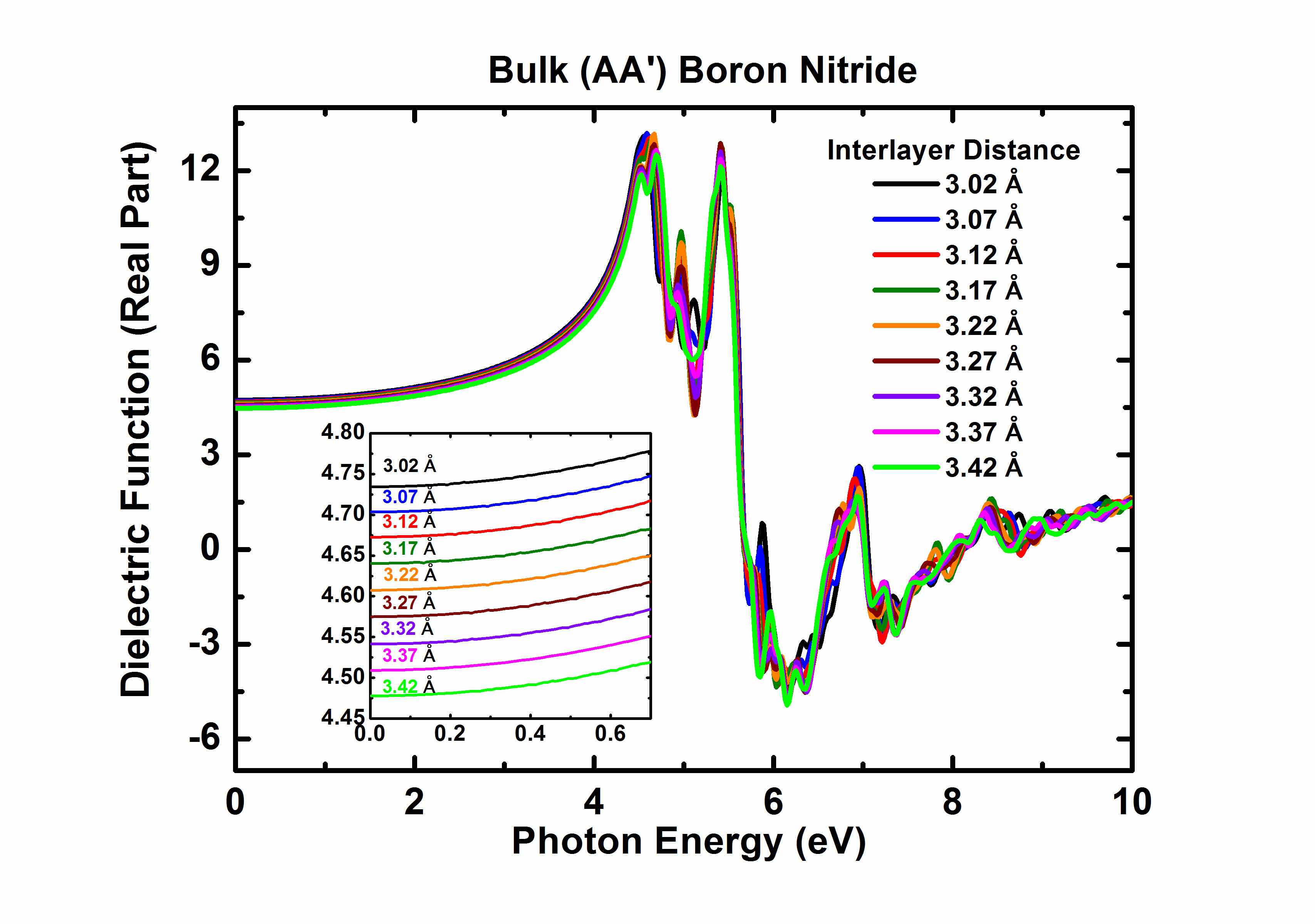}
\end{center}
\caption{
Real part of the dielectric function as a function of energy $\omega$ obtained for different interlayer spacing in bulk AA$'$ stacked hexagonal boron nitride. The AA$'$ stacking have alternating B and N atoms in adjacent vertical layers.
}
\label{dielectricconstant}
\end{figure*}

One of the striking features of the experiment is an increase of the dielectric constant as a function of pressure. 
Our DFT calculations for bulk confirm this behavior showing increases on the order of 3\% for the applied pressures up to 2.5~GPa as shown in Fig.~\ref{dielectricconstant}.
When the pressure is modified from $P=0$ to $P=2.5$~GPa,
the EXX+RPA equilibrium interlayer distance is squeezed by $\sim 0.2~\angstrom$ as we will discuss in the following subsection.

%

First principles calculations based on density functional theory (DFT) were performed using Quantum Espresso \cite{giannozzi2009quantum} under LDA  
within the plane wave basis set. Norm-conserving Vanderbilt pseudopotentials were used. 
The structures are fully optimized without any 
symmetry constraint by using conjugate gradient method.
The convergence criterion for the force on each ion is
taken to be less than 0.005~eV$/\angstrom$ while the energy is converged with a tolerance of 10$^{-5}$~eV.

In an insulating BN, the optical properties are dominated with the direct inter band contributions to the absorptive or imaginary part of dielectric function \cite{fox2001optical}. The matrix elements for a given interband transition $\beta \rightarrow \alpha$ for a set of plane wave Bloch function: 

\begin{widetext}
\begin{equation}
\left| {\psi_{k,n} } \right\rangle  = e^{iG \cdot r} u_{k,n}  = \frac{1}{{\sqrt V }}\sum\limits_G {a_{n,k,G} e^{i(k + G) \cdot r} } 
\end{equation}

\noindent is given by:

\begin{equation}
\hat{M}_{\alpha ,\beta }  = \left( {\sum\limits_G {a_{n,k,G}^* a_{n',k,G} G_\alpha  } } \right)\left( {\sum\limits_G {a_{n,k,G}^* a_{n',k,G} G_\beta  } } \right)
\end{equation}
\end{widetext}

These matrix elements account only for the interband transitions, i.e the electric-dipole approximation where the momentum transfer is zero. The imaginary part of the dielectric tensor $\varepsilon_{2\alpha ,\beta }$ is a response function that comes from a perturbation theory within adiabatic approximation. All of the possible transitions from the occupied to the unoccupied states without local field effects are given by the imaginary dielectric function:

\begin{widetext} 
\begin{equation}
\varepsilon _{2\alpha ,\beta }  = \frac{{4\pi e^2 }}{{\Omega N_k m^2 }}\sum\limits_{n,n'} {\sum\limits_k {\frac{{\hat{M}_{\alpha ,\beta } }}{{(E_{k,n'}  - E_{k,n} )^2 }}} } \left\{ {\frac{{f(E_{k,n} )}}{{E_{k,n'}  - E_{k,n}  + \hbar \omega  + i\hbar \Gamma }} + \frac{{f(E_{k,n} )}}{{E_{k,n'}  - E_{k,n}  - \hbar \omega  - i\hbar \Gamma }}} \right\},
\end{equation}
\end{widetext}

\noindent where $\Gamma$ is the adiabatic parameter which must be zero due to conservation of energy. So, the equation can be rewritten in terms of Dirac delta functions: 

\begin{widetext}
\begin{equation}
\varepsilon_{2\alpha ,\beta }  = \frac{{4\pi e^2 }}{{\Omega N_k m^2 }}\sum\limits_{n,n'} {\sum\limits_k {\frac{{\hat{M}_{\alpha ,\beta } f(E_{k,n} )}}{{(E_{k,n'}  - E_{k,n} )^2 }}} } \left[ {\delta (E_{k,n'}  - E_{k,n}  + \hbar \omega ) + \delta (E_{k,n'}  - E_{k,n}  - \hbar \omega )} \right],
\end{equation}
\end{widetext}

However, interaction with the electromagnetic field which also take place even in the absence of photons, i.e spontaneous emission, brings intrinsic broadening to all excited states that leads to a finite life time ($\Gamma > 0$). In the limit of non-vanishing $\Gamma$, the dielectric tensor takes the Drude-Lorentz form \cite{fox2001optical}:
 
\begin{widetext}
\begin{equation}
\varepsilon _{2\alpha ,\beta }  = \frac{{4\pi e^2 }}{{\Omega N_k m^2 }}\sum\limits_{n,k} {\frac{{df(E_{k,n} )}}{{dE_{k,n} }}\frac{{\eta \omega \hat{M}_{\alpha ,\beta } }}{{\omega^4  - \eta ^2 \omega ^2 }} + \frac{{8\pi e^2 }}{{\Omega N_k m^2 }}\sum\limits_{n \ne n'} {\sum\limits_k {\frac{{\hat{M}_{\alpha ,\beta } }}{{E_{k,n'}  - E_{k,n} }}} } } \frac{{\Gamma \omega f(E_{k,n} )}}{{[(\omega _{k,n'}  - \omega_{k,n} )^2  - \omega^2 ]^2  + \Gamma^2 \omega ^2 }},
\end{equation}
\end{widetext}

\noindent and the real part of the dielectric function is derived from the Kramers-Kronig transformation:

\begin{widetext}
\begin{equation}
\varepsilon_{1\alpha ,\beta }  = 1 + \frac{2}{\pi }\int\limits_0^\infty  {\frac{{\omega '\varepsilon _{2\alpha ,\beta } (\omega ')}}{{\omega'^2  - \omega ^2 }}d\omega'}. 
\end{equation}
\end{widetext}

\noindent Calculations of the dielectric function were carried out using the EPSILON package 
distributed with Quantum Espresso~\cite{giannozzi2009quantum}.


\subsubsection*{DFT deformation with pressure}

\begin{figure*}
\begin{center}
\includegraphics[width=13cm]{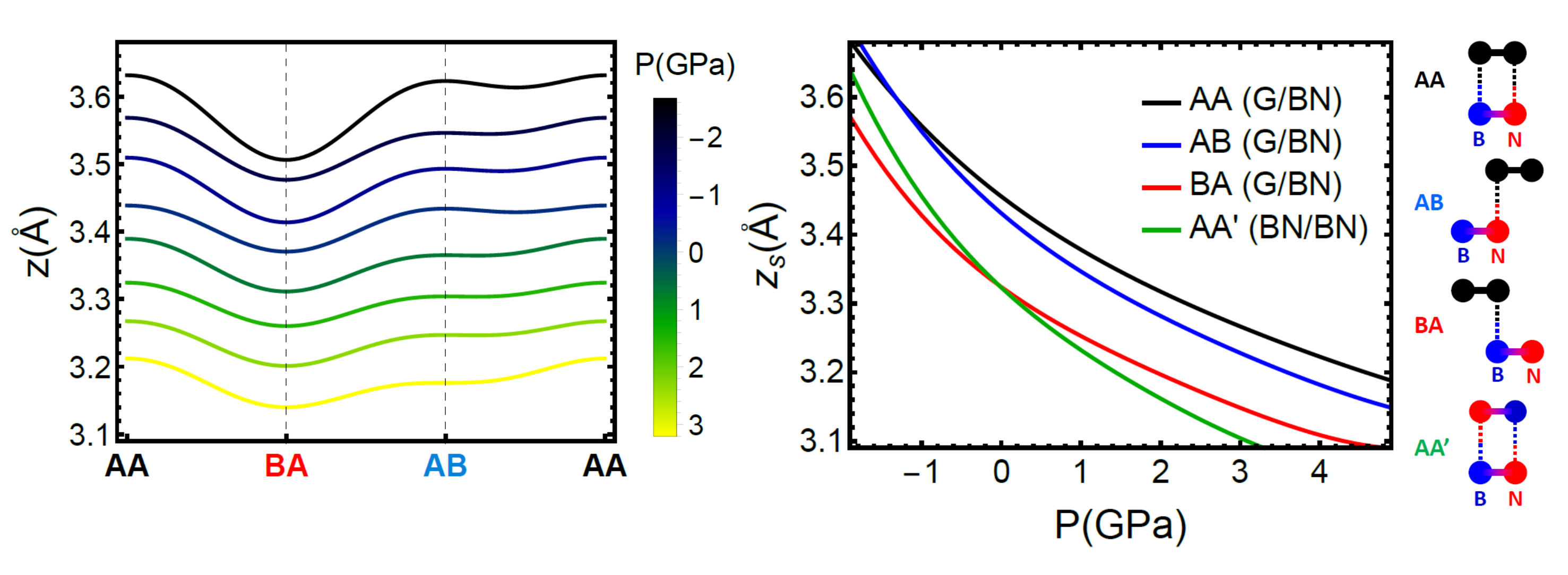} 
%
%
\end{center}
\vspace{-20pt}
\caption{
{\em Left panel:}
Corrugation profiles as a function of pressure neglecting in-plane strains where the interaction with both the substrate and the top layers are accounted for.
{\em Right panel:}
The equilibrium interlayer distances are plotted against pressure for  different stacking configurations in G/BN and for the most stable configuration in bulk hexagonal boron nitride (AA$'$).
}
\label{corrugpott}
\end{figure*}

The mechanical behavior of nearly aligned graphene and boron nitride heterostructure (G/BN) is influenced by the encapsulating top BN layer.
The interlayer interaction of nearly aligned G/BN is described using the assumption that they can be viewed as a collection of different stacking registries to be treated independently, from which the atomic structure of the whole sample can be obtained. 
We obtain the equilibrium spacing for each stacking point $z_{\mathbf{s}}$ $(\mathbf{s}\in \{AA, AB, BA\}$, see Fig. \ref{corrugpott}) by minimizing the total potential energy as a function of the
separation distance between the top and bottom BN layer $d$. We also assume that the elastic resistance of the graphene sheet is negligible in the out-of-plane direction. 
The interlayer potential energy of the graphene sheet considering the effects of bottom and top BN layers is given by:

\begin{widetext}
\begin{equation}
V(z_{AA},z_{AB},z_{BA},d) = \frac{1}{3}\sum_{\mathbf{s}}[V_{\mathbf{s}}(z_{\mathbf{s}}) + \bar{V}(d - z_{\mathbf{s}})]
\end{equation}
\end{widetext}

\noindent where $\bar{V}(d-z) = \frac{1}{3}\sum_{\mathbf{s}} V_{\mathbf{s}}(d-z)$ accounts for the averaging of the interlayer potential between the top BN and the graphene sheet due to their misalignment.
We use for the interlayer interaction potentials the exact exchange and random phase approximation (EXX+RPA) for G/BN and BN/BN interactions as implemented in VASP.~\cite{kresse1999vasp}
%
In equilibrium, the local pressure exerted by the encapsulating layer is equally balanced by that coming from the underlying substrate, i.e. $
P_{\mathbf{s}}(z_{\mathbf{s}}) = \bar{P}(d - z_{\mathbf{s}})$. Defining the interlayer distance between graphene and the top layer $z'_{\mathbf{s}} = d - z_{\mathbf{s}}$, we can write the equilibrium condition as $\frac{1}{3}\sum_{\mathbf{s}} P_{\mathbf{s}}(z_{\mathbf{s}}) = \frac{1}{3}\sum_{\mathbf{s}} \bar{P}(z'_{\mathbf{s}}) = P$, where $P$ is the external pressure applied on the system. Fig.~\ref{corrugpott} illustrates how the corrugation profiles of a graphene sheet change with pressure under the influence of two BN layers.
%
%

\subsubsection*{Electronic coupling parameters and elastic properties of G/BN}

In our treatment of the moir\'e bands in G/BN, we consider modifications in the low-energy regime of graphene electronic structure which are introduced by the interlayer tunneling. The modification is captured by two interlayer coupling terms $V$ and $\tilde{V}$ whose values can be obtained from first-principle calculations. $V$ and $\tilde{V}$ can be calculated from the following expressions \cite{wallbank2015moire}:  

\begin{widetext}
\begin{equation}
V = \frac{1}{2}\bigg[\frac{t_{NC}^{2}}{|\epsilon_{N}|}-\frac{t_{BC}^{2}}{|\epsilon_{B}|}\bigg]\approx 0.04 \text{ meV},\hspace{10pt}
\tilde{V} = \frac{\sqrt{3}}{2}\bigg[\frac{t_{NC}^{2}}{|\epsilon_{N}|}+\frac{t_{BC}^{2}}{|\epsilon_{B}|}\bigg]\approx 10 \text{ meV},
\end{equation} 
\end{widetext}

\noindent where $t_{NC}$ ($t_{BC}$) stands for the interlayer hopping term between C--N (C--B) atoms which highly depends on the interlayer separation, and $\varepsilon_{N}$ ($\varepsilon_{B}$) are the on-site energies of N (B) atoms. Both parameters are calculated from the parametrizations of the interlayer terms in Ref. \cite{jung2014ab}, at a separation of $z_{r} = 3.35\,\angstrom$. 
To fully account for the modification in the electronic couplings, \textit{ab-initio} method in Ref. \cite{jung2015origin} introduces three exponential factors $\{\beta_{AA},\beta_{AB},\beta_{BB}\}$, each quantifies the rate of increase of each sublattice term ($\beta_{i} > 0$). The values of $\beta_{i}$'s are determined by the amount overlap between carbon orbitals with the orbitals of the underlying BN atoms, for which it ranges between 3.0--3.3 $\angstrom^{-1}$ \cite{jung2015origin}.

As sample deformation is expected to play an important role in determining the band features, we model the structural relaxation within Born-von Karman plate theory in which the elastic properties of graphene are fully characterized by two Lam\'e constants: $\lambda_{g} = 3.25 \text{ eV }\angstrom^{-2}$ and $\mu_{g} = 9.57 \text{ eV }\angstrom^{-2}$ \cite{jung2015origin}. One mechanism through which deformation leaves an imprint on the bands is the intralayer effects which are induced by the change of the carbon on-site energy and nearest-neighbor hoppings as each carbon atom is displaced with respect to its neighbors. They give rise to additional terms in the moir\'e couplings which are equivalent to the electrostatic potentials and the pseudomagnetic fields, quantified by $\gamma_{V}\approx 4.0$ eV and $\gamma_{B}\approx 4.5$ eV respectively \cite{ferone2011manifestation}.

\subsection{Theoretical Model}

\begin{figure*}
\begin{center}
\includegraphics[width=0.7\textwidth]{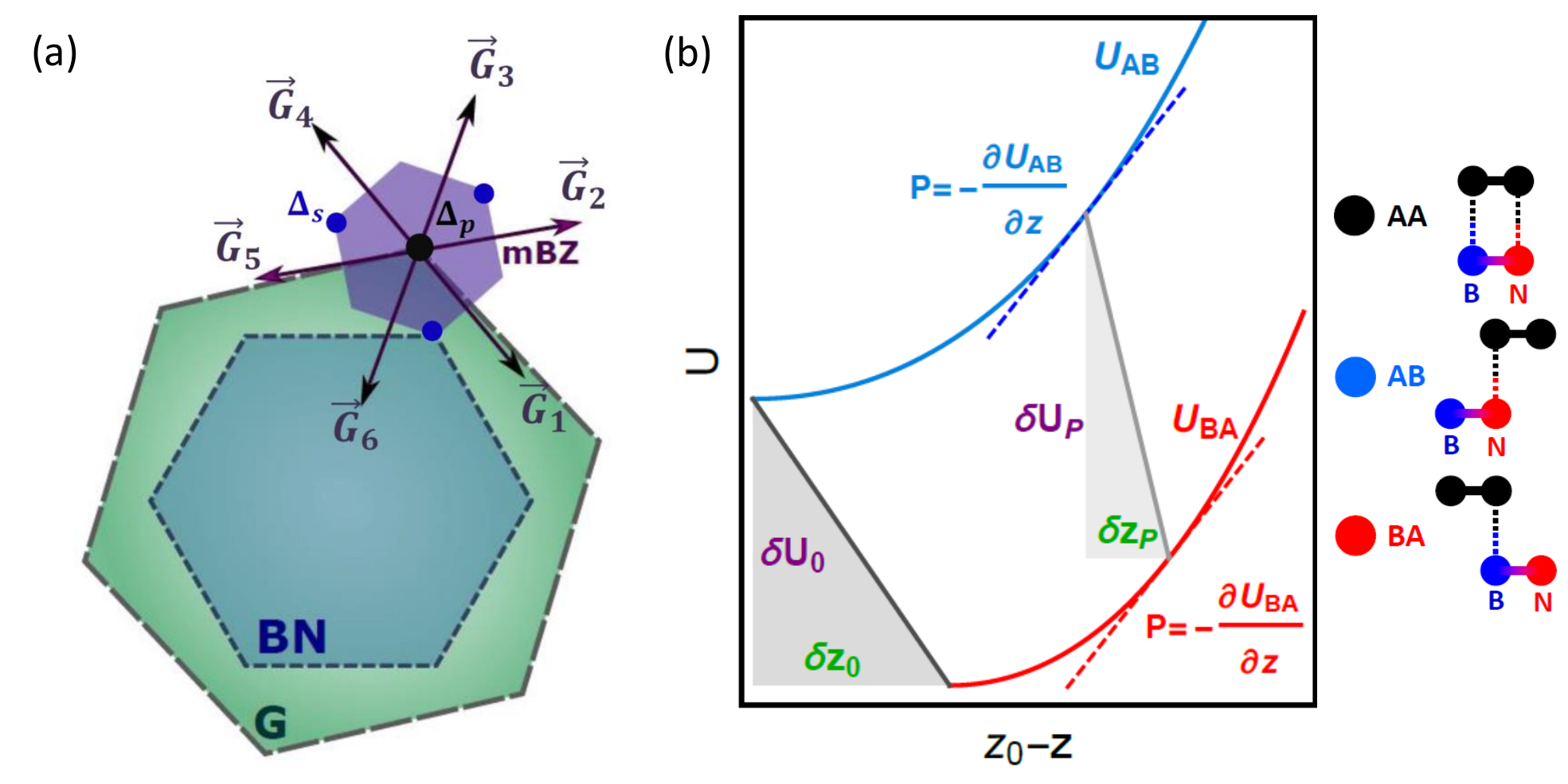}
\end{center}
 \caption{(a) Illustration of the moir\'e Brillouin zone (mBZ) in relation to the reciprocal space of G/BN with slight misalignment $\theta$. The presence of BN substrate introduces a characteristic long-wavelength potential which is one to two-order of magnitude larger than carbon-carbon distance, as defined by $\vec{G}_{m}$. The Dirac cone is located on the black dot, at which BN sublattice asymmetry leads to a primary gap $\Delta_{p}$. The moir\'e potentials lead to gapped secondary Dirac cones $\Delta_{s}$ at the mBZ edges, marked by blue dots. (b) Deformation on graphene is characterized by the potential profiles on the special stacking points: AA (black), AB (blue) and BA (red). 
 When pressure $P$ is applied on the system, the carbon atoms increase their potential energies until a new equilibrium strain profile is reached. On the right panel we show for AB and BA stacking 
how the pressure can influence the in-plane and out-of-plane deformations, respectively proportional to variations in the potential ($\delta U$) and spacing ($\delta z$).}
 \label{potStack}
\end{figure*}

Graphene and BN are two-dimensional materials with hexagonal crystal structures, but with different lattice constants. When graphene is deposited on a BN substrate with near perfect alignment (small twist angles), the slight lattice mismatch  ($\varepsilon = \frac{a_{G}-a_{BN}}{a_{BN}} = -1.7\%$), leads to the formation of a moir\'e structure which is characterized by a wavelength that is one or two orders of magnitude larger than graphene's unit cell. As a consequence, the dynamics of the low energy Dirac electrons of graphene is governed by the corresponding long-wavelength component of the electronic couplings. Similarly, the interlayer potentials and deformations in G/BN are also largely characterized by this long-wavelength theory.  It is by now established that the dominant contributions are captured by the first harmonics of the moir\'e structure
characterized by a set of vectors $\vec{G}_{m}$, that is related with the original graphene's lattice vectors $\vec{g}_{m}$:  

\begin{widetext}
\begin{equation}
\vec{g}_{m} = 
\hat{R}_{2\pi (m-1)/6}\,\big(0,g\big) ,\hspace{5pt}
\vec{G}_{m} = \big[(1+\varepsilon)-\hat{R}_{\theta}\big]\vec{g}_{m}\approx
\varepsilon\vec{g}_{m}-\theta\hat{z}\times\vec{g}_{m},
\label{firstHarmonics}
\end{equation}
\end{widetext}

\noindent where $m \in \{1,2,...,6\}$, $\hat{R}_{\theta}$ denotes a rotation by $\theta$, $g = 4\pi/3a$ is the length of graphene lattice vectors with $a\approx 1.42\angstrom$ stands for the carbon-carbon distance, and the approximate sign indicates the approximation within small twist angle limit ($\theta \ll 1$). 
We also show in Fig.~\ref{potStack}, that $\vec{G}_{m}$ defines the moir\'e Brillouin Zone (mBZ), whose lateral dimension is scaled by $\tilde{\varepsilon}=\sqrt{\varepsilon^{2}+\theta^{2}}$  with respect to graphene BZ that amounts to values $\lesssim 5\%$ for twist angles $\theta \lesssim 2^{\circ}$. In G/BN system which possesses triangular symmetry, we can define two periodic functions within the first harmonics: (1) $f_{1}(\vec{r}) = \sum_{m}\exp(i\vec{G}_{m}\cdot\vec{r})$ which satisfies inversion and hexagonal symmetries, and (2) $f_{2}(\vec{r}) = -i\sum_{m}(-1)^{m}\exp(i\vec{G}_{m}\cdot\vec{r})$ which is asymmetric under inversion.

\subsubsection*{Definition of symmetric and antisymmetric deformations}

\begin{figure*}
\begin{center}
\includegraphics[width=0.82\textwidth]{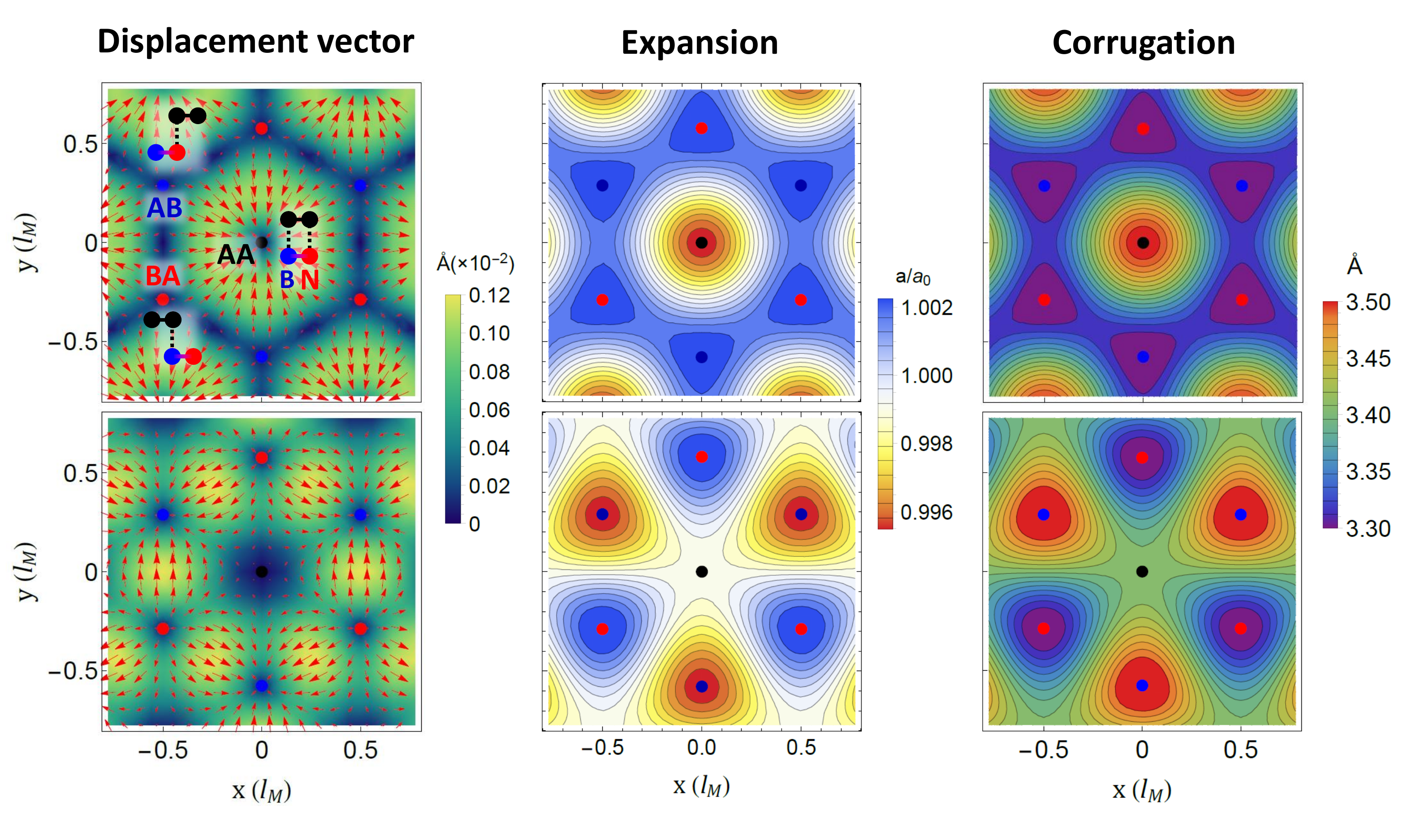}
\end{center}
\vspace{-20pt}
 \caption{Plots of the symmetric (top) and asymmetric (bottom) components of the displacement vector, expansion and corrugation of graphene (defined in Eq. \ref{uvec} and \ref{expand}). To obtain the figures on the top row, we assume $U_{1} = 3$ meV, $z_{1} = 0.01 \angstrom$. In our convention, positive $U_{1}$ places AA at the potential maximum with respect to AB and BA, leading to a shrinking AA region. Positive $z_{1}$ also places AA at the maximum interlayer spacing from the underlying substrate. On the other hand, the bottom row corresponds to $U_{2} = 3$ meV, $z_{2} = 0.01 \angstrom$. Non-zero asymmetric terms account for the inequivalence between AB and BA sites. Positive $U_{2}$ makes BA an energetically favored configuration, while positive $z_{2}$ brings carbon atoms on AB to lie above those on BA. A typical deformation in graphene would be a superposition of both profiles with comparable strengths.}
 \label{deformation}
\end{figure*}

Deformation of the sample results from the minimization of the energy functional, leading to the following equation of motion for the in-plane components:

\begin{widetext}
\begin{equation}
\frac{\varepsilon+\theta(\hat{z}\times\mathds{1})}{\tilde{\varepsilon}^{2}A_{g}}\,
\vec{\nabla}[U_{1}f_{1}(\vec{r}) + U_{2} f_{2}(\vec{r})]
\approx 2\mu_{g}
\bigg[\vec{\nabla}^{2}\vec{u}-\frac{1}{2}(\hat{z}\times\vec{\nabla})
(\partial_{x}u_{y}-\partial_{y}u_{x})\bigg]
+\lambda_{g}\vec{\nabla}(\vec{\nabla}\cdot\vec{u}),
\label{eom}
\end{equation}
\end{widetext}

\noindent Similar to the electronic couplings, the interlayer potential consists of symmmetric and asymmetric components which are quantified by $U_{1}$ and $U_{2}$. We treat the deformation within the first harmonics, which dominates the relaxation profile for small atomic displacements ($|\vec{u}|\ll a$) which accounts for the approximate sign in \ref{eom}. Since the interlayer potential is completely defined by the first harmonics (Eq.~\ref{eom}, LHS), the solution for the displacement vector $\vec{u}(\vec{r}) = u_{x}\,\hat{x} + u_{y}\,\hat{y} + h(\vec{r})\,\hat{z}$ is also constrained by moir\'e periodicity, containing only the first harmonics components. The displacement and the corresponding in-plane expansion of carbon site at $\vec{r}$ is thus given by:

\begin{widetext}
\begin{eqnarray}
\label{uvec}
\vec{u}(\vec{r}) =
\frac{k_{s}|\varepsilon^{3}|}{G^{2}\tilde{\varepsilon}^{2}}\,\bigg[\vec{\nabla}+\frac{\chi_{R}\theta}{\varepsilon}(\hat{z}\times\vec{\nabla})\bigg][U_{1}f_{1}(\vec{r})+U_{2}f_{2}(\vec{r})] + \big[z_{1}f_{1}(\vec{r}) + z_{2}f_{2}(\vec{r})\big]\,\hat{z}, \\
a/a_{0} \approx 1+\frac{1}{2}\text{Tr}[u_{ij}(\vec{r})] = 1 + \frac{1}{2}\bigg[\frac{\partial u_{x}}{\partial x}+\frac{\partial u_{y}}{\partial y}\bigg] = 
1 - 
\frac{k_{s}|\varepsilon^{3}|}{2\tilde{\varepsilon}^{2}}\,[U_{1}f_{1}(\vec{r})+U_{2}f_{2}(\vec{r})],
\label{expand}
\end{eqnarray}
\end{widetext}

\noindent where $U_{1}$ ($U_{2}$) parametrize the inversion-symmetric (asymmetric) components of in-plane deformation, and $z_{1}$ ($z_{2}$) quantifies similar components for the deformation in the out-of-plane direction, i.e. corrugation. Meanwhile, the coefficient $k_{s}$ which relates $\vec{u}(\vec{r})$ with the potentials, and the factor $\chi_{R}$ which sets the magnitude of the curl-like terms are as follows 

\begin{widetext}
\begin{equation}
k_{s} = \frac{2}{3\sqrt{3}\varepsilon^{2}a^{2}(\lambda_{g}+2\mu_{g})}
= 0.029
\,\,\text{meV}^{-1},
\hspace{7pt}
\chi_{R} = \bigg[2+\frac{\lambda_{g}}{\mu_{g}}\bigg]
\approx 2.34,
\label{relaxCoefficients}
\end{equation}
\end{widetext}

\noindent It can be seen in the the subsequent analysis that $k_{s}$ serves as the coefficient which encodes the contribution of in-plane deformation which is parametrized by $U_{1}$ and $U_{2}$, on the resulting gaps. We also assume the following definition for the symmetric and asymmetric components of the deformation:

\begin{widetext}
\begin{equation}
\begin{pmatrix}
U_{0}\\ U_{1}\\ U_{2}
\end{pmatrix}
= \mat{M}
\begin{pmatrix}
U_{AA} \\ U_{AB} \\ U_{BA}
\end{pmatrix}
= \frac{1}{18}
\begin{pmatrix}
6(U_{AA}+U_{AB}+U_{BA})\\
2U_{AA}-U_{AB}-U_{BA}\\
\sqrt{3}(U_{AB}-U_{BA})  
\end{pmatrix},
\label{Ustack}
\end{equation}
\end{widetext}

\noindent in which we take AA stacking point as the origin of our coordinate system, and the matrix elements of $\mat{M}$ are obtained in a way consistent with our definitions of $f_{1}(\vec{r})$ and $f_{2}(\vec{r})$, and in accordance with the convention in the literature \cite{jung2015origin,wallbank2013generic}. The symmetric and antisymmetric profiles are plotted in Fig. \ref{deformation}. We also use similar definition of symmetric and antisymmetric components for the corrugation and moir\'e electronic couplings. Within this choice of coordinates, we can expect from symmetry consideration alone that a sizable primary gap only opens up when there is a relatively large symmetric component in the deformation (the asymmetric terms vanish after doing the mBZ average).  In general, since in our choice of coordinate system the inversion-asymmetric electronic couplings dominate ($\tilde{V} \gg V$), this implies that the dominant contribution to the  global mass term results from symmetric deformations.  Since the secondary Dirac cone gap is not a simple average over the mBZ, we expect both the symmetric and asymmetric contributions to be equally important.

\subsubsection*{Moir\'e couplings and gaps in G/BN}

The dynamics of low-energy electrons in G/BN can be understood as a relativistic Dirac particle subject to three types of perturbations: (1) $H_{0}(\vec{r})$, describing the periodic electrostatic potential, (2) $H_{z}(\vec{r})$, acting as a local mass term which arises from the sublattice asymmetry, and (3) $H_{xy}(\vec{r})$, quantifying the gauge field due to asymmetry in hoppings induced by the interaction with the BN substrate. For graphene electrons on valley K, the equation of motion can be expressed as follows

\begin{widetext}
\begin{align}
\begin{split}
H_{K}(\vec{r}) 
&=\upsilon\vec{p}\cdot\vec{\sigma} +
H_{0}(\vec{r})\mathds{1} +  H_{z}(\vec{r})\sigma_{3}+ \bigg[\frac{\varepsilon(\hat{z}\times \mathds{1})-\theta}{\tilde{\varepsilon}G}\bigg]\vec{\nabla}H_{xy}(\vec{r})\cdot\vec{\sigma},
\label{moireH}
\end{split}
\end{align}
\end{widetext}

\noindent It is important to note that only the $H_{xy}$ contribution is scaled by factors which depend on the twist angle. The term in Eq. \ref{moireH} which is proportional to $\theta$ can be gauged away so that the influence of $H_{xy}$ on the electronic structures is scaled by $\chi_{\theta} = \varepsilon/\tilde{\varepsilon}$. The three pseudospin components can be written in a compact manner using the following matrix notation:

\begin{widetext}
\begin{equation}
\begin{pmatrix}
H_{0}(\vec{r}) & H_{xy}(\vec{r}) & H_{z}(\vec{r})
\end{pmatrix}
=
\begin{pmatrix}
1 & f_{1}(\vec{r}) & f_{2}(\vec{r})
\end{pmatrix}
\mat{W}, 
\end{equation}
\end{widetext}


\noindent where $\mat{W}$ stands for the electronic coupling matrix. Typically each pseudospin term for zero pressure has oscillating amplitudes on the order of $\sim$50~meV. Around AA, DFT parametrizations of the interlayer couplings lead to a dominating inversion-asymmetric electronic coupling ($\tilde{V}\gg V$). In addition, the exponential coefficients for all psuedospin terms also agree within 10\%. Therefore, in this work we develop an analytical model containing only the inversion-asymmetric term $\tilde{V}$, which is scaled according to a single exponential coefficient $\beta$ when pressure is applied on the system. Treating the problem at $\theta = 0^{\circ}$, $\mat{W}$ is given by the following expression:

\begin{widetext}
\begin{equation}
\mat{W} = 
\bigg[ \mathds{1}+
k_{s}
\begin{pmatrix}
0 & 6U_{1} & 6U_{2}\\
0 & U_{1} & -U_{2}\\
0 & -U_{2}& -U_{1}
\end{pmatrix}
\bigg]
\mat{C}\mat{W}_{0}
+
\begin{pmatrix}
0 & 0 & 0\\
-\Gamma_{V}U_{1} & -\Gamma_{B} U_{2}& 0\\
-\Gamma_{V}U_{2} & \Gamma_{B} U_{1} & 0
\end{pmatrix},
\label{rotU}
\end{equation}
\end{widetext}

\noindent which is related to the corrugation matrix $\mat{C}$ and the coupling matrix in a rigid sample $\mat{W}_{0}$:

\begin{widetext}
\begin{equation}
\mat{C} = 
\mat{M}^{-1}
\begin{pmatrix}
e^{-3\beta z_{1}} & 0 & 0 \\ 
0 & e^{-2\sqrt{3}\beta z_{2}} & 0 \\
0 & 0 & e^{2\sqrt{3}\beta z_{2}}\\
\end{pmatrix}
\mat{M},
\hspace{10pt}
\mat{W}_{0}
=
e^{-\beta (z_{0}-z_{r})}
\begin{pmatrix}
0 & 0 & 0 \\ 
0 & 2 & -\sqrt{3}\\
1 & 0 & 0
\end{pmatrix}
\frac{\tilde{V}}{2},
\end{equation}
\end{widetext}

\noindent The coupling terms appearing in $\mat{W}_{0}$ are based on the special ratio between psuedospin components \cite{wallbank2015moire}, while the prefactors for $z_{1}$ and $z_{2}$ are estimated from the full numerical treatment of the Fourier components within the first harmonics. In Eq. \ref{rotU}, the first change in the electronic couplings as captured by the first term, is induced by shifts in the positions of the carbon atoms with respect to the underlying BN substrates. In-plane deformation of graphene brings additional contributions to the electrostatic potentials ($\Gamma_{V}$) and pseudomagnetic fields ($\Gamma_{B}$), both independent on the interlayer spacings.

The primary gap arises primarily from the mass term $\Delta_{p} \sim 2|\omega_{z}|$, while the the secondary gap that appears on the valence band on valley K' at the edges of the mBZ (see Fig. \ref{potStack}): $\vec{K}' = \frac{1}{3}(\vec{G}_{2} + \vec{G}_{3}) = \frac{1}{3}(\vec{G}_{4} + \vec{G}_{5}) = \frac{1}{3}(\vec{G}_{1} + \vec{G}_{6})$ results from an interplay between the pseudospin terms $H_{0}$, $H_{z}$ and $H_{xy}$ \cite{wallbank2013generic}.  These gaps have the following analytical form in a fully relaxed sample:

\begin{widetext}
\begin{eqnarray}
\label{Dp}
\hspace{-22pt}&\Delta_{p}&\hspace{0pt}
= 2\sqrt{3}\tilde{V}e^{-\beta (z_{0}-z_{r})}\bigg[\cosh(2\sqrt{3}\beta z_{2})-e^{-3\beta z_{1}} + k_{s}\big(2U_{1}e^{-3\beta z_{1}}+U_{1}\cosh(2\sqrt{3}\beta z_{2}) + \sqrt{3}U_{2}\sinh(2\sqrt{3}\beta z_{2})\big)\bigg], \\
\label{Ds}
\hspace{-22pt}&\Delta_{s}&\hspace{0pt}
= \frac{\sqrt{3}\tilde{V}}{6}e^{-\beta (z_{0}-z_{r})}\big[7e^{-3\beta z_{1}} + 2\cosh(2\sqrt{3}\beta z_{2}) + 12\sinh(2\sqrt{3}\beta z_{2}) \big] - \Gamma_{G}(\sqrt{3}U_{1}-U_{2}) +\delta_{s},
\end{eqnarray}
\end{widetext}

\noindent where  $\delta_{s}$ is a small additional contribution to the secondary gap which results from the interplay between corrugation and in-plane deformation (assumed here to be zero).  We define $\Gamma_{G} = (\sqrt{3}/2)(\Gamma_V + 2\Gamma_B)$, for $\Gamma_{V,B}= |\varepsilon^{3}|k_{s} \gamma_{V,B}/\tilde{\varepsilon}^{2}$.  The first principle calculations at $\theta = 0^{\circ}$ discussed above, gives a sizable $\Gamma_{G} \approx 5.6 \text{ meV}^{-1}$.  This is important, because for generic band parameters, in the case of $\Gamma_{G} \rightarrow 0$ the bands that form the sDC are possibly mixed with the surrounding moir\'e minibands such that no gap could be observed experimentally.  Since graphene sites have an affinity towards the BA configuration, where one carbon atom sits on top of the boron atom, both $U_{1}$ and $U_{2}$ are expected to acquire positive values $\sim$3 meV, such that the BA domain is enlarged compared to the rigid case. 

The exponential prefactor $e^{-\beta (z_{0}-z_{r})}$ in the primary gap gives the increase in gap arising from the increased coupling between the graphene and BN when the two layers are closer together under pressure.  This multiplies two terms, the first depends only on the corrugations and to leading order is given by the symmetric component of the out-of-plane deformations $\sim  3 \beta z_1$.  The second term proportional to $k_s$ depends on both in-plane and out-of-plane deformations, but for realistic parameters depends only on the symmetric deformation $\sim 3 k_s U_1$.  This shows that to leading order, the primary gap is determined mostly by the symmetric in-plane and out-of-plane deformations and vanishes in the case of rigid graphene.

The secondary gap is finite even for the rigid case without any deformations $\Delta_s = (\sqrt{27}/2) \tilde{V} e^{-\beta (z_{0}-z_{r})}$ that grows with increasing pressure.  The experimental observation that the primary gap is larger than the secondary gap even at zero pressure implies that the second (negative) term in Eq.~\ref{Ds} proportional to $\Gamma_G$ must be of comparable magnitude.  In contrast to the primary gap, the secondary gap is controlled by both the symmetric and antisymmetric components of the deformation i.e. $\sim (U_2 - \sqrt{3}U_1)$ for in-plane deformations, and $\sim ((8 \sqrt{3}/7) z_2 - z_1)$ for corrugations.  In both cases, for the secondary gap to remain constant under pressure, the difference between the asymmetric deformations and symmetric deformations must increase with pressure.       

We note that these experimental observations consistent with DFT expectations.  The EXX-RPA calculations discussed above show that with increasing pressure, (i) The difference between the asymmetric and symmetric components of in-plane deformations increases, while (ii) The asymmetric part of out-of-plane corrugation decreases.  This suggests that the dominant contribution to the flatness of the secondary gap comes from corrugations.  Physically, the AA points (carbon-carbon on BN) start further away form the underlying BN substrate. With increased pressure, they compress more than either the AB (carbon on nitrogen) or BA (carbon on boron).  While all potentials increase with pressure, the BA increases the least making this region expand.  This naturally explains the growing primary gap.  Similarly, under pressure the G/BN spacing decrease at all stacking points (so that the graphene becomes flatter). But the AA points which were further away to begin with decrease faster than either the BA or AB points, which is consistent with the $\Delta_s$ being constant.
\newpage

\subsection{Reverse Engineering}

\subsubsection*{Fixed corrugation, free in-plane deformation} 
\begin{figure*}[h]
\begin{center}
\includegraphics[width=0.72\textwidth]{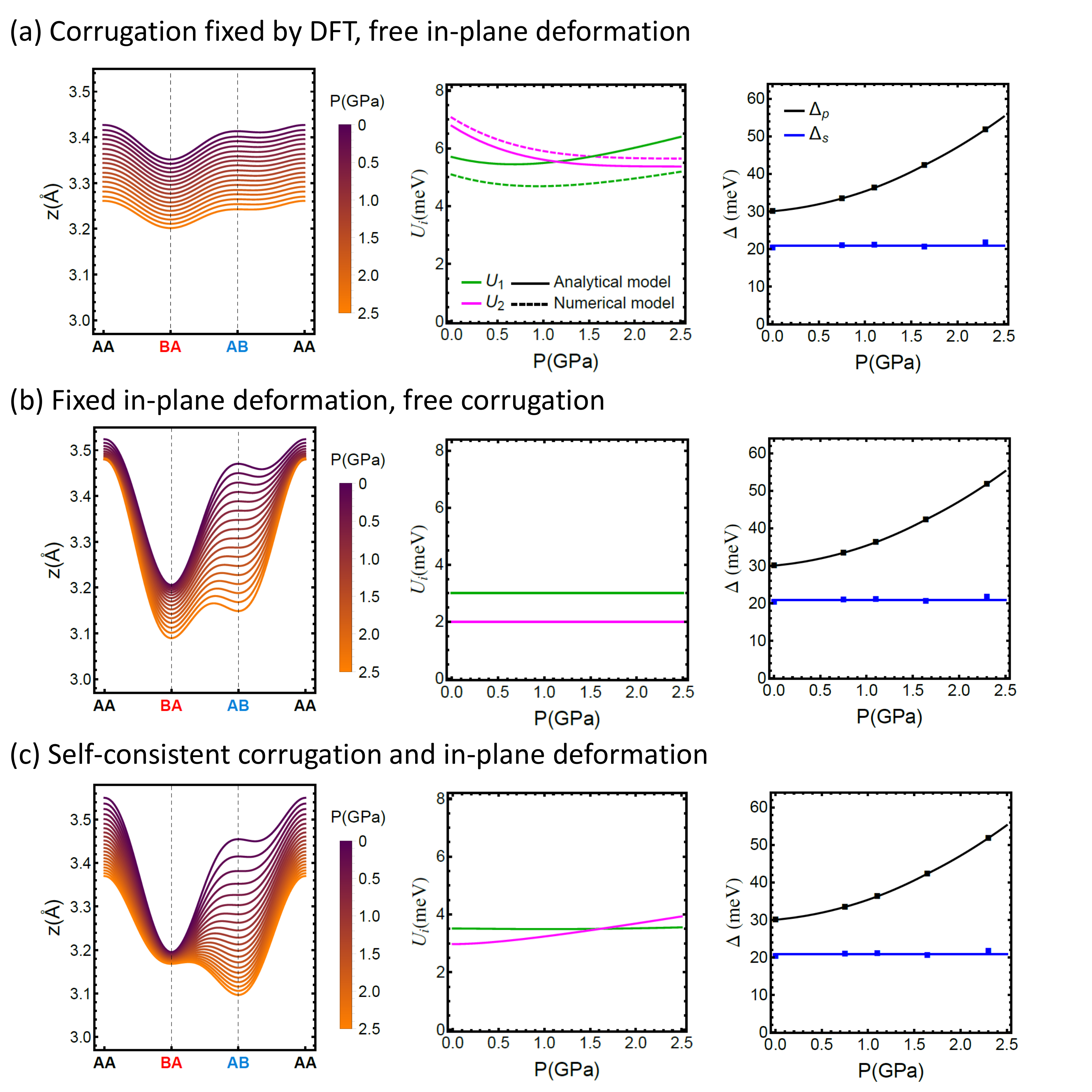}
\end{center}
\vspace{-20pt}
 \caption{
Graphene deformation represented through corrugation profiles and in-plane strain parameters $U_1$ (green) and $U_2$ (magenta), and the associated primary gap $\Delta_p$ (black) and secondary gap $\Delta_s$ (blue) evolving as a function of pressure where the dots represent experimental data. For the analytical model, we assume the following values for the parameters: $\tilde{V} = 18$ meV, $\Gamma_{G} = 10$ meV$^{-1}$, to capture the enhancement due to exchange \cite{jung2015origin}, $\beta = 3.2\,\angstrom^{-1}$ and a reduction in the average spacing of 0.07 $\angstrom/\text{GPa}$ when the problem is treated in a non self-consistent manner. We explore three possible scenarios which would explain the experiment.
(a) Graphene encapsulated between BN when it is aligned with the bottom layer. 
The top BN layer interacts with unequal forces with the graphene sheet at different stacking positions leading to some reduction in the corrugation profile amplitude. The corrugation is obtained from DFT calculations, while the in-plane deformation is calculated within the analytical (solid) and numerical model (dashed, Ref. \cite{jung2015origin}) to reproduce the experimental data. The result is presented in the main text.
(b) In-plane deformation in graphene is assumed to be independent on pressure $(U_{1} = 3\text{ meV, }U_{2} = 2\text{ meV})$, and the corrugation is calculated to fit the experiment. (c) Assuming a uniform pressure across the sample, the corrugation and in-plane deformation are determined self-consistently by pinning their relationship through the interlayer potentials. The solution is obtained by calculating the interlayer potentials consistent with the observed gaps.}
\label{solution}
\end{figure*}

Since the observed secondary gaps are smaller than the primary gaps at all pressures, this implies that $\Gamma_G$ term originating from the in-plane strain-induced pseudomagnetic fields and potentials is large. 
We rely on DFT (EXX+RPA) calculations to determine the corrugation, assuming that the forces exerted by a top BN layer completely fixes the shapes of graphene corrugations under pressure. Since the corrugation is known at all pressures, the in-plane deformation parameters $U_{1}$ and $U_{2}$ are then uniquely determined by the observed primary and secondary gaps.  Here, the graphene sample is relatively flat $\delta z \lesssim 0.1 \angstrom$ and the insensitivity of the secondary gap is accounted by an increase in $U_{1}$ with respect to $U_{2}$ as shown in the main text and in the top row of Fig.~\ref{solution}.

\subsubsection*{Fixed in-plane deformation, free corrugation} 

To explore the role of corrugation in the experimental observation, we consider the opposite scenario where the in-plane deformation is assumed to be constant under pressure.  We fix $U_{1} = 3$ meV and $U_{2} = 2$ meV, comparable to \textit{ab-initio} estimates~\cite{jung2015origin}. The insensitivity of the secondary gap under pressure constrains both the symmetric ($z_{1}$) and asymmetric components ($z_{2}$) of the corrugation, resulting in a significant transformation of the corrugation profile as shown in the middle row Fig.~\ref{solution}.  One striking feature is that the experimental observations can be explained by a decrease in the asymmetric corrugations i.e. the separation between AB and BA becomes smaller under pressure.  Since this contradicts DFT expectations we think this scenario is unlikely.

\subsubsection*{Self-consistent corrugation and in-plane deformation}

Another possible approach to address the gap behaviors is to construct interlayer potentials on G/BN special stacking points, which simultaneously define in-plane and out-of-plane deformation in G/BN and capture the increasing and constant trend in the primary and secondary gaps respectively. We assume a uniform pressure across the sample, and it is expected that there exists a certain potential profile which can account for the observed gaps. Inspired by the DFT results which suggest Lennard-Jones (LJ)-like potential profiles around the minima, we limit our search for the self-consistent potentials $U_{\mathbf{s}}$ which assume the LJ form at each stacking point ($\mathbf{s}\in\{AA,AB,BA\}$), which also leads to an interlayer distance of $z_{\mathbf{s}}(P)$ at pressure $P$:

\begin{widetext}
\begin{equation}
U_{\mathbf{s}}(z) =
\sigma_{\mathbf{s}}\bigg[
\frac{1}{z^{2n_{\mathbf{s}}-1}}-
\bigg(\frac{2n_{\mathbf{s}}-1}{n_{\mathbf{s}}-1}\bigg)
\frac{1}{z_{0,\mathbf{s}}^{n_{\mathbf{s}}}\,z^{n_{\mathbf{s}}-1}}
\bigg]
+ U_{0,\mathbf{s}},
\hspace{10pt}
z_{\mathbf{s}}(P) = \bigg[\frac{1}{2z_{0,\mathbf{s}}^{n_{\mathbf{s}}}} + \sqrt{\frac{1}{2z_{0,\mathbf{s}}^{n_{\mathbf{s}}}}+\frac{4PA_{g}}{(2n_{\mathbf{s}}-1)\sigma_{\mathbf{s}}}}\,
\bigg]^{\frac{1}{n_{\mathbf{s}}}}
\label{Usz}
\end{equation}
\end{widetext}

\noindent The coefficient $\sigma_{\mathbf{s}}$ accounts for the strength of the potentials, and $z_{0,\mathbf{s}}$ denotes the equilibrium distance at zero pressure. The power coefficient $n_{\mathbf{s}}$ determines the behaviors of the profile on the small and large $z$ tails, while $A_{g}$ denotes the area of graphene's unit cell. We therefore try to obtain the best estimates for the potentials which account for the observed experimental trend. The LJ-like potentials which lead to remarkable agreement with the experiment can be found, i.e. $<10\%$ error for each data point, with the fitting parameters are shown in Table~\ref{tabLJ}. 

As expected, this solution suggests the need for asymmetric components of the corrugation to decrease rapidly under pressure.  However, this leads to a striking conclusion: Although BA is initially at a minimum spacing at zero pressure, AB overtakes BA as the closest point from the underlying BN substrate at $P \gtrsim$ 1.5 GPa. This is accompnied by a slight increase in the asymmetric component of the in-plane deformation. Again, in disagreement with the DFT, the self-consistent solution is dominated by the rapid change in the asymmetric component of the corrugations ($z_{AB}-z_{BA}$).  The EXX-RPA \textit{ab initio} potentials (which can also be fit with Lennard-Jones-like functions) are shown in the inset for comparison with the potentials determined self-consistently.

\subsubsection*{Generic features of the deformation}

\begin{table*}
\caption{
Lennard-Jones-like parametrization of the interlayer potentials. We approximate 3 potential parameters for each stacking point $\textbf{s}\in\{AA, AB, BA\}$, given that $n_{\mathbf{s}} = 5$. The resulting potentials $U_{\mathbf{s}}(z)$ are in meV, while $\hat{z}$ is in \angstrom. We also show the potential profiles obtained within EXX+RPA in the inset.}
\vspace{0pt}
\begin{center}
\begin{minipage}{0.27\textwidth}
\includegraphics[width=\textwidth]{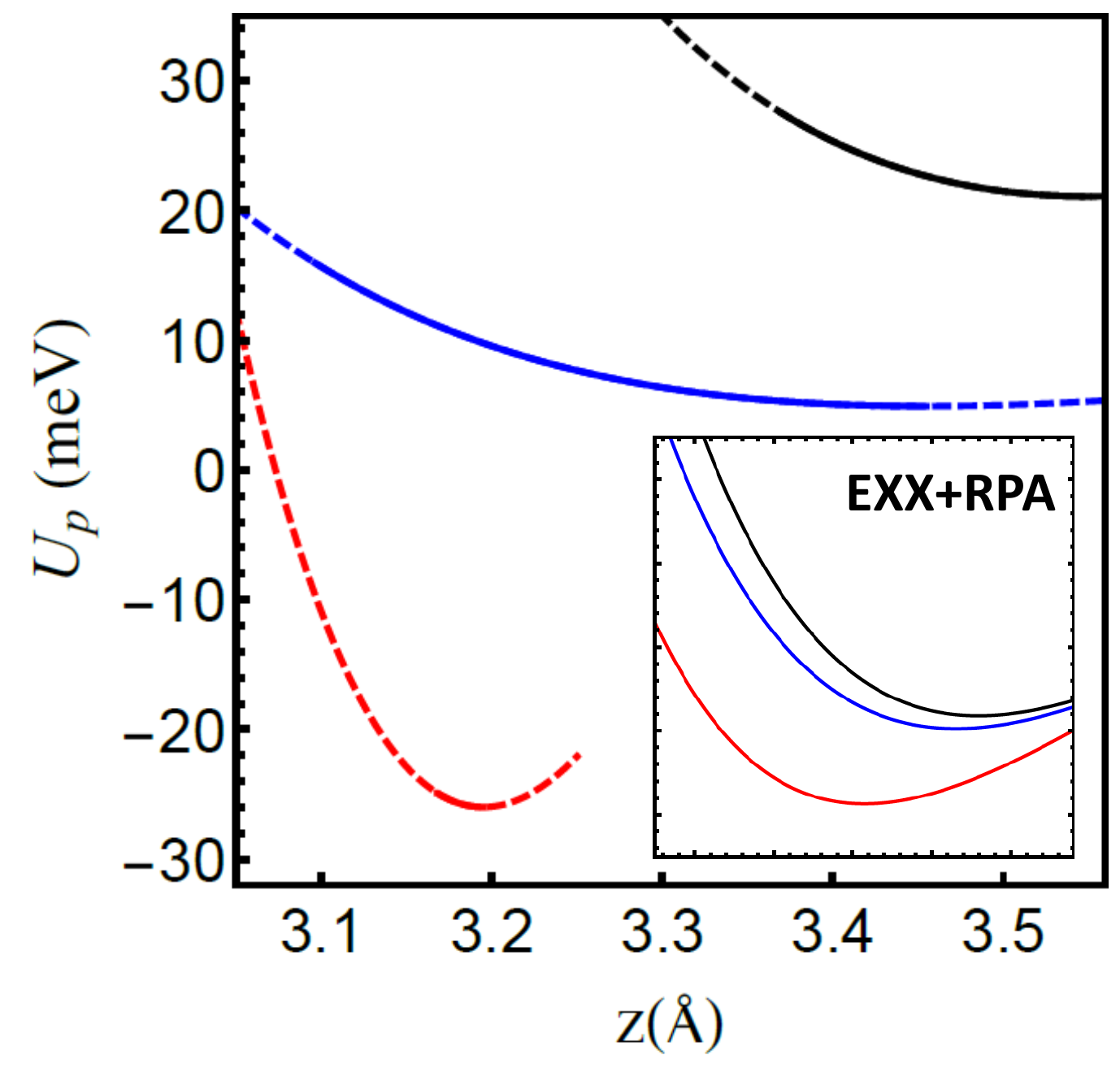}
\end{minipage}
\begin{minipage}{0.55\textwidth}
\begin{center}
{
\begin{tabular}{ C{30pt} | C{70pt} | C{50pt} | C{50pt} N}
$\mathbf{s}$ & $\sigma_{\mathbf{s}}$ (meV $\angstrom^{-5}$) & $z_{0,\mathbf{s}}$ (\angstrom) & $U_{0,\mathbf{s}}$ (meV) &\\[7pt]
\hline
AA & 7.51 & 3.55 & 21.07 &\\[9pt]
AB & 1.75 &  3.45 & 3.19 &\\[9pt]
BA & 22.33 & 3.20  & -- 25.98 &\\[9pt]
\hline
\end{tabular}
}
\end{center}
\end{minipage}
\end{center}
\label{tabLJ}
\end{table*}

While the analysis was done in the simplified semi-analytical model, we have checked that the conclusions remain robust in a full numerical simulation of the first-Harmonic approximation.  The relatively large primary gap suggests there is a strong presence of either in-plane relaxation or corrugation.   The fact that secondary gap is always smaller than the primary gap requires large in-plane deformation at all pressures. Since the increase in the moir\'e couplings as the layers get closer would give larger secondary gaps, the experimental observation of a flat secondary gap behavior implies that one or both of the following must be happening: (1) There is a significant decrease in the corrugation asymmetry $z_{2}$, which places AB closer to BA in the out-of-plane deformation when pressure is applied on G/BN, (2) There is a rapid transformation of the in-plane deformation profile under pressure such that the difference between $U_{1}$ and $U_{2}$ balances the effects of stronger electronic couplings. 

\end{document}